\documentclass[pra,twocolumn,nobalancelastpage,superscriptaddress]{revtex4-1}

\usepackage{amsfonts,amsmath,amssymb,color,graphicx,bm}
\usepackage[shortlabels]{enumitem}
\usepackage[only,llbracket,rrbracket]{stmaryrd}
\usepackage{calc}

\newcommand{\xlabel}[1]{}
\newcommand{\silent}[1]{}

\newcommand{\ket}[1]{\vert#1\rangle}
\newcommand{\bra}[1]{\langle#1\vert}
\newcommand{\vket}[1]{\vert#1)}
\newcommand{\vbra}[1]{(#1\vert}
\newcommand{\str}[2]{S_{\llbracket#1;#2\rrbracket}}
\newcommand{\strb}[2]{\bar{S}_{\llbracket#1;#2\rrbracket}}
\newcommand{\strk}[2]{\bm S_{\llbracket\bm{#1};\bm{#2}\rrbracket}}
\newcommand{\strsemik}[2]{\bm S_{\llbracket{#1};{#2}\rrbracket}}
\newcommand{\ex}[2]{{\llbracket#1;#2\rrbracket}}
\newcommand{\Htwo}[1]{\mathrm{H}^2(#1,\mathrm{U}(1))}
\newcommand{\anyon}[1]{\underline{#1}}

\newcommand{\strDbl}[2]{S_{\ex{#1}{\bar{#2}}-\ex{#1}{#2}}}
\newcommand{\strDblSep}[3]{S_{\ex{#1}{\bar{#2}}{{-}\smash{\raisebox{0.65ex}{\!\!\!{$\scriptstyle#3$}\,}}}\ex{#1}{#2}}}
\newcommand{\strDblBSep}[4]{S_{\ex{#1}{#3}{{-}\smash{\raisebox{0.65ex}{\!\!\!{$\scriptstyle#4$}\,}}}\ex{#1}{#2}}}
\newcommand{\strbDblSep}[3]{\bar{S}_{\ex{#1}{\bar{#2}}{{-}\smash{\raisebox{0.65ex}{\!\!\!{$\scriptstyle#3$}\,}}}\ex{#1}{#2}}}
\newcommand{\strbDblBSep}[4]{\bar S_{\ex{#1}{#3}{{-}\smash{\raisebox{0.65ex}{\!\!\!{$\scriptstyle#4$}\,}}}\ex{#1}{#2}}}
\newcommand{\strkDblSep}[3]{\bm{S}_{\ex{\bm #1}{\bar{\bm #2}}{{-}\smash{\raisebox{0.65ex}{\!\!\!{$\scriptstyle#3$}\,}}}\ex{\bm #1}{\bm #2}}}
\newcommand{\strsemikDblSep}[3]{\bm{S}_{\ex{#1}{\overline{#2}}{{-}\smash{\raisebox{0.65ex}{\!\!\!{$\scriptstyle#3$}\,}}}\ex{\bm #1}{\bm #2}}}

\newtheorem{condition}{Condition}

\begin{document}

\title{Entanglement phases as holographic duals of anyon condensates}

\author{Kasper Duivenvoorden}
\affiliation{JARA Institute for Quantum Information, RWTH Aachen
University, 52056 Aachen, Germany}
\author{Mohsin Iqbal}
\affiliation{JARA Institute for Quantum Information, RWTH Aachen
University, 52056 Aachen, Germany}
\affiliation{Max-Planck-Institute of Quantum Optics, Hans-Kopfermann-Str.\
1, 85748 Garching, Germany}
\author{Jutho Haegeman}
\affiliation{Ghent University, Department of Physics and Astronomy,
Krijgslaan 281-S9, 9000 Gent, Belgium}
\author{Frank Verstraete}
\affiliation{Ghent University, Department of Physics and Astronomy,
Krijgslaan 281-S9, 9000 Gent, Belgium}
\affiliation{Vienna Center for Quantum Science, Universit\"at Wien,
Boltzmanngasse 5, 1090 Wien, Austria}
\author{Norbert Schuch}
\affiliation{Max-Planck-Institute of Quantum Optics, Hans-Kopfermann-Str.\
1, 85748 Garching, Germany}
\begin{abstract}
Anyon condensation forms a mechanism which allows to relate different topological
phases. We study anyon condensation in the framework of Projected
Entangled Pair States (PEPS) where topological order is characterized
through local symmetries of the entanglement.  We show that anyon
condensation is in one-to-one correspondence to the behavior of the
virtual entanglement state at the boundary (i.e., the entanglement spectrum)
under those symmetries, which encompasses both symmetry breaking and symmetry
protected (SPT) order, and we use this to characterize all anyon
condensations for abelian double models through the structure of their
entanglement spectrum.  We illustrate our findings with the $\mathbb Z_4$
double model, which can give rise to both Toric Code and Doubled Semion
order through condensation, distinguished by the SPT
structure of their entanglement. Using the ability of our framework to
directly measure order parameters for condensation and deconfinement, we
numerically study the phase diagram of the model, including direct phase
transitions between the Doubled Semion and the Toric Code phase which are
not described by anyon condensation.  \end{abstract}
\maketitle

\section{Introduction}

The study of topologically ordered phases, their relation, and the
transitions between them has received steadily growing attention in the
last decade. Their
lack of local order parameters, the dependence of the ground space
structure on their topology, and the exotic nature of their anyonic
excitations puts them outside the Landau framework of symmetry breaking and
local order parameters, and thus asks for novel ways of characterizing and
relating different phases, for instance the structure of their ground
space or the nature of their non-trivial excitations (anyons), and the
way in which those are related throughout different phases.

Anyon condensation has been proposed as a mechanism for relating
topological phases~\cite{bais:anyon-condensation}. The main idea is that
some mechanism drives a species $a$ of bosonic anyons to condense into the
vacuum. This, in turn, forces any anyon $b$ which has non-trivial
statistics with $a$ to become confined, as a deconfined $b$ anyon would
have non-trivial statistics with the new vacuum, and moreover leads to the
identification of anyons which differ by fusion with $a$. At the same
time, the relation between anyon types and ground space of a theory
suggests that this condensation is accompanied by a change in the ground
space structure.  The formalism of anyon condensation allows to construct
``simpler'' anyon models from more rich ones, and suggests to think of the
``condensate fraction'' of the condensed anyon as an order parameter for a
Landau-like description of the phase transition. Yet, it is a priori not
clear how such an order parameter should be measured, and existing
approaches describe anyon condensation as a breaking of the
global symmetry of the quantum group or tensor category underlying the
model~\cite{bais:quantum-sym-breaking,bais:hopf-symmetry-breaking-jhep,kitaev:gapped-boundaries,kong:anyon-condensation-tensor-categories}.

Projected Entangled Pair States (PEPS)~\cite{verstraete:mbc-peps} form a
natural framework for the local modelling of topologically ordered
phases~\cite{buerschaper:stringnet-peps,gu:stringnet-peps}. They associate to any
lattice site a tensor which describes both the physical system at that
site, and the way in which it is correlated to the adjacent sites through
entanglement degrees of freedom.  It has been shown that in PEPS,
topological order emerges from a \emph{local} symmetry constraint on the
entanglement degrees of freedom, characterized by a group
action (for so-called double models of groups)~\cite{schuch:peps-sym} or
more generally by Matrix Product Operators for twisted
doubles~\cite{buerschaper:twisted-injectivity} and string-net
models~\cite{sahinoglu:mpo-injectivity,bultinck:mpo-anyons}. In all cases,
both ground states and excitations can be modelled from the very same
symmetries which characterize the local tensors: Group actions and
irreducible representations (irreps) in the former and Matrix Product
Operators with suitable endpoints in the latter
case~\cite{schuch:peps-sym,buerschaper:twisted-injectivity,sahinoglu:mpo-injectivity,bultinck:mpo-anyons}.
Yet, it has been observed that the entanglement symmetry of the tensors is
not in one-to-one correspondence with the topological order in the system:
By
adding a suitable deformation to the fixed point wavefunction, the system
can be driven into a phase transition which is consistent with a
description in terms of anyon
condensation~\cite{schuch:topo-top,haegeman:shadows,marien:fibonacci-condensation,fernandez:symmetrized-tcode}.
This raises the question: What is the exact relation between topological
phase transitions in tensor networks and anyon condensation, and can we
explain this transition ``miscroscopically'' using the local symmetries in
the tensor network description?

In this paper, we derive a comprehensive framework for the explanation,
classification, and study of anyon condensation in PEPS. Our framework
explains and classifies anyon condensation in terms of the different
``entanglement phases'' emerging at the boundary under the action of the
local entanglement symmetry of the tensor, and provides us with the tools
to explicitly study the behavior of order parameters measuring
condensation and confimement of anyons.  More specifically, we show that
the symmetry constraint in the entanglement degrees of freedom of the tensor gives
rise to a corresponding ``doubled'' symmetry in the fixed point of the
transfer operator, this is, in the entanglement spectrum at the boundary.
Anyon condensation can then be understood in terms of the different phases at
the boundary, this is, the symmetry breaking pattern together with a
possibly symmetry-protected phase of the residual unbroken symmetry. We
give necessary and sufficient conditions for the condensation of anyons in
abelian double models in terms of the symmetry at the boundary, and show
that this completely classifies all condensation patterns in double models
of cyclic groups, giving rise to all twisted $\mathbb Z_N$ double models.
We also show that these conditions allow to independently derive the anyon
condensation rules described above, providing a tensor network derivation
of these conditions.  The central idea is to relate anyon condensation and
confinement to the behavior of string order parameters, which in turn can
be related to symmetry breaking and symmetry-protected order, and combine
this with the constraints arising from the positivity of the boundary
state.  

We illustrate our framework by discussing all possible phases
which can be obtained by condensation from a $\mathbb Z_4$ double model,
which can give rise to Toric Code, Doubled Semion, and trivial phases.
Specifically, we show that the Toric Code and Double Semion can exhibit
the same symmetry breaking pattern at the boundary, yet are distinguished
by different SPT orders, corresponding to the condensation of a charge or
a dyon (a combined charge-flux particle), respectively, and thus a
different string order parameter.  Finally, we apply our framework to
numerically study topological phases and the transitions between them
along a range of different interpolations. Specifically, the
interpretation of condensation and confinement in terms of string order
parameters allows us to directly measure order parameters for the
different topological phases, namely condensate fractions and order
parameters for deconfinement, which allow us to study the nature and order
of the phase transitions. Our framework also allows us to set up
interpolations between the Toric Code and Double Semion phase, which are a
priori not related by anyon condensation, and we find that depending on
the nature of the interpolation, we can either find a second-order
simultaneous confinement-deconfinement transition, or a first-order
transition not characterized by anyon condensation.

The paper is structured as follows:  In
Sec.~\ref{sec:peps-definition-anyons}, we introduce PEPS, explain how
topological order and topological excitations are modelled within this
framework, and define condensation and confinement in PEPS models.
Sec.~\ref{sec:classification} contains the classification of anyon
condensation and confinement through the behavior of the boundary: We
start by giving the intuition and the main technical assumption, then
derive the conditions imposed by the symmetry structure and positivity of
the boundary state, and finally show that this classification gives rise
to the well-known anyon condensation rules. In
Sec.~\ref{sec:ZN-and-tw-dbl}, we apply this classification to the case of
$\mathbb Z_N$ quantum doubles and show that it precisely gives rise to all
twisted $\mathbb Z_M$ double models. Finally, in
Sec.~\ref{sec:example-Z4}, we illustrate our framework with a detailed
discussion of the condensation from a $\mathbb Z_4$ double, and study the
corresponding family of models and the transitions between them
numerically.

\section{Symmetries in PEPS and anyons
\label{sec:peps-definition-anyons}}

In this section, we will first introduce the general PEPS framework. We
will then  explain how certain symmetries in PEPS naturally lead to
objects defined on the entanglement degrees of freedom which behave like
anyonic excitations.  The natural question is then to understand the
conditions under which these objects describe observable anyons, or
whether they fail to do so by either leaving the state invariant
(condensation) or by evaluating to zero (confinement) in the thermodynamic
limit.

We will focus our discussion to the case of abelian groups; however,
several of our arguments in fact apply to general groups, and even beyond
that for so-called MPO-injective PEPS; we will discuss these aspects in
Sec.~\ref{sec:conclusions}.

\subsection{PEPS, parent Hamiltonians, and excitations}

\begin{figure}[b]
\centering
\includegraphics[scale=0.7]{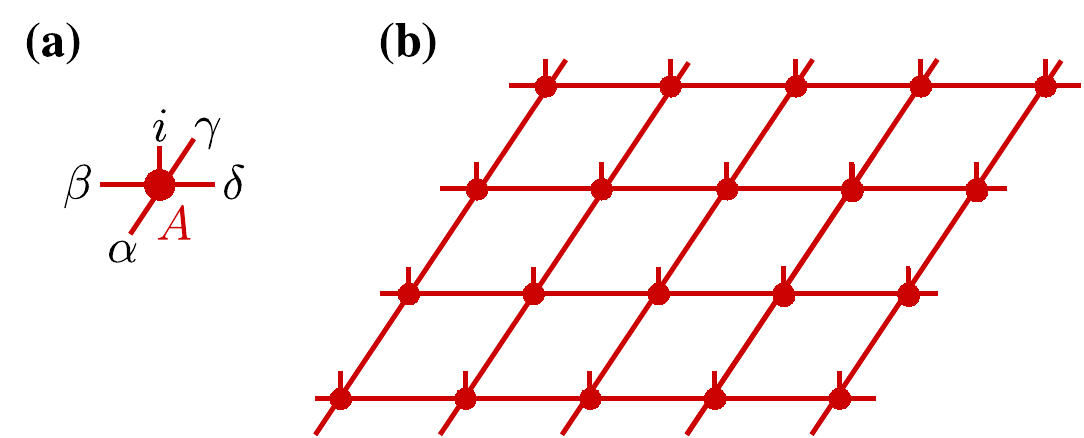}
\flushleft
\caption{
\label{fig:peps-def}
\textbf{(a)} PEPS tensor $A$ with five indices. 
\textbf{(b)} The PEPS wavefunction is built up by contracting the virtual
indices $\alpha,\dots,\delta$ of the PEPS tensors as indicated by
connected lines.
}
\end{figure}

Let us start by introducing Projected Entangled Pair States (PEPS).
We focus on a translational invariant system on a square
lattice with periodic boundary conditions, where we take the system size
to infinity.  PEPS are
constructed from a local tensor $A^i_{\alpha\beta\gamma\delta}$, where
$i=1,\dots,d$ is the \emph{physical} index and
$\alpha,\beta,\gamma,\delta=1,\dots,D$ are the \emph{virtual} indices, and
$D$ is called the bond dimension.  Graphically, they are depicted as a
sphere with five legs, one for each index, cf.~Fig.~\ref{fig:peps-def}a;
equivalently, we can consider $A=\sum A^{i}_{\alpha\beta\gamma\delta}
\ket{i}\bra{\alpha,\beta,\gamma,\delta}$
as a linear map from virtual to physical system.  The tensor $A$ is then
arranged on a square lattice, Fig.~\ref{fig:peps-def}b, and adjacent virtual indices
are contracted (i.e., identified and summed over), which is graphically
depicted by connecting the corresponding legs.  We thus finally obtain a
tensor $c_{i_1\dots i_N}$ which only has physical indices, and thus
describes a quantum
many-body state $\ket\Psi = \sum c_{i_1\dots i_N}\ket{i_1,\dots,i_N}$. 
A useful property of PEPS is the possibility to block sites -- we can take
the tensors on some $k_1\times k_2$ patch and define them as a new tensor
$A'$ with correspondingly larger $D$. This allows us to restrict
statements about properties of localized regions to fixed-size (e.g.,
single-site or overlapping $2\times 2$) patches.

To any PEPS, one can naturally associate a family of parent Hamiltonians
which have this PEPS as their exact zero-energy ground
state~\cite{perez-garcia:parent-ham-2d,schuch:peps-sym}. Such a
Hamiltonian is a sum of local terms $h$, each of which ensures that the
state ``looks locally correct'' on a small patch, i.e., as if it had been
built from the tensor $A$ on that patch.  This is accomplished by choosing
$h\ge0$ such that $h$ is zero on the physical subspace spanned by the
tensors on that patch (for arbitrary virtual boundary conditions) and
positive otherwise; note that by choosing a sufficiently large patch, it
is always possible to find a non-trivial such Hamiltonian (the dimension
of the allowed physical subspace scales with the boundary, 
while the available degrees of freedom scale with the volume).

Clearly, the global PEPS wavefunction is a zero-energy state and thus a
ground state of the parent Hamiltonian $H=\sum h\ge0$.  At the same time,
conditions on $A$ are known under which this ground state is unique (in a
finite volume)~\cite{perez-garcia:parent-ham-2d}: Specifically, it is
sufficient if the map from the virtual to the physical system described
by $A$ (possibly after blocking) is injective; equivalently, this means
that the full auxiliary space can be accessed by acting on the physical
space only, i.e., that one can apply a linear map which ``cuts out'' a
tensor and gives direct access to the auxiliary indices.

\begin{figure}[b]
\includegraphics[width=\columnwidth]{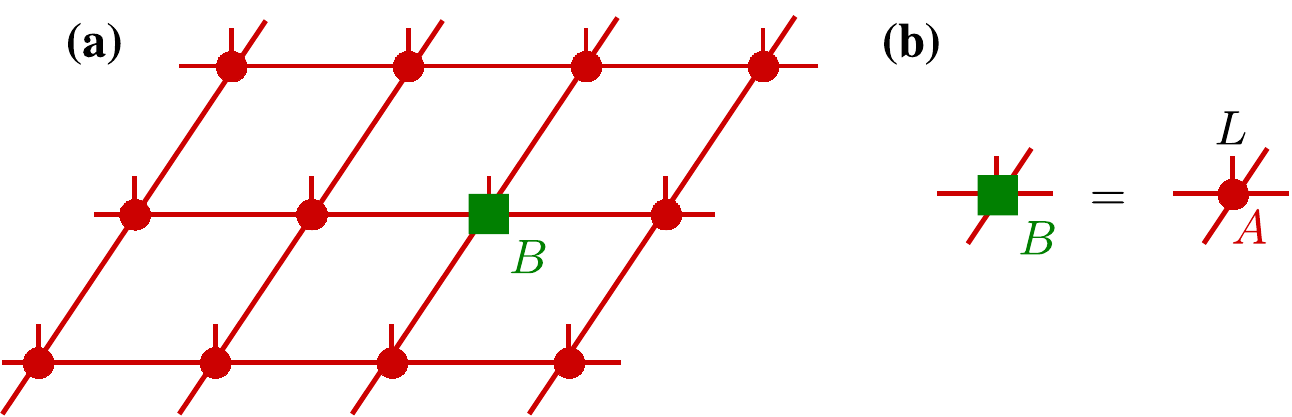}
\caption{
\label{fig:local-excitations}
\textbf{(a)} Replacing one tensor by a different tensor $B$ results
in a localized excitation, i.e., which is not detected by the parent
Hamiltonian anywhere else. \textbf{(b)} An excitation is topologically
trivial if $B$ can be obtained from $A$ by acting with a map $L$ on the
physical system.
}
\end{figure}

Parent Hamiltonians naturally give rise to the notion of localized
excitations, this is, states whose energy differs from the ground state
only in some local regions.  To this end, one replaces some tensors by
``excitation tensors'' $B$, while keeping the original tensor $A$ everywhere
else, cf.~Fig.~\ref{fig:local-excitations}a.  For \emph{injective} PEPS,
these are in fact the only possible localized excitations, since due to
the one-to-one correspondence between virtual and physical system any
tensor $B\ne A$ will yield an increased energy w.r.t.\ the parent
Hamiltonian~\cite{perez-garcia:parent-ham-2d}. 

A key question in the context of this work is when an excitation is
topologically non-trivial.  We will use the following definition: 
An excitation is topologically trivial exactly if it can be created
(with some non-zero probability) by acting locally on the system, i.e., if
there exists a linear (not necessarily unitary) map $L$ on the physical
system which will create that excitation on top of the ground state, this
is, which transforms $A$ to $B$.  It is now straightforward to see that
for an injective PEPS, all localized excitations
(Fig.~\ref{fig:local-excitations}a) are topologically trivial: Injectivity
implies that $A$ (as a map from virtual to physical system) has a
left-inverse $A^{-1}$, and thus $L:=BA^{-1}$ will act as $LA=B$, i.e.,
create the desired excitation locally, as shown in
Fig.~\ref{fig:local-excitations}b.

\subsection{G-injective PEPS and anyonic excitations}

Let us now turn towards PEPS which can support topologically non-trivial
excitations.  To this end, we consider PEPS which are no longer injective,
but enjoy a virtual symmetry under some group action, 
\begin{equation}
\label{eq:g-invariance}
A=A(\bar U_g\otimes \bar U_g  \otimes U_g\otimes U_g)
\end{equation}
with $U_g$ a unitary representation of some finite group $G\ni g$; we will
denote such tensors as $G$-invariant.  Graphically, this is expressed as
\begin{equation}
\label{eq:def-g-invariant}
\raisebox{-2em}{\includegraphics[scale=0.70]{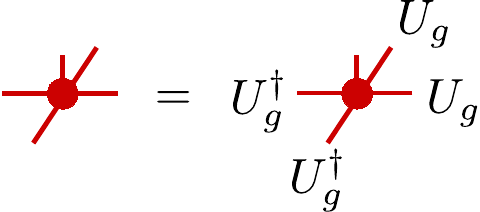}}\quad,
\end{equation}
where we use the convention that matrices act on the indices from left to
right and down to up, such that $\bar U_g$ in Eq.~(\ref{eq:g-invariance})
turns into $U_g^\dagger$.  An important property of $G$-invariance is its
stability under concatenation: When grouping together several $G$-invariant
tensors, the resulting block is still $G$-invariant, as the $U_g$ and
$U_g^\dagger$ on the contracted indices exactly cancel out.  In the
following, we will focus on abelian groups (though various parts of the
discussion generalize to the non-abelian case), and denote the neutral
element by $e\in G$.

If $G$-invariance is the only symmetry of the tensor $A$, i.e., if $A$ is
injective on the subspace left invariant by the symmetry, we call $A$
$G$-injective.  The parent Hamiltonians of $G$-injective PEPS have a
topological ground space degeneracy and can support anyonic
excitations~\cite{schuch:peps-sym}, as we will also discuss in the
following.  We will generally assume that the tensors are $G$-injective,
since otherwise we might be missing a symmetry, likely rendering the
discussion incomplete.

\subsubsection{Electric excitations}

In order to understand how these excitations look like, let us consider
again the possible localized excitations w.r.t.\ the parent Hamiltonian.
As we have seen earlier, any state where one tensor has been replaced by a
different tensor $B$ is by construction a localized excitation.  In the
injective case, any such $B$ could be obtained by acting locally on the
physical degrees of freedom, rendering the excitation topologically
trivial.  However, it is easy to see that this is no longer the case for
$G$-invariant tensors:  Local operations
(Fig.~\ref{fig:local-excitations}b) can only produce tensors $B$
which are again $G$-invariant, i.e., transform trivially under the
action of the symmetry group, since it is exactly the invariant virtual subspace which
is accessible by acting on the physical indices.  In contrast, $B$'s which
transform non-trivially can no longer be created locally, and thus are
topologically non-trivial excitations.  It is natural to label these
excitations by irreducible representations $\alpha(g)\in\mathbb C$ of
the abelian symmetry group $G$, this is, we can write 
\begin{equation}
\label{eq:B-eq-oplus-Balpha}
B=\sum_\alpha B_\alpha\ ,
\end{equation}
where 
\[
\includegraphics[scale=0.70]{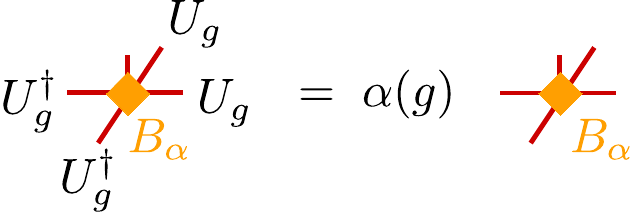}
\]
This is, any such excitation can be understood as a superposition of
excitations with fixed $\alpha$, and we will focus on excitations with a
fixed $\alpha$ in the following.  These excitations will be denoted as
\emph{electric excitations} with \emph{charge $\alpha$}. (For non-abelian
groups, we would require instead that each $B_\alpha$ is supported on the
irrep $\alpha$ of the group action.)

It is straightforward to see that for $G$-injective PEPS, the topological
part of the excitation is fully characterized by $\alpha$: In case
$B_\alpha$ itself is injective on the irrep $\alpha$, this is immediate
since it can be transformed into any other $B_\alpha'$ by locally acting
on the physical index; in case $B_\alpha$ is not injective, the same can
be done by acting on a $3\times 3$ block centered around $\alpha$ (due to
$G$-injectivity, this allows to access all degrees of freedom at the
boundary in the irrep $\alpha$).

In the following, we will focus our attention on electric excitations of
the form 
\begin{equation}
\label{eq:Balpha-eq-C-Ralpha}
\includegraphics[scale=0.70]{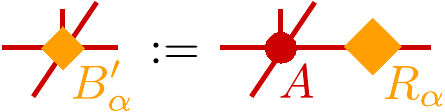}
\end{equation}
where $R_\alpha$ (the yellow diamond) transforms as 
$R_\alpha U_g = \alpha(g) U_g R_\alpha$;
the general case will be discussed in Appendix~\ref{sec:app:dressed}.

An important point to note about electric excitations is that for any
system with periodic boundaries, they must come in pairs (or groups) which
together transform trivially under the symmetry action, i.e., have total
trivial charge, since otherwise the state would vanish.

\subsubsection{Magnetic excitations}

\begin{figure}[t]
\centering
\includegraphics[scale=0.7]{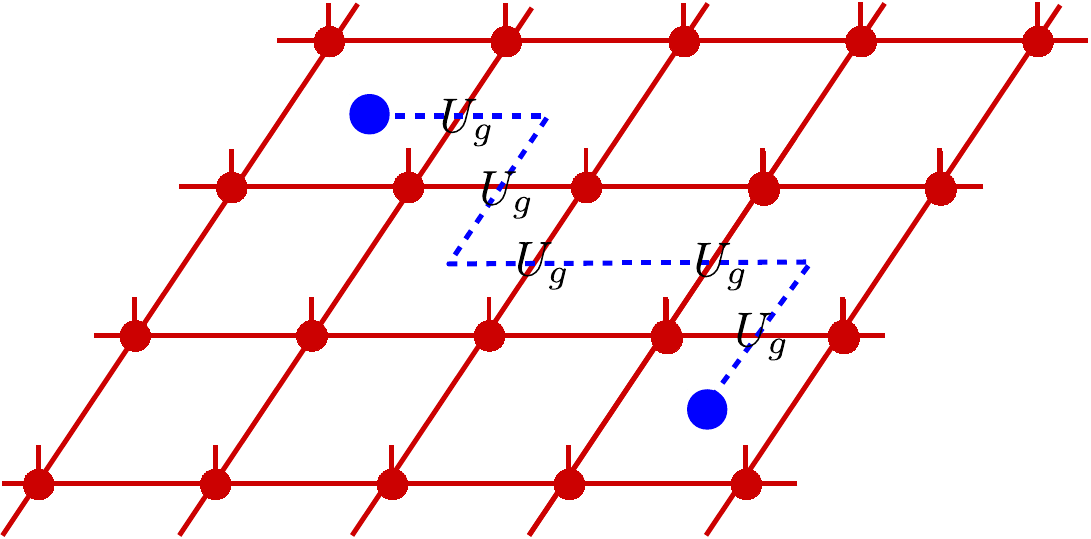}
\flushleft
\caption{
\label{fig:excitation-string-pair}
A string of $U_g$ actions on the lattice can be freely moved
[Eq.~\eqref{eq:pulling-through}], making it invisible to the parent
Hamiltonian except at its endpoints. It therefore describes a pair of
topological excitations.
}
\end{figure}

For injective PEPS, locally changing tensors was the only way to obtain
localized excitations, due to the one-to-one correspondence of physical
and virtual system~\cite{perez-garcia:parent-ham-2d}. For $G$-injective
PEPS, however, there exist ways to non-locally change the tensor network
without creating an excitation, or only creating a localized
excitation~\cite{schuch:peps-sym}. To this end, note that
Eq.~\eqref{eq:def-g-invariant} can be reformulated as
\begin{equation}
\label{eq:pulling-through}
\raisebox{-1.5em}{\includegraphics[scale=0.65]{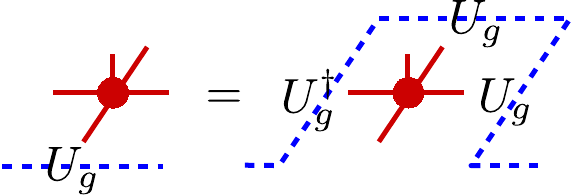}}
\mbox{\ \ and\ \ }
\raisebox{-1.5em}{\includegraphics[scale=0.65]{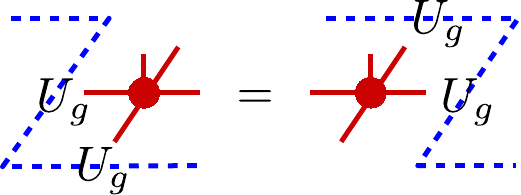}}
\end{equation}
and rotated versions thereof.
This has the natural interpretation of the $U_g$ and $U_g^\dagger$ forming
strings (symbolized by the dashed blue lines above), which can be freely
moved through the lattice (``pulling though condition'').  (Whether $U_g$
or $U_g^\dagger$ has to be used depends on the orientation of the string
relative to the lattice~\cite{schuch:peps-sym}.) Thus, any string of
$U_g$'s is naturally invisible to the parent Hamiltonian, as it can be
moved away from any patch the parent Hamiltonian acts on. Indeed, if
$G$-injectivity holds, one can use the equivalence of physical and virtual
system on the invariant subspace to prove that such strings are the only
non-local objects which cannot be detected by the parent
Hamiltonian~\cite{schuch:peps-sym}. This yields a natural way to build
localized excitations by placing a string of $U_g$'s with
open ends on the lattice, as illustrated in
Fig.~\ref{fig:excitation-string-pair}: Any such string can only be
detected at its endpoints, thereby forming a localized excitation. These
excitations are topological by construction, since by
acting on the endpoints alone, we are not able to create such a string.
At the same time, using $G$-injectivity one can prove that the endpoints
can always be detected in a finite system. Thus, we arrive at a second
type of topological non-trivial excitations, namely strings of $U_g$'s
with an endpoint,
\[
\includegraphics[scale=0.65]{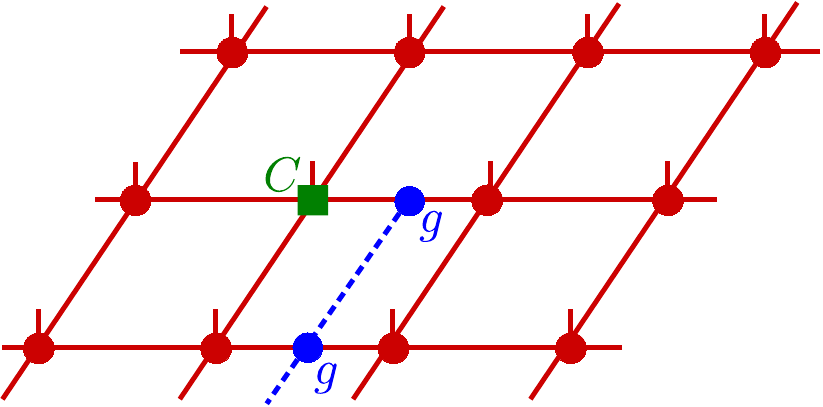} \quad .
\] 
(We have followed the notation introduced in
Fig.~\ref{fig:excitation-string-pair}, where blue dots denote $U_g$ or
$U_g^\dagger$, and the dashed blue line highlights the string formed.)
Again, $C$ is an arbitrary $G$-invariant tensor which can be used to dress
the endpoint with an arbitrary topologically trivial excitation; under
blocking, it can always be assumed to only sit on a single site as shown.
Again, given periodic boundaries any such string must end in a second
anyon (or more generally the strings emerging from several anyons can fuse
as long as the corresponding group elements multiply to the identity).

We will denote these excitations as \emph{magnetic excitations} with
\emph{flux $g$}.

\subsubsection{Dyonic excitations}

Beyond electric and magnetic excitations, it is also possible to combine
the two into a so-called \emph{dyon} which is of the form
\begin{equation}
\label{eq:general-anyon}
\raisebox{-3em}{\includegraphics[scale=0.65]{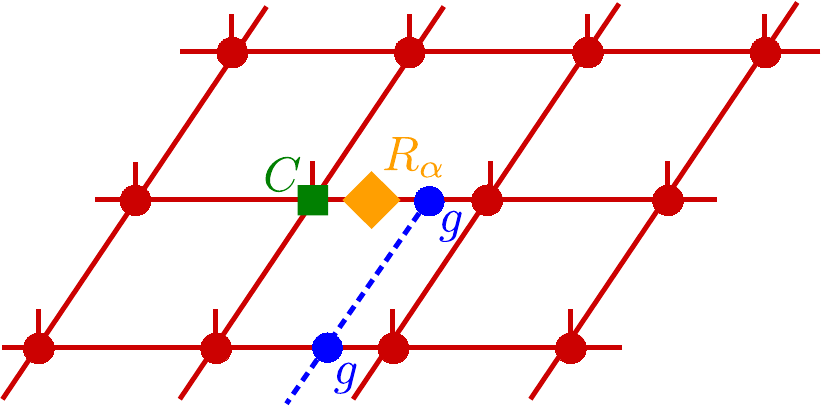}}
\end{equation}
Note that we have made the choice that the irrep $R_\alpha$ sits on the
same leg at which the $U_g$-string ends. While this choice is arbitrary,
it is related to any other endpoint, e.g.\ one where the string ends on
the leg before $R_\alpha$, by a \emph{local} $U_g$-string, i.e., a pair of
magnetic excitations, which can be created locally and can thus be
accounted for by an appropriate choice of $C$, or even incorporated in
$R_\alpha$.

A general anyonic excitation is thus up to local modifications labeled by
a tuple $g$ and $\alpha$; we denote the anyon by $\ex{g}{\alpha}$, and an
anyon string with the two conjugate anyons $\ex{g}{\alpha}$ and
$\ex{g}{\bar{\alpha}}$ at its endpoints by $\strDbl{g}{\alpha}$.

\subsubsection{Braiding statistics}

Let us briefly comment on the braiding statistics of these excitations;
we refer to Ref.~\cite{schuch:peps-sym} for details.
Any physical procedure for moving anyons will result in the $U_g$-string
being pulled along the path. Thus, a half-exchange of two identical anyons
transforms 
\[
\includegraphics[scale=0.6]{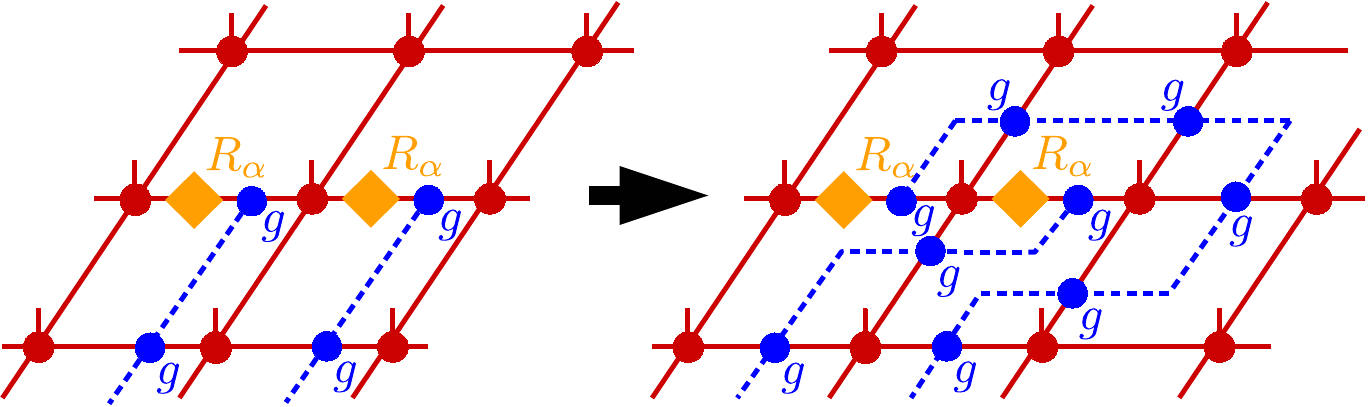}
\]
(where for simplicity we have set $C=A$, as it transforms trivially).
Straightening the string by pulling it through the right excitation
requires to commute $g$ with $R_\alpha$, which gives rise to a phase
$\alpha(g)$; since the resulting two crossing strings are identical to two
non-crossing strings, we thus obtain a overall phase of $\alpha(g)$ due to
the half exchange. 

Similarly, full exchange of two different anyons $\ex{g}{\alpha}$ and
$\ex{h}{\beta}$ gives rise to two such exchanges, and thus to a mutual
statistics $\alpha(h)\beta(g)$ for a full exchange.

We therefore see that the strings defined this way indeed exhibit the same
statistics as $D(G)$, the quantum double model of
$G$~\cite{kitaev:toriccode,schuch:peps-sym}.

\subsection{Virtual level vs.\ observable excitations: Condensation and confinement}

\subsubsection{Anyon condensation and confinement}

It is suggestive to assume that this is the complete picture, and
$G$-injective PEPS always exhibit an anyon theory given by the quantum
double $D(G)$. However, it has by now been understood that this is not the
case~\cite{schuch:topo-top}: By adding a physical deformation $\Lambda$ to
the tensor, $A\rightarrow \Lambda A$, one can drive the system towards a
product state, eventually crossing a phase transition.  E.g., in the toric
code this induces string tension (or more precisely loop fugacity), which
eventually leads to the breakdown of topological
order~\cite{castelnovo:tc-tension-topoentropy}.

This is directly related to the question as to whether the objects which
we have just identified as anyonic excitations on the \emph{virtual} level
actually describe observable anyons in the thermodynamic limit, and in the
limit of large separation between the individual anyons. While, as we have
argued, one can prove~\cite{schuch:peps-sym} that the endpoints of a
virtual string $\strDbl{g}{\alpha}$ correspond to observable excitations,
this only applies in a finite volume.  However, it is perfectly possible
that---depending on the choice of $A$---new behavior emerges in the
thermodynamic limit, which is reflected in a non-trivial environment
imposed on a virtual anyon string $\strDblSep{g}{\alpha}{\ell}$ (with
$\ell$ the separation between the endpoints) which can prevent it from
describing an observable anyonic excitation as $\ell\rightarrow\infty$.
This can happen in at least two distinct ways: Either the environment
transforms trivially under $\strDblSep{g}{\alpha}{\ell}$, in which case the
PEPS with $\strDblSep{g}{\alpha}{\ell}$ still describes the ground
state, or the environment is orthogonal to $\strDblSep{g}{\alpha}{\ell}$,
in which case the state has norm zero and is thus unphysical.

\begin{figure}[t]
\centering
\includegraphics[scale=0.5]{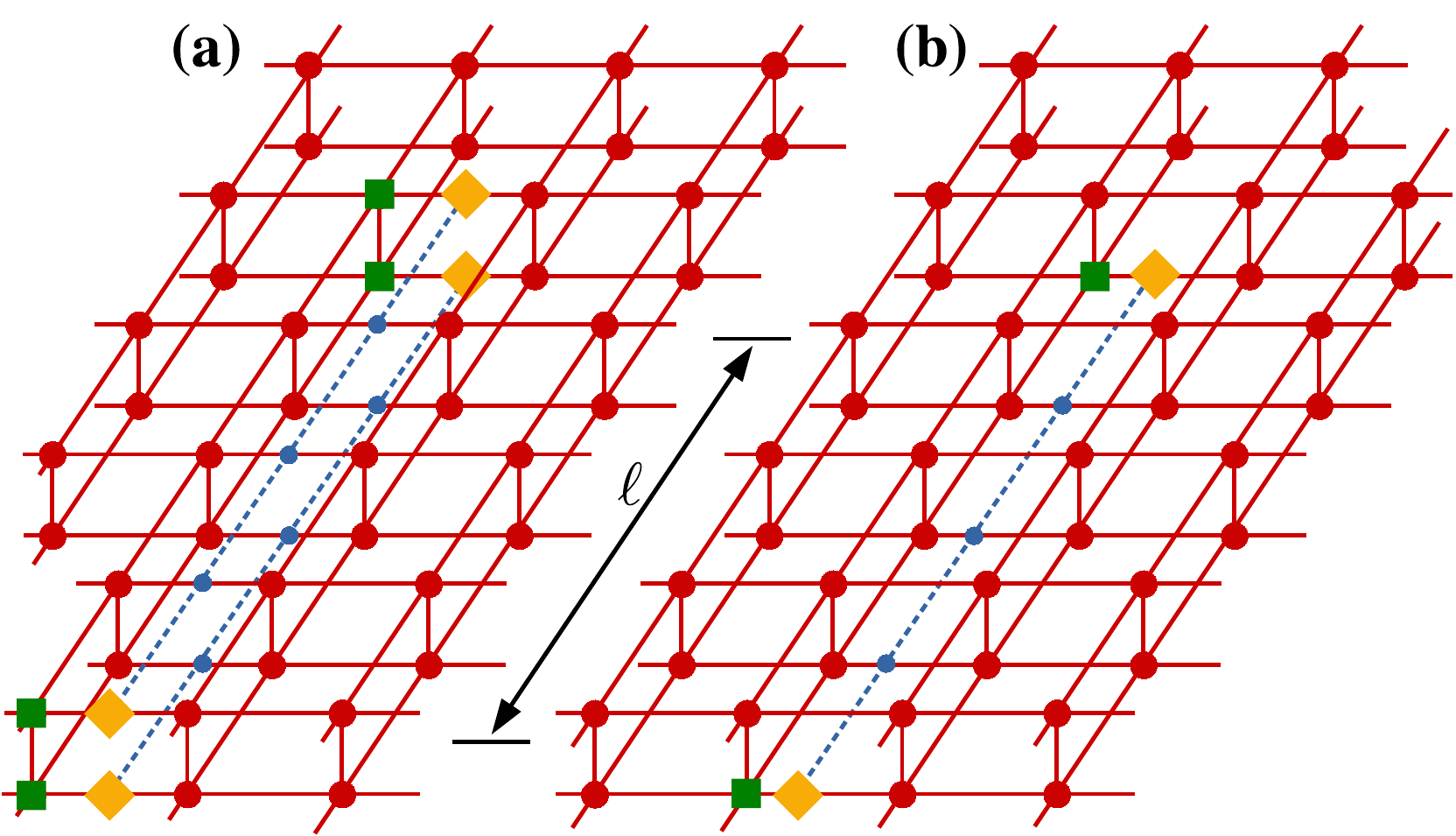}
\caption{
\label{fig:cond-conf}
Tensor networks for detecing \textbf{(a)} confinement (the network for the
norm of the state evaluates to zero) and \textbf{(b)}~condensation (the
network for the overlap with the ground state evaluates to non-zero, both
as $\ell\rightarrow\infty$) for a general anyon pair of the form
Eq.~(\ref{eq:general-anyon}). 
}
\end{figure}

We will thus distinguish two different ways in which non-trivial virtual
excitations $\str{g}{\alpha}$ might fail to describe observable anyonic
excitations:
\\[0.5ex]
\emph{1. Confinement:} The state $\ket{\psi[\strDblSep{g}{\alpha}{\ell}]}$
of the system with an anyon string does not describe a properly
normalizable quantum state, i.e., 
\begin{equation}
\label{eq:confinement-def}
\langle\psi[\strDblSep{g}{\alpha}{\ell}]
\ket{\psi[\strDblSep{g}{\alpha}{\ell}]}\rightarrow 0
\mbox{\ as\ }N,\ell\rightarrow\infty\ 
\end{equation}
where first the system size $N$ and then the separation $\ell$ is taken to
infinity.  The expectation value in Eq.~(\ref{eq:confinement-def})
corresponds to the tensor network in Fig.~\ref{fig:cond-conf}a, this is,
the expectation value of the string operator
$\strbDblSep{g}{\alpha}{\ell}\otimes\strDblSep{g}{\alpha}{\ell}$ in the
double layer ket+bra tensor network.
\\[0.5ex]
\emph{2.\ Condensation:} $\ket{\psi[\strDblSep{g}{\alpha}{\ell}]}$ is not 
orthogonal to the ground state $\ket\psi$ in the thermodynamic limit,
\begin{equation}
\label{eq:condensation-def}
\langle\psi\ket{\psi[\strDblSep{g}{\alpha}{\ell}]}\ne0
\quad\mbox{as}\quad N,\ell\rightarrow\infty\ ,
\end{equation}
i.e., the individual endpoints are not distinguished any more from the
ground state by a topological symmetry, and thus differ from it at most in
local properties.  The corresponding tensor network is shown in
Fig.~\ref{fig:cond-conf}b and corresponds to the expectation value of
the string operator $\openone\otimes\strDblSep{g}{\alpha}{\ell}$.  

In the remainder of this paper, we will explore the conditions under which
condensation and confinement occurs in PEPS models, and provide a
classification of the possibly ways in which this can happen.

\subsubsection{Condensation, confinement, and string order parameters}

In order to understand condensation and confinement of
anyons in PEPS models, we need to assess the behavior of overlaps 
$\langle\psi[\strbDblBSep{g'}{\alpha'}{\bar\alpha'}{\ell}]
\ket{\psi[\strDblSep{g}{\alpha}\ell]}$,
corresponding to string operators $\strDblSep{g}{\alpha}\ell\otimes
\strbDblBSep{g'}{\alpha'}{\bar\alpha'}{\ell}$ on the virtual level,
cf.~Fig.~\ref{fig:cond-conf}, in the thermodynamic limit and as
$\ell\rightarrow\infty$. In what follows, we will assume $C=C'=A$ for
simplicity; we discuss how to adapt the arguments to the general case in
Appendix~\ref{sec:app:dressed}.  

It is instrumental to introduce the transfer operator 
\[
\mathbb T:=
\raisebox{-5em}{\includegraphics[scale=0.6]{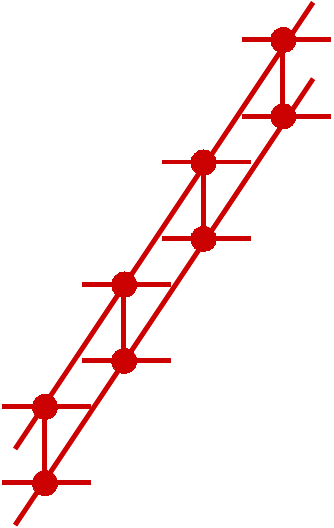}}
\]
which is a completely positive map (from left to right) acting on a
one-dimensional chain of $D$-level systems; if we disregard complete
positivity, we can equally think of $\mathbb T$ as a map on a 1D chain of
$\mathbb C^D\otimes \mathbb C^D$ systems.  In the following, we will
restrict to the case of hermitian $\mathbb T$ (corresponding e.g.\ to a
system with combined reflection and time-reversal symmetry), which in
particular implies that the left and right fixed points of $\mathbb T$ are
equal.  

Let us now see how the symmetry of the tensor $A$ is reflected in the
transfer operator.  $G$-invariance of the $A$ is inherited by $\mathbb T$,
which thus enjoys the symmetries $[\mathbb T,U^{\otimes N}\otimes
\openone]=[\mathbb T,\openone \otimes \bar{U}^{\otimes N}]=0$ (with
$N\rightarrow\infty$ the system size); this is, $\mathbb T$ carries an
on-site $\bm G :=G\times G$ symmetry with representation ${\bm U}_{\bm g} =
U_g\otimes U_{g'}$, with $\bm g\equiv (g,g') \in
\bm G$.  The irreps of $\bm G$ are given by ${\bm\alpha}((h,h'))=
{\bm\alpha}((h,e)) {\bm\alpha}((e,h'))\equiv \alpha(h)\bar{\alpha}'(h')$,
where $\alpha(\cdot):={\bm\alpha}((\cdot,e))$ and
$\alpha'(\cdot):={\bar{\bm\alpha}}((e,\cdot))$ are irreps of $G$; there is
thus a correpondence between irreps of $\bm G$ and pairs of irreps of $G$,
and we will write $\bm \alpha = (\alpha,\alpha')$.  The trivial irrep
will be denoted by $1$. Finally, we define
$\strkDblSep{g}{\alpha}{\ell}:= \strDblSep{g}{\alpha}{\ell}\otimes
\strbDblBSep{g'}{\alpha'}{\bar{\alpha}'}{\ell}$. Generally,
we will stick to the convention that we use boldface letters for objects
living on ket+bra.

In terms of the transfer operator, we can now re-express our quantities of
interest for condensation and confinement as expectation values of
$\strkDblSep{g}{\alpha}{\ell}$ in some left and right fixed
points $(\rho_L|$ and
$|\rho_R)$ of $\mathbb T$
\[
\langle\psi[\strDblBSep{g'}{\alpha'}{\bar\alpha'}{\ell}]
\ket{\psi[\strDblSep{g}{\alpha}{\ell}]}
=
(\rho_L|
\strkDblSep{g}{\alpha}{\ell}
|\rho_R)\;,
\]
where we assume $(\rho_L|\rho_R)=1$.
[We use round brackets $|\cdot)$ to denote vectors on the joint ket+bra
virtual level.] The $|\rho_\bullet)$ can also be understood as 
operators acting between ket and bra level, in which case we will denote
them by $\rho_\bullet$.  Specifically, $\rho_L\rho_R$ has been shown to
exactly reproduce the entanglement spectrum of a bipartition of the
system~\cite{cirac:peps-boundaries}, and thus any statement about the
$\rho_\bullet$ translates into a property of the entanglement spectrum.
Note that $\strkDblSep{g}{\alpha}{\ell}$ is formed exactly by a string of
symmetry operations and terminated by irreps of the doubled symmetry group
$\bm G\equiv G\times G$, i.e., a string order parameter, and it is thus
suggestive to understand the condensation and confinement of anyons by
studying the possible behavior of string order parameters for the group
$\bm G$.

\section{\label{sec:classification}
Classification of string order parameters and condensation}

The following section presents the core result of the paper: 
We classify all different behaviors which the string operators
$\strkDblSep{g}{\alpha}{\ell}$ in a $G$-invariant PEPS can exhibit by
relating them to the classification of symmetry-protected (SPT) phases in
one dimension, as given by the fixed point of the transfer operator.
We start in Sec.~\ref{sec:class:intuition} by explaining the intuition why
the classification of anyon behaviors should be related to the
classification of 1D phases.  In Sec.~\ref{sec:class:fptstruct} we explicitly
state the technical assumptions made (specifically, the form of the fixed
point space). 
Secs.~\ref{sec:class:symbreaking}--\ref{sec:class:positivity}
contain the classification: In Sec.~\ref{sec:class:symbreaking}, we 
study the structure of symmetry breaking of the fixed point space and
show that the endpoints 
$\strkDblSep{g}{\alpha}{\ell}$ decouple as $\ell\rightarrow\infty$,
allowing us to restrict to semi-infinite strings in the following; 
in Sec.~\ref{sec:class:symbr-decoupling}, we derive the constraints
imposed by the symmetry breaking on the anyons and show how it allows to 
decouple anyon pairs;
in
Sec.~\ref{sec:class:spts}, we make the connection between the behavior of
anyons and the SPT structure of the fixed points, and in
Sec.~\ref{sec:class:positivity}, we show that there exists an additional
non-trivial restriction on the SPTs which can appear as fixed points of
$\mathbb T$, and thus to the possible anyon behavior, arising from the
(complete) positivity of $\mathbb T$. Finally, in
Sec.~\ref{sec:subsection-anyon-condensation-rules}, we show that the
conditions derived in the preceding sections precisely give rise to the
known anyon condensation rules.

\subsection{Intuition
\label{sec:class:intuition}}

Let us first present the intuition behind this classification.  To this
end, we use that we are interested in gapped phases and thus the system is
short-range correlated: This suggests that the fixed point of the transfer
operator $\mathbb T$ is short range correlated as well, and thus has the
same structure as the ground state of a local Hamiltonian with the
identical symmetry $[\mathbb T,{\bm U}_{\bm g}^{\otimes N}]=0$.  

Let us now consider the different phases of such a Hamiltonian.  We first
restrict to the the regime of Landau theory, where phases are classified
by order parameters, i.e., irreps of the symmetry group. Depending on the
phase, different irreps will have zero or non-zero expectation values,
which implies condensation [for a non-zero expectation value of an
irrep $(\alpha,e)$ with $\alpha\ne e$] and confinement [for a vanishing
expectation value of an irrep $(\alpha,\alpha)$] of charges,
corresponding to broken diagonal or unbroken non-diagonal symmetries,
respectively. On the other hand, assuming a mean-field ansatz (which is
exact in a long-wavelength limit), we find that strings of group actions
either create a domain wall (for a broken symmetry) or act trivially (for
an unbroken symmetry), relating the symmetry breaking patterns also to the
condensation and confinement of magnons.  We thus see that the
condensation and confinement of electric and magnetic excitations
corresponds to Landau-type symmetry breaking in the fixed point of the
transfer operator, as observed in Ref.~\cite{haegeman:shadows}. As we will
see in the following, this picture becomes more rich when we go beyond
Landau theory and allow for SPT phases: These phases are not captured by
mean-field theory and are rather characterized by
the behavior or string order parameters, i.e., strings of group actions
terminated by order parameters, which give rise to condensation and
confinement of dyonic excitations.

\subsection{The assumption: Matrix Product fixed points
\label{sec:class:fptstruct}}

We start by stating our main technical assumption:
\emph{The fixed point space of $\mathbb T$ (possibly after blocking) is
spanned by a set of injective Matrix Product States (MPS), which are related
by the action of the symmetry group.}

Let us be more specific. Let ${\bm i}=(i,i')$ denote a joint ket+bra index
of the blocked transfer operator.  Then, we assume there exists a set of
matrices $M^{\bm i,\bm c}$ which describe distinct MPS
\[
\vket{\rho_{\bm c}}=
\raisebox{-0.75em}{\includegraphics[scale=.9]{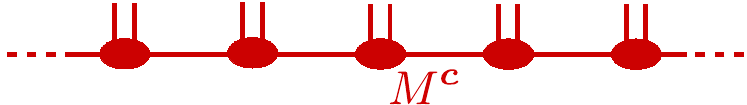}}
\]
on a finite chain with periodic boundary conditions.
We require that these MPS fulfill the following conditions:
\begin{enumerate}
\item
\label{item:assumptions:approx}
The $\vket{\rho_{\bm c}}$ span the full fixed point space of $\mathbb T$. (This
is, evaluating any quantity of interest either in the fixed point space of
$\mathbb T$ or in $\mathrm{span}\{\vket{\rho_{\bm c}}\}$ yields the same result
in the thermodynamic limit.)
\item 
The $\vket{\rho_{\bm c}}$ are injective, i.e., $\mathbb E_{\bm c}^{\bm c}$ has a
unique eigenvalue with maximal magnitude, where $\mathbb E_{\bm c}^{\bm c'}:=
\sum_{\bm i} M^{{\bm i},\bm c} \bar{M}^{ {\bm i},\bm c'}$ is the mixed transfer operator
of the MPS. W.l.o.g., we choose to normalize  $M^{ {\bm i},\bm c}$
such that $\lambda_{\mathrm{max}}(\mathbb E_{\bm c}^{\bm c})=1$.
\item  For each $\bm c$ and $\bm g$, there is a $\bm c'$ such that
$\bm{U_g}\ket{\rho_{\bm c}}=\ket{\rho_{\bm c'}}$, and for each pair $\bm
c$, $\bm c'$, there is a corresponding $\bm g$. (Here and in the following,
we use $\bm {U_g}$ as a shorthand for the global symmetry action ${\bm
U}_{\bm g}^{\otimes N}$ whenever the meaning is clear from the context.)
\end{enumerate}
\vspace*{0.7ex}
Note that we make no assumption that the $\rho_{\bm c}$ are positive, and
in fact in many cases the fixed point space cannot be spanned by positive
\emph{and} injective MPS. 

Assumption~\ref{item:assumptions:approx} is the main technical assumption
here. Note that to some extent a similar assumption underlies the
classification of phases of 1D
Hamiltonians~\cite{chen:1d-phases-rg,schuch:mps-phases}, where the ground
space is approximated by MPS as well: While this is motivated by the known
result that MPS can approximate ground states of finite systems
efficiently~\cite{hastings:arealaw,verstraete:faithfully,schuch:mps-entropies},
also in that scenario it is yet unproven whether this rigorously implies
that MPS are sufficiently general to classify phases in the thermodynamic
limit. 

Assumptions 2 and 3 can be replaced by the weaker assumption that the
fixed point space is spanned by \emph{some} MPS, together with the
assumption that we are not missing any symmetries. Specifically, given an
MPS with periodic boundary conditions, it can be brought into a standard
form (possibly involving blocking of sites)  where it can be understood as
a superposition of distinct injective MPS $\vket{\rho_{\bm c}}$ (possibly with
size-dependent amplitudes)~\cite{perez-garcia:mps-reps,cirac:mpdo-rgfp}.
While the $\vket{\rho_{\bm c}}$ are not necessarily fixed points of
the transfer operator themselves, such as in the case of an
antiferromagnet where the transfer operator acts by permuting the $\vket{\rho_{\bm
c}}$, they will be fixed points of the transfer operator obtained after
suitable blocking.
Since, as we will see in a moment, cross-terms between different
$\vket{\rho_{\bm c}}$ vanish when computing physical quantities of interest, we
can instead work with a fixed point space spanned by the
$\vket{\rho_{\bm c}}$~\footnote{%
Note that this does not imply that the fixed point space is actually
spanned by the ${\vket{\rho_{\bm c}}}$. In fact, it is easy to see that this
would require extra conditions such as rotational invariance, since e.g.\
a transfer operator projecting onto a GHZ-type state would have a unique
fixed point (the GHZ state) which is not an injective MPS.}, corresponding
to Assumption 2.

\begin{figure}[b]
\includegraphics[scale=0.6]{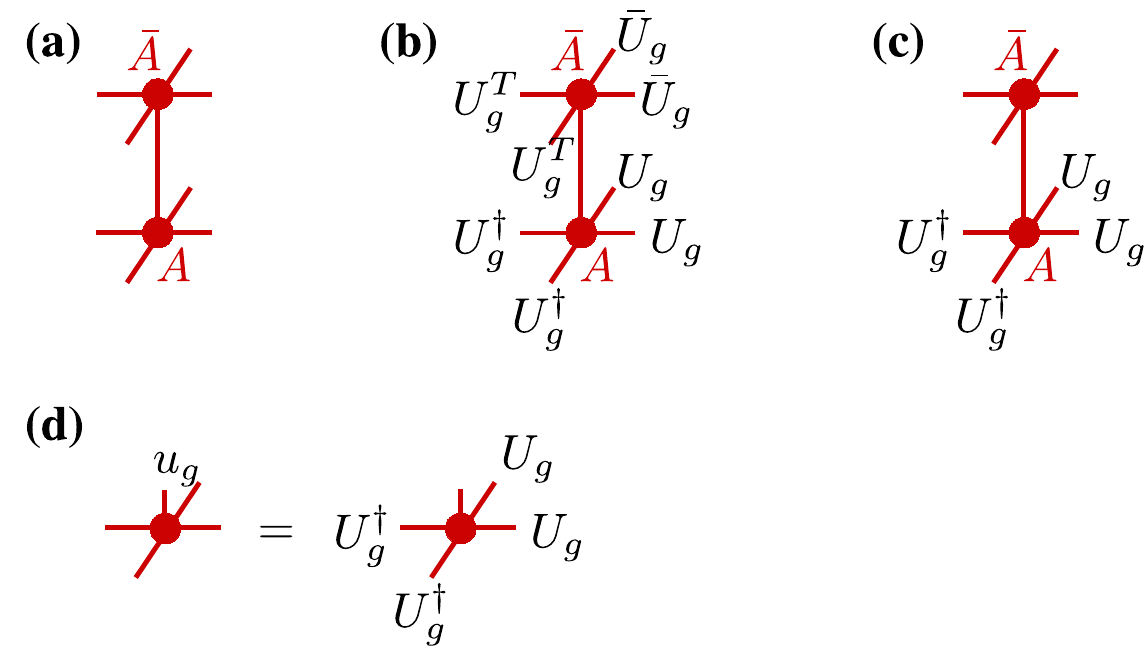}
\caption{
\label{fig:symaction-types}
Possible symmetries in the single-site tensor of the transfer operator.
\textbf{(a)} Joint ket+bra single-site tensor. \textbf{(b)}~Single-layer
symmetry, related to topological order, Eq.~(\ref{eq:g-invariance}).
\textbf{(c)} Double-layer symmetry.  \textbf{(d)} Local encoding of a
global physical symmetry $u_g$ related to the symmetry (c).  Degeneracy of
the transfer operator under such a joint symmetry implies breaking of the
physical symmetry~\cite{rispler:peps-symmetrybreaking}.
}
\end{figure}

Assumption 3 can be justified by requiring that any degeneracy is due to
some symmetry of the transfer operator---otherwise, it would be an
accidental degeneracy and thus not stable against perturbations.  Since
the transfer operator has itself a Matrix Product structure, any symmetry
of the transfer operator must be encoded locally, i.e., it will show up as
a symmetry of the single-site ket+bra object shown in
Fig.~\ref{fig:symaction-types}a~\cite{perez-garcia:inj-peps-syms}.  There
can be two distinct types of such symmetries: Those which act identically
on ket and bra layer, shown in Fig.~\ref{fig:symaction-types}b for on-site
symmetries, and those which only act on one layer, shown in
Fig.~\ref{fig:symaction-types}c.  (Symmetries which act on the two layers
in distinct ways can be split into a product of the former two symmetries,
cf.~the argument at the beginning of Sec.~\ref{sec:class:symbreaking}.)
Symmetries which only act on one layer correspond to topological
symmetries of the PEPS tensor, such as those of
Eq.~(\ref{eq:g-invariance}), and thus need to be incorporated into the
description from the very beginning.  Symmetries acting identically on ket
and bra layer, on the other hand, give rise to a non-trivial physical
symmetry action through the identity in Fig.~\ref{fig:symaction-types}d and
thus correspond to a global physical symmetry of the system; since their
corresponding symmetry sectors are degenerate in the transfer operator, they are
susceptible to physical perturbations which lead to symmetry
breaking~\cite{rispler:peps-symmetrybreaking}, and we can therefore
assume that the system is in one of the symmetry-broken sectors, in which
all fixed points are related by the action of the topological symmetry.
This in particular includes breaking of translational symmetry, which
warrants that we can obtain injective tensors by blocking sites.  Note
that it is conceivable that different symmetry-broken sectors are
described by a different condensation scheme (a simple example can be
obtained by coupling different deformations of the system to an Ising
model).

\subsection{Symmetry breaking structure
\label{sec:class:symbreaking}}

In this section, we clarify the symmetry breaking structure of the fixed
point space, and show that the relevant expecation values do not depend on
which vector in the fixed point space we choose. 

To this end, consider the set of $\vket{\rho_{\bm c}}$ satisfying the
three assumptions just laid out.  For each $\bm c$, let $\bm H_{\bm
c}:=\{\bm h\in \bm G\,:\;\bm U_{\bm h}\vket{\rho_{\bm c}}=\vket{\rho_{\bm
c}}\}$. It is clear that $\bm H_{\bm c}\subset \bm G$ is a subgroup of
$\bm G$; furthermore, for $G$ abelian  $\bm
H_{\bm c}$ is independent of $\bm c$, since for any $\bm h\in \bm H_{\bm
c}$ and $\bm g\in \bm G$ s.th.\ 
$\bm U_{\bm g}\ket{\rho_{\bm c}}=\ket{\rho_{\bm c'}}$,
\[
\ket{\rho_{\bm c'}} = \bm U_{\bm g}\ket{\rho_{\bm c}} =
 \bm U_{\bm g}\bm U_{\bm h}\ket{\rho_{\bm c}} =
 \bm U_{\bm h}\bm U_{\bm g}\ket{\rho_{\bm c}} = 
 \bm U_{\bm h}\ket{\rho_{\bm c'}}\ ,
\]
and we write $\bm H\equiv \bm H_{\bm c}$.

What is the structure of $\bm H$?  To this end, consider arbitrary
$\gamma_{\bm c}$ s.th.\ $\rho:=\sum \gamma_{\bm c} \rho_{\bm c}\ge0$.  For
any $\bm h=(h,h') \in \bm H$, we have that $\rho = U_h\rho
U_{h'}^\dagger$, and thus
\[
\rho^2 = \rho\rho^\dagger = (U_h\rho U_{h'}^\dagger)(U_{h'}\rho
U_{h}^\dagger) = U_h\rho^2 U_h^\dagger\ ,
\]
and thus $[\rho^2,U_h]=0$.  Since $\rho\ge0$, this implies that
$[\rho,U_h]=0$ as well, or
\begin{equation}
\label{eq:sec3-symbr:Uh-rho-Uh}
\rho = U_h\rho U_h^\dagger\ ,
\end{equation}
and similarly $\rho = U_{h'}\rho U_{h'}^\dagger$. Now choose $\gamma_{\bm
c}$ s.th.\ $\rho=\sum\gamma_{\bm c}\rho_{\bm c}=\mathbb{T^\infty}(\openone)$,
the fixed point of $\mathbb T$ obtained when starting from $\openone$, and
pick some $\bm c_0$.  Then, for sufficiently small $\epsilon\ge0$,
$\sigma':=\openone+\epsilon(\rho_{\bm c_0}+\rho_{\bm c_0}^\dagger)\ge0$ and 
$\sigma'':=\openone+i\epsilon(\rho_{\bm c_0}-\rho_{\bm c_0}^\dagger)\ge0$,
and thus $\rho':=\mathbb T^\infty(\sigma')$ and $\rho'':=\mathbb
T^\infty(\sigma'')$ are both positive fixed points and therefore satisfy 
Eq.~(\ref{eq:sec3-symbr:Uh-rho-Uh}), which implies that also 
$\rho_{\bm c_0}=\tfrac{1}{2\epsilon}[(\rho'-\rho)-i(\rho''-\rho)]$ 
satisfies
$\rho_{\bm c_0} = U_{h}\rho_{\bm c_0} U_{h}^\dagger$. We thus find that
whenever $(h,h')\in \bm H$, we must also have that $(h,h)\in \bm H$ and
$(h',h')\in \bm H$.  

Now consider a general element $(k\ell,k)\in \bm H$. Then, $(k,k)\in \bm
H$, and thus $(\ell,e)=(k\ell,k)\cdot(k,k)^{-1}\in \bm H$. It follows that
$K\ni k$ and $L\ni \ell$ form groups, and since $(\ell,e)\in\bm
H\,\Rightarrow (\ell,\ell)\in\bm H$, $L\subset K$. 
\begin{condition} 
\label{condition:1-structure-of-H}
The conserved symmetry $\bm H$ is
isomorphic to a direct product $K\times
L$ with $L\subset K$, where $K$ labels the the diagonal and $L$ the
off-diagonal symmetry, i.e., $H\ni\bm h = (k\ell,k)$ with $k\in K$ and
$\ell\in L$. 
\end{condition}
To distinguish it from the ket/bra product, we will denote the
diagonal/off-diagonal product by $\bm H = K \boxtimes L$.

Let us now consider the evaluation of a anyonic string order parameter
$\strkDblSep{g}{\alpha}{\ell}$ inside general left and right boundary
conditions $\sum \lambda_{\bm c}^{l/r}|\rho_{\bm c})$.
This results in a sum over terms of the form
\begin{equation}
\label{eq:sec3:Occ'-def}
O_{\bm c}^{\bm c'}:=
\raisebox{-2.8em}{\includegraphics[scale=0.6]{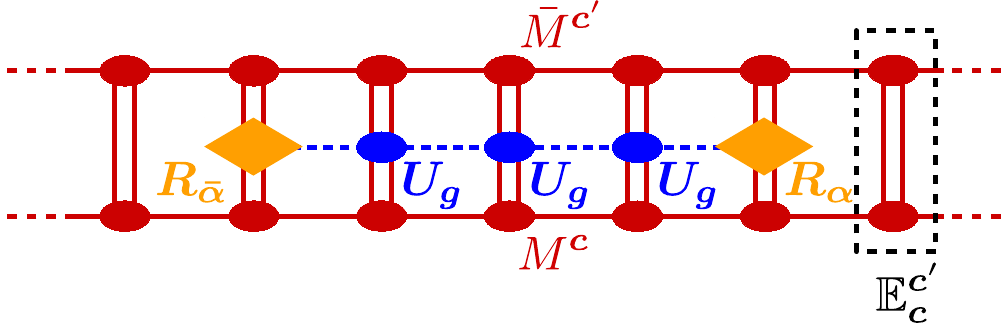}}\ ,
\end{equation}
where we supress the dependency of $O_{\bm c}^{\bm c'}$ on $\bm\alpha$ and
$\bm g$.
In case $\bm c\ne \bm c'$, the largest eigenvalue of the mixed transfer operator
$\mathbb E_{\bm c}^{\bm c'}$ is strictly smaller than one (a
straightforward application of Cauchy-Schwarz, see e.g.\ Lemma~8 of
Ref.~\cite{fernandez-gonzalez:uncle-long}), and thus, 
$O_{\bm c}^{\bm c'}\rightarrow 0$ exponentially as
$N\rightarrow\infty$, i.e., only terms with $\bm c=\bm c'$ survive in
the thermodynamic limit.  In case $\bm c=\bm c'$, we use that
$\vket{\rho_{\bm c}} = \bm U_{\bm h}\vket{\rho_{{\bm c}_0}}$ for some
fiducial ${\bm c}_0$ with $\bm h\in \bm G$,
and thus 
\begin{equation}
\label{eq:sec3:Occ-indep-of-c}
O_{\bm c}^{\bm c}=
\raisebox{-3.6em}{\includegraphics[scale=0.6]{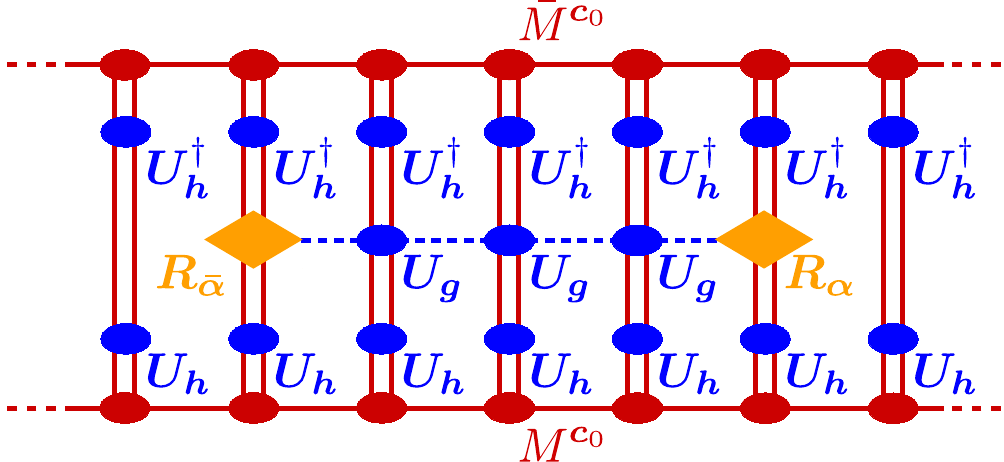}}
\end{equation}
and since $\bm U_{\bm g}$ and $\bm U_{\bm h}$ commute, and the phases from
commuting $\bm U_{\bm h} {\bm R_{\bm \alpha}}=\bm \alpha(\bm h) 
{\bm R_{\bm \alpha}} \bm U_{\bm h}$ and 
$\bm U_{\bm h} {\bm R_{\bar{\bm \alpha}}}=
\bar{\bm\alpha}(\bm h) {\bm R_{\bar{\bm \alpha}}} \bm U_{\bm h}$ cancel out, we
find that $O_{\bm c}^{\bm c}=O_{\bm c_0}^{\bm c_0}$. We thus find that the
expectation value for any string is the same regardless of the boundary
conditions, and we will therefore omit the subscript $\bm c_0$ from now
on and write $\rho\equiv \rho_{\bm c_0}$ and $M\equiv M^{c_0}$ (in fact,
we will most of the time also omit the label $M$ of the tensor).

After these considerations, we are left with the following question: 
Given a symmetry $\bm H\subset \bm G$, $\bm H=K\boxtimes L$, and an
invariant fixed point $\vket\rho$ given by an injective MPS with tensor
$M$, what are the the different possible ways in which strings describing
the behavior of anyons can behave regarding condensation and confinement.

\subsection{Behavior of string order parameters I:
Symmetry breaking and decoupling
\label{sec:class:symbr-decoupling}}

Let us now consider what happens when we separate the two ends of a string 
$\strkDblSep{g}{\alpha}{\ell}$. 
Evaluated in the fixed point MPS $\vket{\rho}\equiv\vket{\rho_{\bm c_0}}$,
this corresponds to 
\begin{equation}
\label{eq:sop-in-mps}
\begin{array}{l}
\hspace*{-0.4cm}(\rho|\strkDblSep{g}{\alpha}{\ell}\vket\rho=\\[0.6em]
\hspace*{0.5cm}\raisebox{-1.2em}{\includegraphics[scale=0.6]{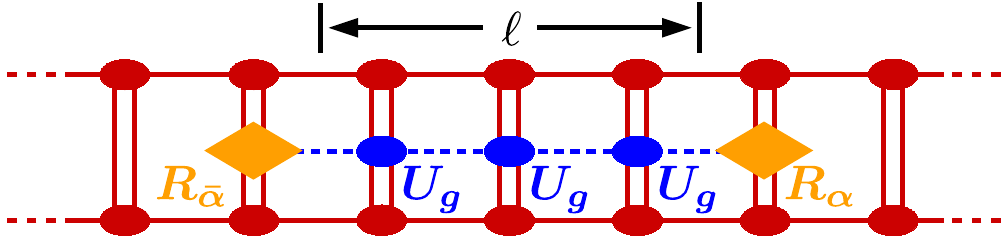}}\quad .
\end{array}\vspace*{1em}
\end{equation}
We now distinguish two cases: If $\bm g\notin\bm H$, then
$\bm{U_g}^{\otimes N}\ket{\rho_{\bm c_0}} = \ket{\rho_{\bm c'}}$ with $\bm
c'\ne\bm c_0$, and since different representations of an injective MPS are
related by a local gauge
transformation~\cite{perez-garcia:mps-reps,cirac:mpdo-rgfp}, 
it holds that
\begin{equation*}
\raisebox{-0.8em}{\includegraphics[scale=0.7]{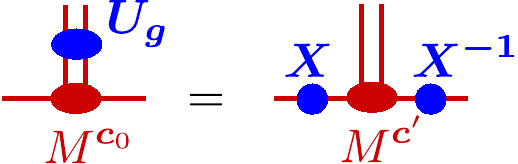}}\quad .
\end{equation*}
and thus
\[
\raisebox{-3.2em}{\includegraphics[scale=0.7]{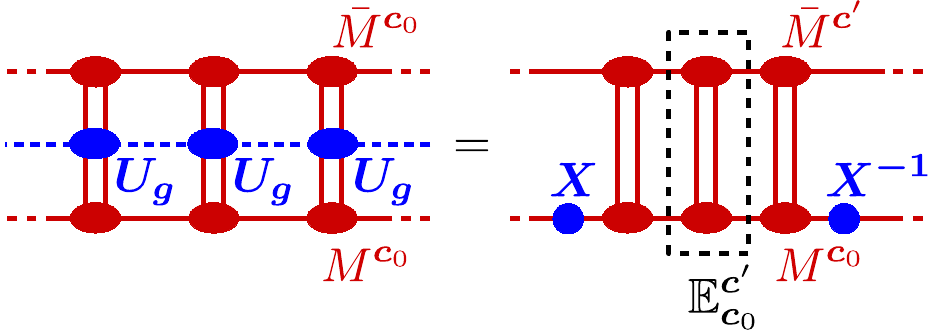}}\quad,
\]
and since $\lambda_{\max}(\mathbb E^{\bm c'}_{\bm c_0})<1$,
$(\rho|\strkDblSep{g}{\alpha}{\ell}\vket\rho\rightarrow0$ as
$\ell\to\infty$. We thus obtain
\begin{condition}
\label{condition:2-domainwall-strings-vanish}
$(\rho|\strkDblSep{g}{\alpha}{\ell}\vket\rho\rightarrow0$
 unless $\bm g\in\bm H$.  In particular, all anyons $\ex{g}{\alpha}$
with $g\not\in K$ are confined.
\end{condition}

On the other hand, if $\bm g\in\bm H$, 
$\bm{U_g}^{\otimes N}\ket{\rho_{\bm c_0}} = \ket{\rho_{\bm c_0}}$ and thus
there exist $\bm V_{\bm g}$ such that
\begin{equation}
\label{eq:Ug-Vg-def}
\raisebox{-1em}{\includegraphics[scale=0.7]{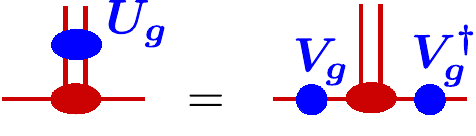}}
\quad ,
\end{equation}
where $\bm V_{\bm g}$ forms a projective representation of $\bm
H$ which can be chosen unitary by a suitable gauge of the
MPS~\cite{sanz:mps-syms}.  Injectivity of the MPS further implies that its
transfer operator
$\mathbb E\equiv \mathbb E_{\bm c_0}^{\bm c_0}$ has a unique fixed point 
\[
\mathbb E^{N-\ell-2}=\
\raisebox{-1.2em}{\includegraphics[scale=0.6]{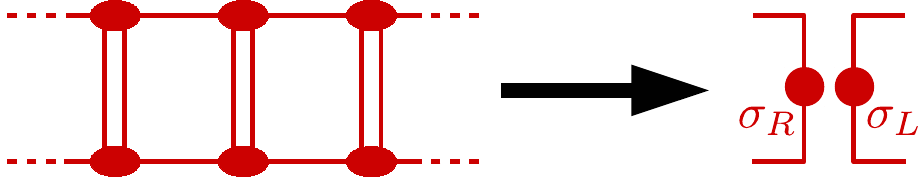}}
\]
(w.l.o.g., we choose $\sigma_R,\sigma_L\ge0$, and normalization 
implies $\mathrm{tr}[\sigma_L\sigma_R]=1$),
and using Eq.~\eqref{eq:Ug-Vg-def}, this implies that
\[
\raisebox{-1.2em}{\includegraphics[scale=0.6]{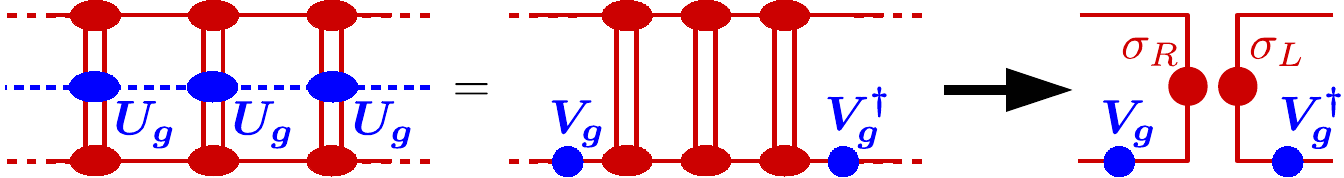}}\quad .
\]
Also, since $[\mathbb E,\bm V_{\bm g}\otimes \bar{\bm V}_{\bm g}]=0$,
uniqueness of the fixed point of $\mathbb E$ implies that $\bm{V_g}
\sigma_{\bullet}\bm{V_g}^\dagger = \sigma_{\bullet}$, where $\bullet=L,R$,
and the ordering of the indices of $\sigma_{\bullet}$ is chosen
accordingly.  With this, we can rewrite Eq.~(\ref{eq:sop-in-mps}) as 
\begin{equation}
\label{eq:sop-endpoints-separate}
(\rho|\strkDblSep{g}{\alpha}{\ell}\vket\rho\rightarrow
\langle \strk{g}{\bar{\alpha}}^*\rangle
\langle \strk{g}{\alpha}\rangle
\end{equation}
where
\begin{equation}
\label{eq:sop-in-mps-ring}
\langle \strk{g}{\alpha}\rangle := \ 
\raisebox{-1.2em}{\includegraphics[scale=0.6]{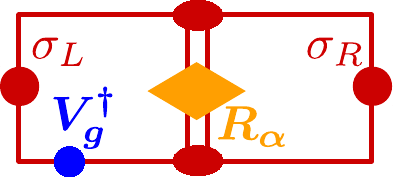}}\quad ,
\end{equation}
and correspondingly $\langle \strk{g}{\bar{\alpha}}^*\rangle$.  This
implies that the expectation value of any string order parameter decouples
into a product of two expectation values corresponding to semi-infinite
strings, and in order to study condensation and confinement, it is thus
sufficient to to consider the behavior of $\langle
\strk{g}{\alpha}\rangle$, Eq.~(\ref{eq:sop-in-mps-ring}). In order to
highlight the role played by the two layers, we will sometimes also write
$\langle \str{g}{\alpha}\otimes\strb{g'}{\alpha'}\rangle :=\langle
\strk{g}{\alpha}\rangle$, with $\bm g=(g,g')$, $\bm\alpha = (\alpha,\alpha')$.

\subsection{Behavior of string order parameters II: Symmetry protected
phases and group cohomology
\label{sec:class:spts}}

We will now study the behavior of string order parameters $\langle
\strk{g}{\alpha}\rangle$, Eq.~(\ref{eq:sop-in-mps-ring}), with $\bm g\in
\bm H$ more closely and show that they are directly related to the
classification of symmetry-protected phases through group cohomology. The
crucial point here is that, following Eq.~\eqref{eq:Ug-Vg-def}, a physical
symmetry action $\bm U_{\bm g}$ can be replaced by a virtual symmetry
action $\bm V_{\bm g}$, where the $V_{\bm k}$ form a \emph{projective
representation} of the symmetry group, i.e.,
$\bm V_{\bm g}\bm V_{\bm h}=\omega(\bm g,\bm h)\bm V_{\bm{gh}}$, where
$\omega:\bm H\times \bm H\rightarrow\mathrm{U}(1)$ is a so-called
$2$-cocycle -- i.e., it satisfies $\omega(\bm g, \bm {hk})\omega(\bm h,\bm
k)=\omega(\bm g,\bm h)\omega(\bm{gh},\bm k)$ due to associativity --
which, up to gauge choices $V_{\bm g}\sim e^{i\phi_{\bm g}}V_{\bm g}$ is
classified by the \emph{second cohomology group} $\Htwo{\bm H}$; this
discrete classification of the $\bm V_{\bm g}$ is what is underlying the
classification of symmetry-protected phases in one
dimensions~\cite{pollmann:symprot-1d,chen:spt-order-and-cohomology,schuch:mps-phases}.

The $2$-cocycle $\omega$ also encodes what happens when we commute $\bm
V_{\bm g}$ and $\bm V_{\bm h}$: 
\[
\bm V_{\bm g}\bm V_{\bm h} = \omega(\bm g, \bm h) \bm V_{\bm{gh}} = 
\omega(\bm g, \bm h) \bm V_{\bm{hg}}  =
\frac{\omega(\bm g, \bm h)}{\omega(\bm h, \bm g)}
 \bm V_{\bm h}\bm V_{\bm g} \ .
\]
Here, $\tfrac{\omega(\bm g,\bm h)}{\omega(\bm h,\bm g)}=:\nu_{\bm h}(\bm
g)$ is called the \emph{slant product}~\cite{propitius:phd-thesis} of
$\omega$ with $\bm h$; for abelian groups, it forms a one-dimensional
representation of $\bm H$, $\nu_{\bm h}(\bm g_1)\nu_{\bm h}(\bm g_2)=\nu_{\bm
h}(\bm g_1\bm g_2)$~\footnote{%
This can be seen using the cocycle conditions and the fact that the group
is abelian as follows:
\begin{align*}
\hspace*{1.5em}\frac{\nu_{\bm h}(\bm g_1)
\nu_{\bm h}(\bm g_2)}{\nu_{\bm h}(\bm g_1\bm g_2)}
&=\frac{\omega(\bm g_1,\bm h)\omega(\bm g_2,\bm h)
	    \omega(\bm h,\bm g_1\bm g_2)}{
      \omega(\bm h,\bm g_1)\omega(\bm h,\bm g_2)
	    \omega(\bm g_1\bm g_2,\bm h)}  \,
 \frac{\omega(\bm h \bm g_1,\bm g_2)}{\omega(\bm h \bm g_1,\bm g_2)}
\\
&=\frac{\omega(\bm g_1,\bm h)\omega(\bm g_2,\bm h)
	    \omega(\bm h,\bm g_1\bm g_2)
	    \omega(\bm h \bm g_1,\bm g_2)
}{
      \omega(\bm h, \bm g_1\bm g_2)
	    \omega(\bm g_1,\bm g_2)
	    \omega(\bm h,\bm g_2)
	    \omega(\bm g_1\bm g_2,\bm h)}
\\
&=\frac{\omega(\bm g_1,\bm h)\omega(\bm g_2,\bm h)
	    \omega(\bm h,\bm g_1\bm g_2)
	    \omega(\bm h \bm g_1,\bm g_2)
}{
      \omega(\bm h, \bm g_1\bm g_2)
	    \omega(\bm h,\bm g_2)
	    \omega(\bm g_1,\bm g_2\bm h)  
	    \omega(\bm g_2,\bm h)
}
\\
&=\frac{\omega(\bm g_1,\bm h)
	    \omega(\bm h \bm g_1,\bm g_2)
}{
	    \omega(\bm h,\bm g_2)
	    \omega(\bm g_1,\bm g_2\bm h)  
}
\\
&=\frac{\omega(\bm g_1,\bm h)
	    \omega(\bm g_1 \bm h,\bm g_2)
}{
	    \omega(\bm h,\bm g_2)
	    \omega(\bm g_1,\bm h\bm g_2)  
} =1 \ .
\end{align*}
}. Note that we can always construct (non-unique) representations
$\gamma$ and $\gamma'$ of $G$ such that
$\gamma(g)\overline{\gamma'(g')}=\nu_{\bm h}((g,g'))$ for $(g,g')\in \bm H$: To
this end, let $\gamma(g):=\nu_{\bm h}((g,e))$ for $g\in L$, extend 
$\gamma$ to a representation of $g\in K$ (formally, this corresponds to an 
induced representation), and define $\gamma'(g):=\overline{\nu_{\bm
h}((g,g)/\gamma(g)}$; finally, both $\gamma$ and $\gamma'$ can be extended
independently to representations of $G$.

We will now derive conditions on $\bm g$ and $\bm \alpha$ under which 
$\langle\strk{g}{\alpha}\rangle$  must be zero and demonstrate how in
the remaining cases, it can be made non-zero by an appropriate choice of
$\bm{R_\alpha}$, and we find that this is in one-to-one correspondence to
the inequivalent $2$-cocycles, i.e., elements of $\Htwo{\bm H}$; the no-go
part part of this discussion has been first given in
Ref.~\cite{pollmann:spt-detection-1d} in the context of string order
parameters for SPT phases.
To this end, let us consider an MPS with a specific projective
representation $\bm V_{\bm g}$ with corresponding $\omega(\bm g,\bm h)$, and
consider a string order parameter 
$\strk{g}{\alpha}$ evaluated in that MPS, 
\begin{equation}
\langle \strk{g}{\alpha}\rangle = \ 
\raisebox{-1.2em}{\includegraphics[scale=0.6]{figs/mps_sop_fptring}}\quad .
\end{equation}
We now insert a resolution of the identity
$\bm{U_h}\bm{U_h}^\dagger$ before $\bm{R_\alpha}$ and use
$\bm{R_\alpha}\bm{U_h} = {\bm\alpha}({\bm h}) \bm{U_h}\bm{R_\alpha}$, which
gives
\[
\includegraphics[scale=0.6]{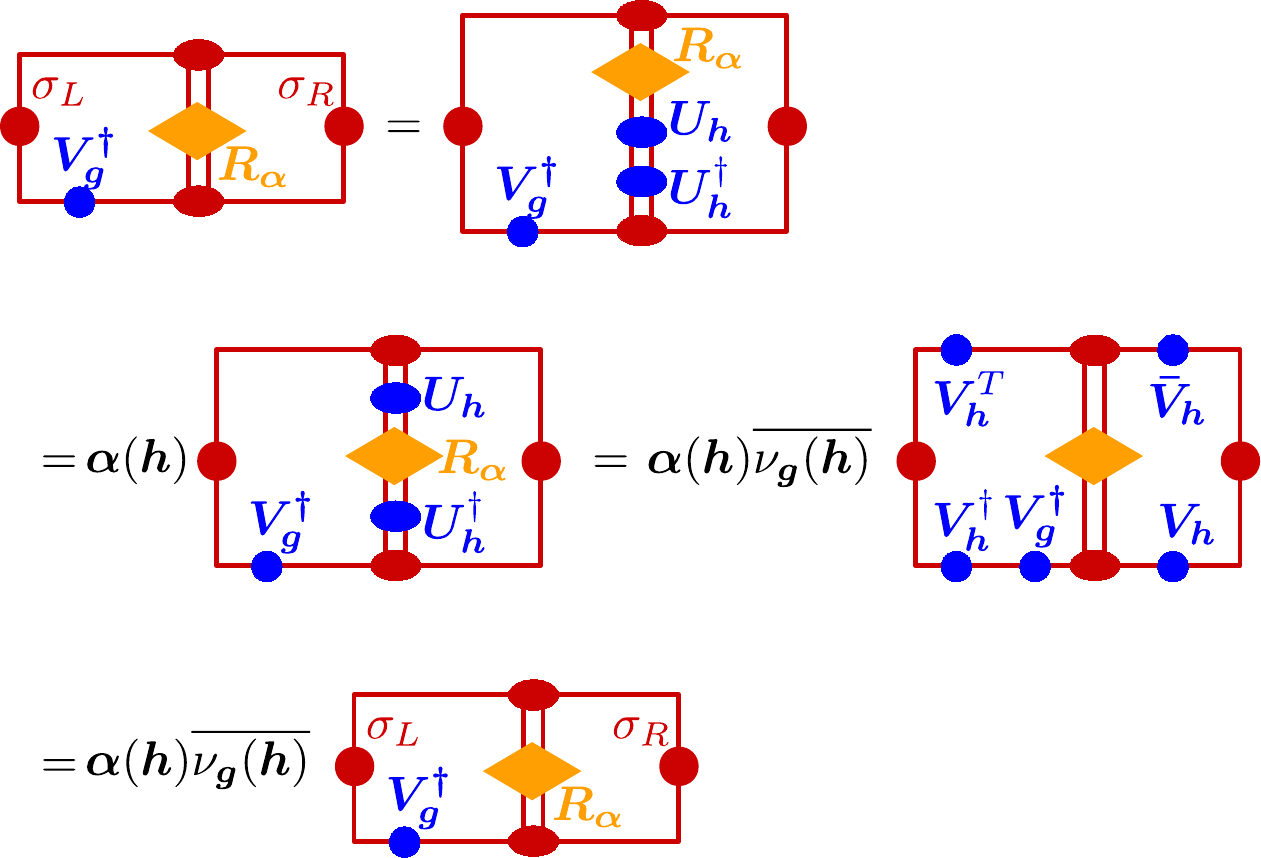}
\]
We thus find that for $\langle\strk{g}{\alpha}\rangle$ to be non-zero, it
must hold that $\bm\alpha = \nu_{\bm g}$, the irreducible representation
obtained as the slant product of the $2$-cocycle $\omega$. Conversely, by
choosing  $\bm{R_\alpha}$ such that 
\begin{equation}
    \label{eq:explicit_Ralpha}
\raisebox{-1.2em}{\includegraphics[scale=0.6]{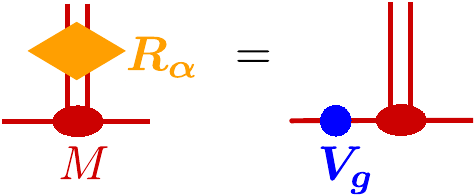}}
\end{equation}
-- which is always possible due to the injectivity of $M$ -- we have that
\[
\includegraphics[scale=0.6]{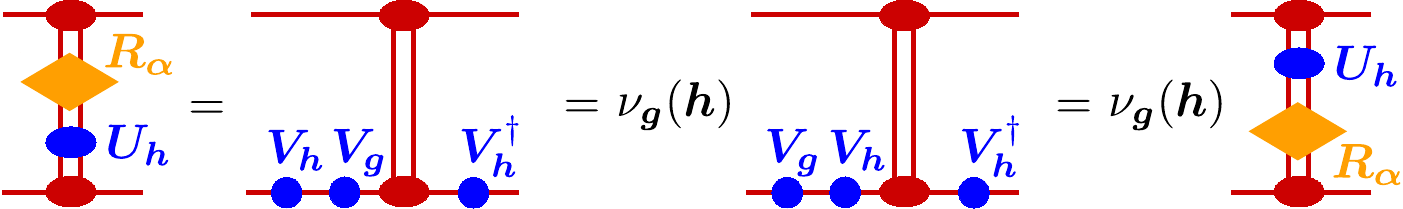}
\]
i.e., $\bm{R_\alpha}$ transforms as ${\bm\alpha}\equiv \nu_{\bm g}$ on
$\bm H$ as required, and
\[
\langle \strk{g}{\alpha}\rangle\! = 
\raisebox{-1.2em}{\includegraphics[scale=0.6]{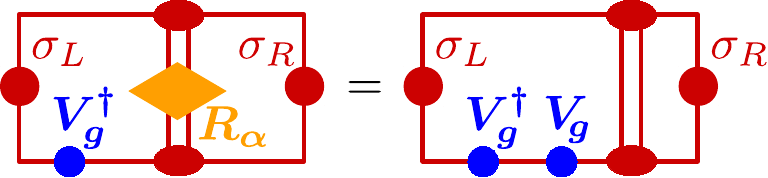}}
\ = 1\,.
\]
It remains to see how $\bm R_\alpha$ transforms under the action of the
full symmetry group $\bm G$, and more specifically that the construction
can be generalized to any irrep $\bm \alpha$ of $\bm G$ with restriction
$\bm \alpha\vert_{\bm H}\equiv \nu_{\bm g}$; this,  together with how to
separate $\bm R_{\alpha}$ into independent ket and bra actions, is
discussed in Appendix~\ref{app:explicit-endpoints}.

We thus see that the behavior of string order parameters is in one-to-one
correspondence with the different SPT phases appearing in the fixed point
of the transfer matrix: For a given SPT phase, a string order parameter
$\langle\strk{g}{\alpha}\rangle$ can only be non-zero if $\alpha=\nu_g$,
and at the same time, it is always possible to set up the endpoint ${\bm
R}_{\bm\alpha}$ of the string order parameter such that $\langle
\strk{g}{\alpha}\rangle$ actually is non-zero.

\begin{condition}
\label{condition:3-alpha-nu_g}
A string operator $\strk{g}{\alpha}$ with
$\langle\strk{g}{\alpha}\rangle\ne0$ exists if and only if $\bm\alpha(\bm
h)=\nu_{\bm g}(\bm h)$ for all $\bm h\in \bm H$, with $\bm\alpha(
(h,h'))=\alpha(h)\bar\alpha'(h')$, $\bm g=(g,g')$, and $\nu_{\bm
g}(\bm h)=\omega(\bm h,\bm g)/\omega(\bm g,\bm h)$, where $\omega$ is the
$2$-cocycle classifying the fixed point of the transfer operator.
\end{condition}
Clearly, the same derivation for the other endpoint of the string,
$\langle \strk{g}{\bar{\alpha}}^*\rangle$, yields exactly the same
condition.

Note that Conditions~\ref{condition:2-domainwall-strings-vanish}
and~\ref{condition:3-alpha-nu_g} together show that the ``amount of
topological order'' -- this is, the number of anyons -- is related to
``symmetry breaking gap'' between ket and bra,  $|K|/|L|$, where $\bm
H=K\boxtimes L$: Deconfined anyons $\ex{g}{\alpha}$ satisfy $(g,g)\in \bm
H$, i.e., $g\in K$, and $(\alpha,\alpha)=\nu_{(g,g)}$, which fixes
$\alpha$ on $L$ and thus leaves $|G|/|L|$ possibilities to extend it to
$G$, yielding a total of $|K||G|/|L|$ deconfined anyons.  Out of those,
pairs $\ex{g}{\alpha}$ and $\ex{gk}{\alpha\beta}$ are indistinguishable if
$(gk,g)\in\bm H$, i.e., $k\in L$, and $(\alpha\beta,\alpha)=\nu_{(gk,g)}$,
which fixes $\beta$ on $K$, leaving $|G|/|K|$ possible extensions; the
size of each set of indistinguishable anyons is thus $|L||G|/|K|$.  The
total number of anyons -- the ratio of these numbers -- is thus
$(|K|/|L|)^2$, and the total quantum dimension is $|K|/|L|$, the
``symmetry breaking gap'' between ket and bra.

\subsection{Constraints from positivity
\label{sec:class:positivity}}

The condition that $\langle\strk{g}{\alpha}\rangle\ne0$ iff $\bm\alpha=
\nu_{\bm g}$ (Condition~\ref{condition:3-alpha-nu_g}) has been derived for
a general fixed point of MPO form.  However, as we have seen in
Sec.~\ref{sec:class:symbreaking}, we can w.l.o.g.\ take the fixed point to
be positive semidefinite, which gave rise to the structure of the unbroken
symmetry subgroup $\bm H$ (Condition~\ref{condition:1-structure-of-H}).
As we will see now, positivity induces yet another constraint, namely on
the $2$-cocycles realizable in the fixed point.

To this end, consider a \emph{positive} fixed point $\rho\ge0$ with an SPT
characterized by some $2$-cocycle $\omega:\bm H\times \bm H\rightarrow
\mathbb C$, and consider some $g'$, $\alpha$, and $\alpha'$ such that
$\langle\strsemik{(e,g')}{(\alpha,\alpha')}\rangle\ne0$.  Then, also for
the other endpoint 
$\langle\strsemik{(e,g')}{(\alpha,\alpha')}^*\rangle\ne0$, and thus
[following Eq.~(\ref{eq:sop-endpoints-separate})] 
\begin{align*}
0 & < \big|(\rho|\strsemikDblSep{(e,g')}{(\alpha,\alpha')}{\ell}\vket\rho\big|^2
\\
&=
\big\vert\mathrm{tr}[ 
    \strDblSep{e}{\alpha}{\ell}
    \rho 
    \strbDblBSep{g'}{\alpha'}{\bar{\alpha}'}{\ell}
    \rho]\big\vert^2
\\
&=
\big\vert\mathrm{tr}[(\sqrt{\rho} 
    \strDblSep{e}{\alpha}{\ell}
    \sqrt{\rho})
    \,(\sqrt\rho 
    \strbDblBSep{g'}{\alpha'}{\bar{\alpha}'}{\ell}
    \sqrt\rho)
]\big\vert^2
\\
&\stackrel{(*)}{\le}
\mathrm{tr}[(\sqrt{\rho} 
    \strDblSep{e}{\alpha}{\ell}
    \sqrt{\rho})(\ldots)^\dagger]
    \\
    &\hspace*{6em}\times
    \mathrm{tr}[(\sqrt\rho 
    \strbDblBSep{g'}{\alpha'}{\bar{\alpha}'}{\ell}
    \sqrt\rho)(\ldots)^\dagger ]
\\
&=  (\rho|\strsemikDblSep{(e,e)}{(\alpha,\alpha)}{\ell}\vket\rho
    \\
    &\hspace*{6em}\times
    (\rho|\strsemikDblSep{(g',g')}{(\alpha',\alpha')}{\ell}\vket\rho\ ,
\end{align*}
where we have used Cauchy-Schwarz in $(*)$ [here, $(\ldots)$ denotes the
preceding term].  Following Eq.~(\ref{eq:sop-endpoints-separate}), this
implies 
$\langle\strsemik{(e,e)}{(\alpha,\alpha)}\rangle\ne0$, and thus (from
Condition~\ref{condition:3-alpha-nu_g})
$\alpha(h)=\alpha(h)\bar\alpha(e)=\nu_{(e,e)}((h,e)) \equiv 1$ 
for $(h,e)\in\bm H$.
At the same time, 
$\langle\strsemik{(e,g')}{(\alpha,\alpha')}\rangle\ne0$ implies that 
$\alpha(h)=\alpha(h)\bar{\alpha}'(e)=\nu_{(e,g')}((h,e))$, and thus
\[
1=\nu_{(e,g')}((h,e)) = 
    \frac{\omega((h,e),(e,g'))}{\omega((e,g'),(h,e))}\ ,
\]
i.e.:
\begin{condition}
\label{condition:4-proj-reps-commute}
The projective representations of ket and bra symmetry actions must
commute,  
\begin{equation}
\label{eq:projrep-ket-bra-commute}
\bm V_{(g,e)}\bm V_{(e,g')} = 
\bm V_{(e,g')} \bm V_{(g,e)}\ ,
\end{equation}
or $\nu_{(g,e)}( (e,g'))=1$, where $(g,e),(e,g')\in\bm H$.
\end{condition}

\subsection{\label{sec:subsection-anyon-condensation-rules}
Anyon condensation rules}

Let us now show that the
Conditions~\ref{condition:1-structure-of-H}--\ref{condition:4-proj-reps-commute}
exactly give rise to the anyon condensation rules mentioned in the
introduction:
\begin{enumerate}
\item 
\label{condrules:selfboson}
Only self-bosons can condense.
\item 
\label{condrules:confined}
Anyons become confined if and only if they have mutual non-bosonic
statistics with some condensed anyon.
\item 
\label{condrules:identified}
Non-confined anyons which differ by a condensed anyon become
indistinguishable.
\end{enumerate}

\subsubsection{Only self-bosons can condense.}

Consider a condensed anyon $\ex{g}{\alpha}$, 
$\langle\strsemik{(g,e)}{(\alpha,1)}\rangle\ne0$. This requires
$(g,e)\in\bm H=K\boxtimes L$, i.e., $g\in L$, and moreover
$\alpha(h)=\nu_{(g,e)}((h,h'))\ \forall\,(h,h')\in \bm H$, and thus 
$\alpha(g)=\nu_{(g,e)}((g,e))=1$, i.e., $\ex{g}{\alpha}$ is a self-boson.

\subsubsection{Anyons become confined if and only if they have mutual
non-bosonic statistics with some condensed anyon.}

Let us first show that an unconfined anyon $\ex{k}{\beta}$, 
$\langle\strsemik{(k,k)}{(\beta,\beta)}\rangle\ne0$, must have mutual
bosonic statistics with all condensed anyons $\ex{g}{\alpha}$,
$\langle\strsemik{(g,e)}{(\alpha,1)}\rangle\ne0$.
$\langle\strsemik{(k,k)}{(\beta,\beta)}\rangle\ne0$ implies $k\in K$ and
$\beta(h)\overline{\beta(h')}=\nu_{(k,k)}((h,h'))$ for $(h,h')\in\bm H$,
i.e., $\beta(h)=\nu_{(k,k)}((h,e))$ for $h\in L$. On the other hand,
$\langle\strsemik{(g,e)}{(\alpha,1)}\rangle\ne0$ implies $(g,e)\in\bm H$,
i.e., $g\in L$, and $\alpha(h)=\nu_{(g,e)}((h,h))$ for $h\in K$. We thus
have that
\[
\alpha(k)\beta(g) = \nu_{(g,e)}((k,k)) \nu_{(k,k)}((g,e)) = 1\ ,
\]
since $\nu_{\bm g}(\bm h)\nu_{\bm h}(\bm g)=1$.

Conversely, consider a confined anyon $\ex{k}{\beta}$, 
$\langle\strsemik{(k,k)}{(\beta,\beta)}\rangle=0$: we will show that this
implies the existence of a condensed anyon $\ex{g}{\alpha}$, 
$\langle\strsemik{(g,e)}{(\alpha,1)}\rangle\ne0$, which has mutual
non-bosonic statistics, $\alpha(k)\beta(g)\ne1$, by explicitly
constructing such an anyon $\ex{g}{\alpha}$. 
$\langle\strsemik{(k,k)}{(\beta,\beta)}\rangle=0$ implies that either (i)
$k\notin K$ or (ii) there exists $(h,h')\in\bm H$ s.th.\
$\beta(h)\overline{\beta(h')}\ne \nu_{(k,k)}((h,h'))$.

Let us first consider case (i), $k\notin K$. Let $g=e$ [thus $(g,e)\in\bm
H$], and choose an irrep $\alpha$ of $G$ s.th.\
$\alpha(h):=\nu_{(g,e)}((h,h'))= \nu_{(e,e)}((h,h'))\equiv 1$ for
$(h,h')\in\bm H$ -- this is, $\ex{g}{\alpha}$ is condensed. On the other
hand, since $k\notin K$ we can always choose $\alpha$ s.th.\ $\alpha(k)\ne
1$ (as the extension of the irrep from $K$ to $G$ is non-unique), and
thus, $\alpha(k)\beta(g)\ne 1$, i.e., the anyons have mutual non-bosonic
statistics.

Now consider case (ii): $k\in K$, but there exists some $(h_0,h_0')\in\bm
H$ s.th.\ $\beta(h_0)\overline{\beta(h_0')}\ne\nu_{(k,k)}((h_0,h_0'))$.
Define $g:=h_0h_0'^{-1}\in L$.  Since hermiticity implies
$\nu_{(k,k)}((h,e))=\overline{\nu_{(k,k)}((e,h))}$ (as can be shown by
relating $\nu_{(k,k)}$ to the behavior of string order parameters)
and thus
$\nu_{(k,k)}((h,h))=\nu_{(k,k)}((h,e))\nu_{(k,k)}((e,h))=1$, we have that
\begin{align*}
\beta(g) & = \beta(g)\beta(h_0')\overline{\beta(h_0')} 
=\beta(h_0)\overline{\beta(h_0')}\\
& \ne \nu_{(k,k)}((h_0,h_0')) 
= \nu_{(k,k)}((g,e)) \nu_{(k,k)}((h_0',h_0')) \\
&= \nu_{(k,k)}((g,e))\ .
\end{align*}
Further, let $\overline{\alpha'(h')}:=\nu_{(g,e)}((e,h'))\equiv 1$ for
$h\in L$, and extend it to the trivial irrep $\alpha'\equiv 1$ of $G$.
Then, $\alpha(h):=\nu_{(g,e)}((h,h))/\overline{\alpha'(h)}$, $h\in K$, can be
extended to an irrep $\alpha$ of $G$ s.th.
$\alpha(h)=\alpha(h)\overline{\alpha'(h')} = \nu_{(g,e)}((h,h'))$ for all
$(h,h')\in\bm H$, i.e., $\ex{g}{\alpha}$ is condensed. Finally,
$\alpha(k)\beta(g)\ne \nu_{(g,e)}((k,k)) \nu_{(k,k)}((g,e)) = 1$, i.e.,
$\ex{g}{\alpha}$ and $\ex{k}{\beta}$ have mutual non-bosonic statistics.

\subsubsection{Non-confined anyons which differ by a condensed anyon become
indistinguishable.}

Let $\ex{g}{\alpha}$ be condensed, i.e., $g\in L$ and
$\alpha(h)=\nu_{(g,e)}((h,h'))\ \forall(h,h')\in\bm H$, and
$\ex{k}{\beta}$ unconfined, $k\in K$ and
$\beta(h)\overline{\beta(h')}=\nu_{(k,k)}((h,h'))\ \forall(h,h')\in \bm
H$. Then, $(gk,k)\in \bm H$, and 
\begin{align*}
\alpha(h)\beta(h)\overline{\beta(h')} 
    &= \nu_{(g,e)}((h,h'))\nu_{(k,k)}((h,h')) 
\\
    &= \overline{\nu_{(h,h')}((g,e))}\overline{\nu_{(h,h')}((k,k))} 
\\
    &= \overline{\nu_{(h,h')}((gk,k))} 
    = \nu_{(gk,k)}((h,h'))\ ,
\end{align*}
i.e., the anyons $\ex{k}{\beta}$ and $\ex{gk}{\alpha\beta}$ become
indistinguishable.

\section{Anyon condensation in $D(\mathbb Z_N)$ and twisted double models
\label{sec:ZN-and-tw-dbl}}

We will now show that in the case of cyclic groups, $G=\mathbb Z_N$, this
allows for a full classification of all condensation patterns, and that
these condensation patterns give rise exactly to all twisted quantum
doubles $D^{\omega_3}(\mathbb Z_M)$, where the twist $\omega_3$ is given
by a $3$-cocycle of $\mathbb Z_M$~\footnote{%
Note that the same cannot hold for all abelian groups: Condensing from an
abelian group gives another abelian model, while twisting an abelian
model can give rise to non-abelian models~\cite{propitius:phd-thesis}.
}.
In what follows, we will write the
groups additively with neutral element zero, and addition is understood modulo
the order of the group.

\subsection{Allowed phases at the boundary}

Let us first study the effect of the above conditions on the possible
SPT phases at the boundary, and thus the possible condensation patterns.
As we have seen, the symmetry $\bm G=\mathbb Z_N\times \mathbb Z_N$
of the transfer operator is broken down to a symmetry $\bm H=\mathbb
Z_{qt}\boxtimes \mathbb Z_q=\{(g,g+th)\,:\,g=0,\dots,qt-1,\,h=0,\dots,q-1\}$.
Let us now consider the restriction imposed by
Eq.~(\ref{eq:projrep-ket-bra-commute}) on the second cohomology
classification of the projective representations of $\bm H$,
$\Htwo{\bm H}=\mathbb Z_q$~\cite{propitius:phd-thesis}. To this end, given
an element $n\in \Htwo{\bm H}$, $n=0,\dots,q-1$, we choose a projective
representation
\begin{equation}
\label{eq:Vgg-Vg0-def}
\bm V_{(g,g)} = X^{g}\mbox{\ and\ } \bm V_{(ht,0)} = Z^{hn}\ ,
\end{equation}
of $\bm H$, $g=0,\dots,qt-1$, $h=0,\dots,q-1$, where $X$ and $Z$ are such that
$ZX=\mu XZ$, $\mu=\exp(2\pi i/q)$ (e.g.\ $X$ a cyclic shift and $Z$ a
diagonal $q\times q$ matrix), and where
$V_{(g+ht,g)}:=V_{(g,g)}V_{(ht,0)}$.  It is straightforward to check that
these yield $q$ inequivalent (and thus all) projective representation,
e.g.\ by comparing the gauge-invariant commutator $\omega(
(t,0),(1,1))/\omega( (1,1),(t,0)) = \mu^{n}$.  We now have that
$V_{(0,h't)}=V_{(h't,h't)}V_{((q-h')t,0)}$ and thus
Eq.~\eqref{eq:projrep-ket-bra-commute} reads 
\[
V_{(ht,0)}\,V_{(h't,h't)}V_{((q-h')t,0)} = 
V_{(h't,h't)}V_{((q-h')t,0)}\,V_{(ht,0)}
\]
which using Eq.~\eqref{eq:Vgg-Vg0-def} is equivalent to $\mu^{hn\cdot
h't}=1$, or (since $h$ and $h'$ are arbitrary) $\mu^{nt} = 1$.
This is the case whenever $nt$ is a multiple of $q$, i.e.,
$n=k\tfrac{q}{\mathrm{gcd}(t,q)}$. Since at the same time, $0\le n<q$, we find that
$k=0,1,\dots,\mathrm{gcd}(t,q)-1$. This is, out of the $q$
different SPT phases under the symmetry group $\bm H$, only 
$\mathrm{gcd}(t,q)$ are allowed due to positivity constraints.

\subsection{Explicit construction of all twisted $\mathbb Z_t$ doubles
and completeness of classification for $\mathbb Z_N$}

We will now show that for cyclic groups, this classification is complete.
To this end, we will first show how to obtain all \emph{twisted} quantum
doubles of $\mathbb Z_t$ by anyon condensation from a $\mathbb Z_N$
double, and subsequently use this construction to derive explicit PEPS
models for all cases consistent with
Conditions~\ref{condition:1-structure-of-H}--\ref{condition:4-proj-reps-commute}.

\subsubsection{Twisted doubles of $\mathbb Z_t$ from anyon condensation}

In the following, we will describe how to construct all so-called
\emph{twisted quantum doubles} $D^{\omega_r}(\mathbb Z_t)$ of $\mathbb Z_t$
by anyon condensation from a quantum double of some $\mathbb Z_N$, and
derive the structure of the SPT at the boundary.  Here, $\omega_r\equiv
\omega_r(g,h,\ell)$ is a so-called $3$-cocycle, labelled by an element $r$ of
third cohomology group $\mathrm{H}^3(\mathbb Z_t,\mathrm{U}(1))=\mathbb
Z_t$, $r=0,\dots,t-1$. We will just state the corresponding results in the
main text and postpone the proofs to Appendix~\ref{appendix:twisted-Zt};
for more details on twisted double models and $3$-cocycles, we refer the
reader to the appendix or Ref.~\cite{propitius:phd-thesis}.

Let $r=0,\dots,t-1$ label an element of $\mathrm{H}^3(\mathbb
Z_t,\mathrm{U}(1))=\mathbb Z_t$, set $q=t/\text{gcd}(t,r)$, and let
$N:=qt$. We now define tensors with non-zero elements
\begin{align}
\label{eq:twdbl:app_MPO1} 
M(a) := \sqrt{\frac{q}{t}} &
    \hspace*{-1em}
    \begin{split}
    \includegraphics{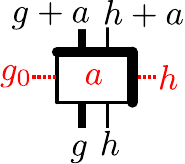} 
    \end{split}\ 
\omega(a,g,h-g_0)\ \ ,
\\ 
\label{eq:twdbl:app_MPO2} 
N(a) := \sqrt{\frac{q}{t}} &
    \hspace*{-1em}
    \begin{split}
    \includegraphics{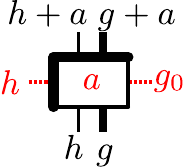} 
    \end{split}\
\overline{\omega(a,g,h-g_0)}
\end{align} 
for $a=0,\dots,t-1$.  Here, the range of the thick vertical indices is
$0,\dots,t-1$ and that of the thin (horizontal and vertical) indices is
$0,\dots,q-1$, and $h=0,\dots,q-1$, $g=0,\dots,t-1$, and
$g_0=g\,\mathrm{mod}\,q$. Depending on the index, variables are understood
modulo $t$ or $q$.  The \mbox{$3$-cocycle} $\omega\equiv\omega_r$ is defined by
\[
\omega(a,g,d)  = \exp\left[
    \frac{2\pi i r d}{t^2}(a+g-(a+g)\,\mathrm{mod}\,t)
\right]  \ ,
\]
where there is no modular arithmetics in the exponential except for
the $\mathrm{mod}\:t$.
Eqs.~(\ref{eq:twdbl:app_MPO1},\ref{eq:twdbl:app_MPO2}) determine the
amplitude of all non-zero elements of $M(a)$ and $N(a)$, respectively, while all
tensor elements inconsistent with the labels of the indices are zero.

The PEPS tensor for the model is now defined as 
\[
A =\frac{t}{q^2} \sum_a \ 
\raisebox{-3em}{\includegraphics[scale=0.8]{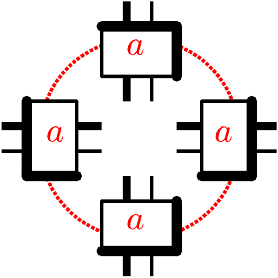}}
\quad,
\]
where the inner legs correspond to the physical and the outer legs to the
virtual indices. As we show in
Appendix~\ref{app:twdbl:subsec-equiv-twdbl}, the PEPS defined by this
model describes a twisted $\mathbb Z_t$ double model with twist $\omega$.
As also shown in the appendix, $A$ satisfies $A^\dagger=A=A^2$, which implies
that $AA^\dagger=A$ and thus the transfer operator is of the form
\begin{equation}
\label{eq:twdbl:top-fpt}
 \begin{split}
   \includegraphics{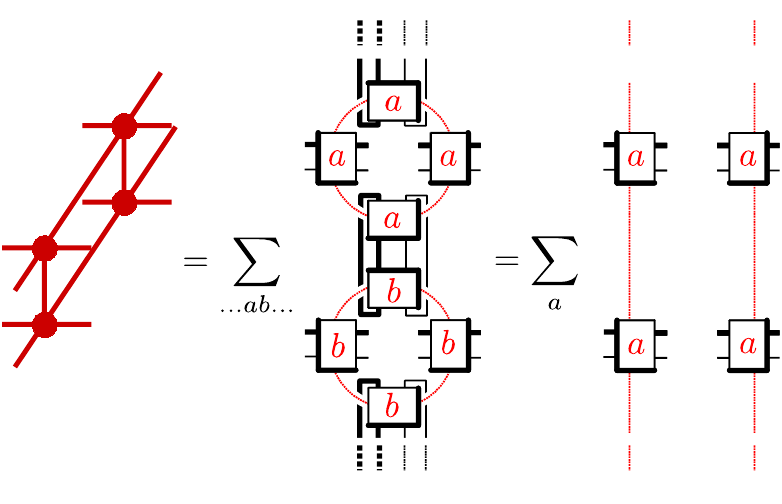}
 \end{split} \ \ ,
\end{equation}
where the second equality holds since
\[
 \begin{split}
   \includegraphics{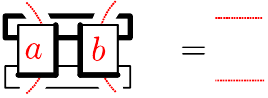}
 \end{split} \ \delta_{ab} \ \ .
\]
We thus find that the left and right fixed points of the transfer operator
are again described by the same tensor.  As it turns out, for any fixed
$a$ they describe an injective MPS, and thus, the boundary exhibits $t$
symmetry broken sectors labelled by $a=0,\dots,t-1$.

The tensor $A$ has a $\mathbb Z_{qt}$ symmetry with generator
$S_{(i_1,i_2),(j_1,j_2)} =
\delta_{i_1+1,j_1}\delta_{i_2+1,j_2}\omega(1,i_1,i_2-i_1)$ (with $i_1$,
$j_1$ $\mathrm{mod}\,t$ and $i_2$, $j_2$ $\mathrm{mod}\,q$), which follows
from the local condition
\begin{equation}
\label{eq:twdbl:S-UU}
 \raisebox{-0.3em}{\includegraphics{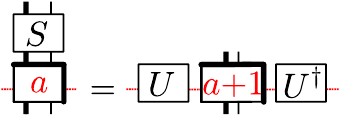}}
  \quad,
\end{equation}
where $U_{ij}\equiv U(a)_{ij} = \delta_{ij}\overline{\omega(1,a,i)}$. 
Together with its ``twin'' equation
\begin{equation}
\label{eq:twdbl:S-VV}
 \raisebox{-1.8em}{\includegraphics{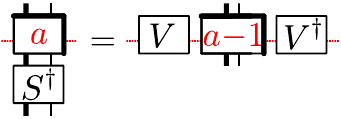}}
\quad,
\end{equation}
$V_{ij}\equiv V(a)_{ij} = e^{i\phi_a}\omega(1,a-1,i)\delta_{i,j-1}$,
Eq.~(\ref{eq:twdbl:S-UU}) allows us to verify that the symmetry
$\mathbb{Z}_{qt}\times\mathbb{Z}_{qt}$ in the fixed point
[Eq.~(\ref{eq:twdbl:top-fpt})] is broken to $\mathbb
Z_{qt}\boxtimes\mathbb Z_{q}$, with generators $G_1=S\otimes S^\dagger$
and $G_2=S^t\otimes \openone$. The element $n\in\Htwo{\mathbb
Z_{qt}\boxtimes \mathbb Z_{q}}=\mathbb Z_q$ labelling the virtual symmetry
action -- determined by the commutation relation of the virtual
representations $P_1(a)=U(a)V(a+1)$ and $P_2(a)=\prod_{i=a}^{a+t-1}U(i)$
of $G_1$ and $G_2$ -- is given by $n=\tfrac{rq}{t}$.

Overall, given $t$ and $r=0,\dots,t-1$, and $q=t/\mathrm{gcd}(t,r)$, we
thus have constructed a PEPS with bond dimension $qt$ and virtual $\mathbb
Z_{qt}$ symmetry which describes a twisted $\mathbb Z_t$ quantum double
with twist $r\in\mathrm{H}^3(\mathbb Z_t,\mathrm{U}(1))=\mathbb Z_t$. In the fixed
point of the transfer operator, the symmetry is broken down to 
$\mathbb Z_{qt}\boxtimes \mathbb Z_q$, and the cocycle $n\in\Htwo{\mathbb
Z_{qt}\boxtimes \mathbb Z_q}=\mathbb Z_q$ characterizing the
virtual symmetry action in the fixed point is given by $n=\tfrac{rq}{t}$.

\subsubsection{\label{sec:twdouble-embed-symmetry}
Completeness of the
Conditions~\ref{condition:1-structure-of-H}--\ref{condition:4-proj-reps-commute}}

Let us now show that this construction allows us to obtain PEPS models for
cyclic $G$ for any case compatible with
Conditions~\ref{condition:1-structure-of-H}--\ref{condition:4-proj-reps-commute}.
Concretely, those conditions imply that given a virtual symmetry $\mathbb
Z_N$ in the tensor, it can be broken down to any $\mathbb Z_{qt}\boxtimes
\mathbb Z_{q}$ symmetry where $qt|N$  (``$|$'' denotes ``divides''), and
furthermore, the label $n$ of the cocycle characterizing the fixed points
must be a multiple of $\tfrac{q}{\mathrm{gcd}(q,t)}$,
$n=k\tfrac{q}{\mathrm{gcd}(q,t)}$.

To this end, define $q':=\mathrm{gcd}(q,t)$, and let $\alpha=q/q'$,
$n'=n/\alpha=k\tfrac{q/\alpha}{\mathrm{gcd}(q,t)}=k=0,\dots,q'-1$. Next, let
$\beta=\mathrm{gcd}(n',q')$, and define $n''=n'/\beta$ and $q''=q'/\beta$. With
$\gamma=\alpha\beta$, we then have that $q=\gamma q''$ and $n=\gamma n''$.
Since $q''|q'$ and $q'|t$, $x:=t/q''$ is integer.    Then, the construction for the
twisted double with $\tilde t=t=xq''$ and twist $\tilde r=xn''$ described in
the preceding section yields 
\[
\tilde q = \frac{t}{\mathrm{gcd}(\tilde r,t)} 
    = \frac{xq''}{\mathrm{gcd}(x n'',xq'')}  
    = \frac{q''}{\mathrm{gcd}(n'',q'')} 
    = q''\ ,
\]
and the $2$-cocycle of the fixed point is characterized by 
$\tilde n = \tfrac{\tilde r\tilde q}{t} = \tfrac{xn''q''}{xq''} = n''$.

We thus know how to create a model with parameters $t$, $q''=q/\gamma$,
and $n''=n/\gamma$; let us denote its tensor by $A^i_{k_1,k_2,k_3,k_4}$, with
$k_s=0,\dots,q''t-1$, and the generator of the $\mathbb Z_{q''t}$ symmetry
by $S$; w.l.o.g., we choose a basis $|k_s)$ such that $S=\sum
|k+1)(k|$.  We will now show how from this model, we can create
a PEPS with parameters $t$, $q$, and $n$, and overall symmetry $\mathbb
Z_M$ with $M=qt$.  (In a second step, we will then generalize this to any
$\mathbb Z_N$ with $qt|N$.) To this end, we extend the bond space to a
$M=\gamma (q''t)$-dimensional space, $k_s\leadsto(\ell_s,k_s)$,
$\ell_s=0,\dots,\gamma-1$, and construct the new tensor $\tilde A$ by
tensoring each virtual index of $A$ independently with an equal weight
superposition of all $|\ell_s)$, i.e., 
\[
\tilde A^i_{(\ell_1,k_1),(\ell_2,k_2),(\ell_3,k_3),(\ell_4,k_4)}
= A^i_{k_1,k_2,k_3,k_4}\ \mbox{for all $\ell_s$\ .}
\]
As the generator of the $\mathbb Z_M$ symmetry we choose the regular
representation in $\mathbb Z_M$ with basis $|\ell q''t+k)$, i.e., $\tilde
S:|\ell,k)\mapsto|\ell+\lfloor k/q''t\rfloor,k+1)$; since each $\ell_s$
index is in a uniform superposition $\sum |\ell_s)$, $\tilde S$ acts
exactly as $S$ on the non-trivial degrees of freedom $k_s$ of $\tilde A$,
while leaving $\sum |\ell_s)$ invariant.  The resulting tensor has thus a
$\mathbb Z_M$ symmetry which is broken to $\mathbb Z_{\gamma
q''t}\boxtimes \mathbb Z_{\gamma q''}=\mathbb Z_{qt}\boxtimes \mathbb
Z_{q}$ in the fixed point, with $q:=\gamma q''$.  The element
$n\in\Htwo{\mathbb Z_{\gamma q''t}\boxtimes \mathbb Z_{\gamma
q''}}=\mathbb Z_{\gamma q''}$ is determined by the commutation phase of
the virtual representations of the two generators $\tilde
S\otimes\bar{\tilde S}$ and $\tilde S\otimes\openone$ in the fixed point
MPS, which equal those of $S\otimes\bar S$ and $S\otimes\openone$,  and
which is thus $\exp[2\pi i \tfrac{n}{\gamma q''}] = \exp[2\pi i
\tfrac{n''}{q''}]$; we therefore have $n\equiv \gamma n''$, as claimed.

To obtain the most general case, we still need to show how to go from a
$\mathbb Z_M$ to a $\mathbb Z_N$ symmetry (with $M|N$) which in the fixed
point is broken down to at least $\mathbb Z_N\times \mathbb Z_N$, and
possibly further. To this end, let $\sigma:=N/M$, 
denote the original tensor again by $A^i_{k_1,k_2,k_3,k_4}$ with
$k_s=0,\dots,M-1$, extend the indices as $(k_s,\ell_s)$ with
$\ell_s=0,\dots,\sigma-1$, and define the new tensor 
\[
\tilde A^i_{(k_1,\ell_1),(k_2,\ell_2),(k_3,\ell_3),(k_4,\ell_4)}
= A^i_{k_1,k_2,k_3,k_4}\delta_{\ell_1=\ell_2=\ell_3=\ell4}\ ,
\]
where $\delta_{\ell_1=\ell_2=\ell_3=\ell4}=1$ if all $\ell_s$ are equal,
and zero otherwise. Further, define $\tilde S = S^{1/\sigma}\otimes
\sum_{\ell=0}^{\sigma-1}\ket{\ell+1}\bra{\ell}$ (with addition modulo
$\sigma$). Clearly, $\tilde S$ generates a representation of $\mathbb Z_M$
(which is faithful if $S$ was faithful). Further, the additional degrees
of freedom labelled by $\ell$ yield two independent GHZ states (i.e.,
correlated block-diagonal structures) in ket and bra level in the fixed
point, which are cyclicly permuted by the action of $\tilde S$: The
$\mathbb Z_N\times \mathbb Z_N$ symmetry is thus at least broken to
$\mathbb Z_M\times \mathbb Z_M$, with the model in each symmetry broken
sector described by the original PEPS, and the $\mathbb Z_M$ symmetry
action generated by $\tilde S^\sigma = S\otimes \openone$.  

Together, this concludes the construction of an explicit example for all
cases consistent with
Conditions~\ref{condition:1-structure-of-H}--\ref{condition:4-proj-reps-commute}.

\section{Example: Condensation of $D(\mathbb Z_4)$ and the double semion
model
\label{sec:example-Z4}}

In the following, we will discuss some examples for anyon condensation in
doubles $D(\mathbb Z_N)$.  As a warm-up, we will start with the Toric Code
model $D(\mathbb Z_2)$, and then discuss in detail the possible
condensations in $D(\mathbb Z_4)$, where we will see how condensing a dyon
-- corresponding to a non-trivial SPT at the boundary -- can give rise to
the doubled semion model which cannot be described as a double model of a
group.

Given the double $D(\mathbb Z_N)$, its excitations $\ex{g}{\alpha}$ are
labelled by group elements $g=0,\dots, N-1$ and irreps $\alpha=\exp(2\pi i
k/N$)$, k=0,\dots,N-1$, where $\alpha(g)\equiv \alpha^g$. (We will again
write the group additively with neutral element $0$.) The
self-statistics for a half-exchange of two $\ex{g}{\alpha}$ particles is
$\alpha^g$, and the phase acquired through the full exchange of
$\ex{g}{\alpha}$ and $\ex{h}{\beta}$ is given by $\alpha^h\beta^g$. Fusing
particles $\ex{g}{\alpha}$ and $\ex{h}{\beta}$ results in
$\ex{g+h}{\alpha\beta}$.

As derived in Sec.~\ref{sec:subsection-anyon-condensation-rules}, in order
for a particle $\ex{g}{\alpha}$ to condense, it must be have bosonic
self-statistics, i.e., $\alpha^g=1$.  This leads to the identification
of $\ex{g}{\alpha}$ with the vacuum $\ex{0}{1}$, and subsequently to the
identification all pairs $\ex{h}{\beta}$ and $\ex{h+g}{\beta\alpha}$.
Moreover, all particles $\ex{h}{\beta}$ which braid non-trivially with
$\ex{g}{\alpha}$, $\alpha^h\beta^g\ne1$, become confined.

\subsection{\label{sec:subsec:cond-tcode}
Warm-up: Condensation of the Toric Code}

\begin{figure}[b]
\centering
\includegraphics[scale=0.95]{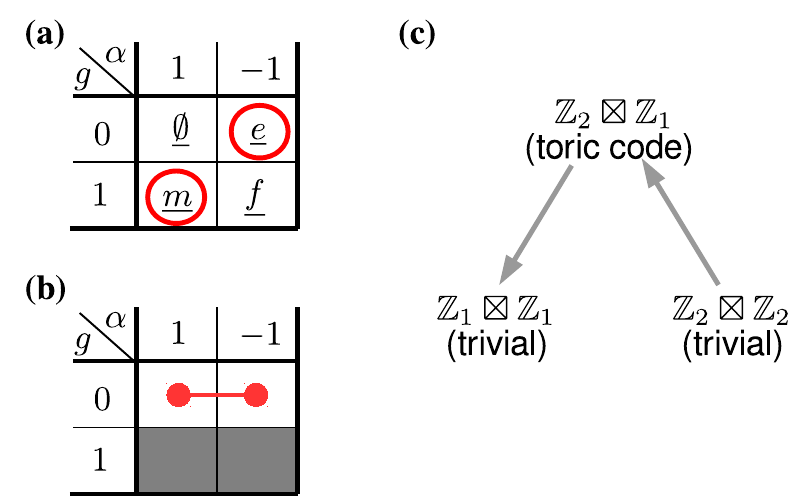}
\caption{
\label{fig:tc-condensation}
\textbf{(a)} Anyon table for the toric code. $\anyon{e}$ and $\anyon{m}$
are self-bosons and can condense. 
\textbf{(b)} Condensation of $\anyon{e}$ results in confinement of
particles $\ex{1}{*}$ (gray) and thus yields a trivial model.
\textbf{(c)} Condensation scheme for the Toric Code.  The diagram
lists the preserved symmetry $\bm H$ and the corresponding phases. 
Note that the two trivial phases correspond to condensing either
$\anyon{e}$ (left) or $\anyon{m}$ (right).
}
\end{figure}

Let us start by considering the Toric Code model $\mathbb D(\mathbb Z_2)$.
It has four particles: The vacuum $\anyon{\emptyset} = \ex{0}{1}$, the
magnetic particle $\anyon{m}=\ex{1}{1}$, the electric particle
$\anyon{e}=\ex{0}{-1}$, and the fermion $\anyon f\equiv \anyon{e}\times\anyon{m}
= \ex{1}{-1}$; they can be visualized in a two-dimensional grid with $g$
and $\alpha$ as row and column labels, respectively,
Fig.~\ref{fig:tc-condensation}a. $\anyon e$ and $\anyon m$ (marked red)
have bosonic self-statistics $\alpha^g$ and can therefore condense.
Fig.~\ref{fig:tc-condensation}b illustrates the condensation of the
$\anyon{e}$ particle: $\anyon e$ is identified with the vacuum (indicated by
connected dots), and since both $\anyon{m}$ and $\anyon{f}$ have
non-trivial mutual statistics $\alpha^h\beta^g$ with $\anyon{e}$ (as
$\alpha=-1$, $g=0$ for $\anyon e$ and $h=1$ for $\anyon m$, $\anyon f$),
they become confined (indicated by grayed out boxes).

Let us now study the condensation in terms of the symmetry of the transfer
operator.  We have $\bm G=\mathbb Z_2\times \mathbb Z_2$. The possible
symmetry breaking patterns $\bm H=K\boxtimes L$ are given by $\bm
H=\mathbb Z_2\boxtimes \mathbb Z_2$, $\bm H=\mathbb Z_2\boxtimes \mathbb
Z_1$, and $\bm H=\mathbb Z_1\boxtimes \mathbb Z_1$, respectively. This is
shown in Fig.~\ref{fig:tc-condensation}c, where the horizontal layers are
arranged according to their ``ket-bra symmetry breaking gap'' $|K|/|L|$
corresponding to the number of anyons in the model, and the arrows
point in the direction of decreased symmetry.  

Let us now consider the three possibilities case by case.
\begin{enumerate}
\item 
$\bm H=\mathbb Z_2\boxtimes \mathbb Z_1$: This is the topological case. On
the one hand, we have following Cond.~\ref{condition:3-alpha-nu_g} that
$\langle\str{g}{\alpha}\otimes\strb{g}{\alpha}\rangle\ne0$ for all $g$ and
$\alpha$, since $(g,g)\in \bm H$ and 
the restriction of $(\alpha,\alpha)$ to $\bm H$ is
$(\alpha,\alpha)((h,h))=\alpha(h)\bar{\alpha}(h)=1=\nu_{(h,h)}$ [as
$\Htwo{\bm H}$ is trivial]; this is, all particles $\ex{g}{\alpha}$ are
unconfined. On the other hand,
$\langle\str{g}{\alpha}\otimes\strb{g'}{\alpha'}\rangle=0$ whenever either
$g\ne g'$ [as $(g,g')\notin \bm H$] or $\alpha\ne\alpha'$ [as then
$(\alpha,\alpha')((h,h))\not\equiv 1$]; this is, no particles are condensed.
\item 
$\bm H=\mathbb Z_1\boxtimes \mathbb Z_1$: This is the trivial phase in
which $\anyon{e}=\ex{0}{-1}$ is condensed (and thus $\anyon{m}=\ex{1}{1}$
is confined). Firstly, since all symmetries are broken, 
$\langle\str{g}{\alpha}\otimes\strb{g'}{\alpha'}\rangle=0$ whenever $g\ne0$ or
$g'\ne0$, which implies that $\anyon{m}$ and
$\anyon{f}=\anyon{e}\times\anyon{m}=\ex{1}{-1}$ are confined. On the other
hand, 
$\langle\str{0}{\alpha}\otimes\strb{0}{\alpha'}\rangle\ne0$, since
$(0,0)\in\bm H$ and $(\alpha,\alpha')$ restricted to $\bm H$ is trivially
the identity, and thus equals $\nu_{(0,0)}$.  
\item 
$\bm H=\mathbb Z_2\boxtimes \mathbb Z_2$: This is another trivial phase, in
which $\anyon{m}=\ex{1}{1}$ is condensed and $\anyon{e}=\ex{0}{-1}$
is confined. Firstly, note that while $\Htwo{\mathbb Z_2\boxtimes \mathbb
Z_2}=\mathbb Z_2$, we have that $qt=q=2$ and $\mathrm{gcd}(t,q)=1$,
i.e., only the trivial cocycle is allowed due to positivity. We have that 
$\langle\str{g}{1}\otimes\strb{g'}{1}\rangle\ne0$ since
$(g,g')\in \bm H$ and $(1,1)=\nu_{(g,g')}$, implying that $\anyon{m}$
is condensed.  On the other hand,
$\langle\str{g}{\alpha}\otimes\strb{g'}{\alpha'}\rangle=0$ whenever
$\alpha\ne1$ or $\alpha'\ne 1$, since $(\alpha,\alpha')(
(g,g'))=\alpha(g)\overline{\alpha'}(g')\not\equiv 1$ for some $(g,g')\in\bm H$,
i.e., $\anyon{e}$ and $\anyon{f}$ are confined.
\end{enumerate}

\subsection{Condensation of $D(\mathbb Z_4)$}

\begin{figure}[t]
\includegraphics[width=\columnwidth]{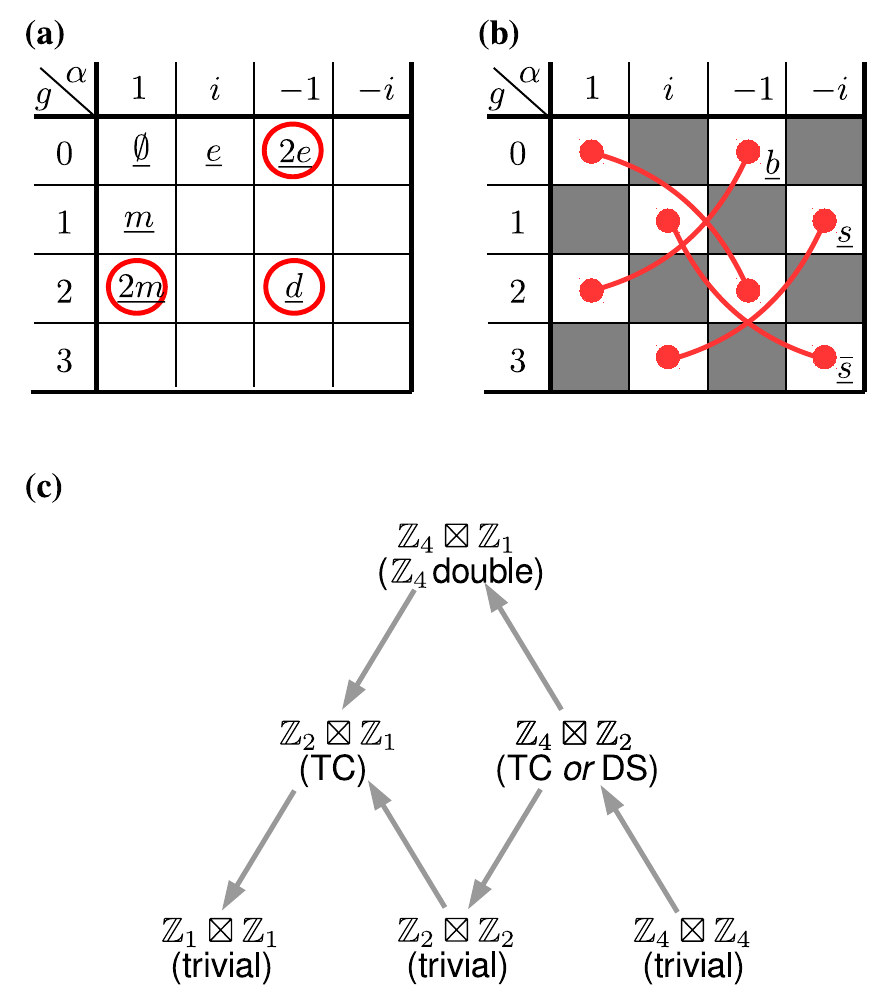}
\caption{\label{fig:example-z4}
\textbf{(a)} Anyon table for the $D(\mathbb Z_4)$ quantum double model.
The first row, first column, and the dyon $\anyon{d}$ have bosonic
self-statistics and can be condensed.  \textbf{(b)} Effective anyon model
obtained by condensing $\anyon{d}$: The gray anyons become confined, while
the remaining ones become identified as indicated. The resulting anyon
theory is a doubled semion model with semions $\anyon{s}$ and $\anyon{\bar
s}$ with self-statistics $i$.  
\textbf{(c)} Full condensation scheme for $D(\mathbb Z_4)$.  The diagram
lists the preserved symmetry $\bm H$ and the corresponding phase (TC=toric
code, DS=double semion).  The case $\bm H=\mathbb Z_4\boxtimes \mathbb
Z_2$ can describe two different topogical phases (TC and DS), depending on
the SPT phase of the fixed point.
}
\end{figure}

Let us now turn to our second example, the double $D(\mathbb Z_4)$. The
anyon table is given in Fig.~\ref{fig:example-z4}a; here, we find three
bosons (marked red), namely $\anyon{2e}$, $\anyon{2m}$, and the dyon
$\anyon{d}=\anyon{2e}\times\anyon{2m}$. It is straighforward to work out
the particle tables obtained by condensation: While condensing
$\anyon{2e}$ or $\anyon{2m}$ leads to two inequivalent toric codes,
condensing $\anyon{d}$ -- as shown in Fig.~\ref{fig:example-z4}b -- leads
to the so-called double semion model, with particles $\anyon{s}$ and
$\anyon{\bar s}$, which have (anti-)semionic self-statistics $g^\alpha=\pm
i$, and which fuse with themselves to the vacuum and with each other to
the non-trivial boson $\anyon{b}=\ex{0}{-1}\equiv\ex{2}{1}$. The double
semion model is not a regular double model but can be obtained by twisting
$D(\mathbb Z_2)$ with a non-trivial $3$-cocycle of $\mathbb Z_2$, and is
thus the simplest example of a twisted model obtained by condensing a
regular double.

Let us now study the possible symmetry breaking pattern $\bm H$ of
$D(\mathbb Z_4)$, shown in Fig.~\ref{fig:example-z4}c. We find six
possibilities.
\begin{enumerate}
\item 
$\bm  H = \mathbb Z_4\boxtimes \mathbb Z_1$.
This is the $D(\mathbb Z_4)$ phase; the discussion is analogous to the
case 1 for the Toric Code in Sec.~\ref{sec:subsec:cond-tcode} above.
\item 
$\bm  H = \mathbb Z_2\boxtimes \mathbb Z_1$.  This is a toric code phase
in which the $\anyon{2e}$ particle has been condensed.  We have that
$\langle \str{g}{\alpha}\otimes \strb{g'}{\alpha'}\rangle=0$ unless
$(g,g')\in\bm H$, i.e., $g=g'=0$ or $g=g'=2$, which implies that
$\ex{1}{*}$ and $\ex{3}{*}$ are all confined, and $\ex{2}{*}$ is
uncondensed. On the other hand, $\langle \str{g}{\alpha}\otimes
\strb{g}{\alpha'}\rangle\ne0$ iff $\alpha(g)\overline{\alpha'}(g)=1$ (as
there is only a trivial cocycle), and thus $\ex{1}{-1}$ is condensed, and
$\ex{0}{i}\equiv\ex{0}{-i}$ and $\ex{2}{1}\equiv \ex{2}{-1}$ form the
electric and magnetic particle of the Toric Code, respectively.
\item 
$\bm  H = \mathbb Z_4\boxtimes \mathbb Z_2$.
This is the first case with non-trivial $\Htwo{\bm H}=\mathbb Z_2$, and
thus exhibits two distinct condensed phases with identical symmetry
breaking pattern.\\
The phase with trivial cocycle corresponds to a Toric Code phase in which
the $\anyon{2m}$ particle has been condensed. 
First, $\langle \str{g}{\alpha}\otimes \strb{g'}{\alpha'}\rangle=0$ 
whenever $\alpha^h\overline{\alpha'}{}^{h'}\not\equiv 1$ for some $(h,h')\in \bm
H$, i.e.\ unless $\alpha=\alpha'=\pm1$, and thus $\ex{*}{\pm i}$ are
confined, while $\ex{*}{-1}$ is not condensed. On the other hand, 
$\langle \str{g}{\pm1}\otimes \strb{g'}{\pm1}\rangle\ne0$  iff $(g,g')\in
\bm H$: Thus, $\ex{2}{1}$ condenses, and $\ex{1}{1}\equiv\ex{3}{1}$ and
$\ex{0}{-1}\equiv\ex{2}{-1}$ form the new magnetic and electric particles,
respectively.

Let us now turn towards the phase with non-trivial cocycle. As we will
see, it corresponds to a double semion model with the condensation pattern
indicated in Fig.~\ref{fig:example-z4}.  It is straightforward to check
that for the non-trivial cocycle of $\Htwo{\bm H}=\mathbb Z_2$,
$\nu_{(g,g')}((h,h'))=i^{gh}(-i)^{g'h'}$ (e.g., by checking it on the
generators).  Then, $\langle \str{g}{\alpha}\otimes
\strb{g'}{\alpha'}\rangle=0$ whenever
$\alpha^h\overline{\alpha'}{}^{h'}\not\equiv \nu_{(g,g')}((h,h'))$ for
some $(h,h')\in \bm H$, i.e.\ unless $\alpha=\pm i^g$ and $\alpha'=\pm
i^{g'}$ (with the identical choice of $\pm$). This implies that all
$\ex{g}{\pm i^{g+1}}$ are confined, and only anyons $\ex{g}{i^g}$ can
condense. Since $\langle \str{g}{\alpha}\otimes \strb{0}{1}\rangle\ne0$
in addition requires $(g,0)\in \bm H$, we find that it is $\ex{2}{-1}$
which condenses.

\item 
$\bm  H = \mathbb Z_1\boxtimes \mathbb Z_1$.
This is a trivial phase where all $\anyon e$ particles have been
condensed; it is fully analogous to 
case 2 for the Toric Code in Sec.~\ref{sec:subsec:cond-tcode}.
\item 
$\bm  H = \mathbb Z_2\boxtimes \mathbb Z_2$.
This is a trivial phase where $\anyon{2e}$ and $\anyon{2m}$ have been
condensed.  We have that $\langle \str{g}{\alpha}\otimes
\strb{g'}{\alpha'}\rangle=0$  unless $g,g'\in\{0,2\}$, i.e., $\ex{1}{*}$ and
$\ex{3}{*}$ have been confined. It is also zero unless
$\alpha^g\overline{\alpha'}{}^{g'}=1$ for all $g,g'\in\{0,2\}$ (there is only the
trivial cocycle), and thus, $\ex{*}{\pm i}$ is confined as well. For all
remaining cases, $\langle \str{g}{\alpha}\otimes
\strb{g'}{\alpha'}\rangle\ne0$, and thus, all other particles are
condensed.
\item 
$\bm  H = \mathbb Z_4\boxtimes \mathbb Z_4$.
This is a trivial phase where all $\anyon m$ particles have been
condensed; it is fully analogous to 
case 3 for the Toric Code in Sec.~\ref{sec:subsec:cond-tcode}. Note that
there is again only the trivial cocycle.
\end{enumerate}

\subsection{Numerical study\label{sec:Z4-numerics}}

In the following, we provide numerical results on different topological
phases which can be obtained through condensation from a $D(\mathbb Z_4)$
double model, and the transitions between them. To this end, we have
constructed a three-parameter family interpolating between different fixed
point models, including the $D(\mathbb Z_4)$ phase, both Toric Code
phases, the double semion phase, and a trivial phase.  Here, we will limit
ourselves to a brief overview of the results; an in-depth discussion of
the specific wavefunction family considered as well as the numerical
methods used, together with additional results, will be presented
elsewhere~\cite{iqbal:preparation}.

Let us start by introducing the family of tensors used: 
\begin{equation}
\label{eq:numerics:def-3par-family}
A\left(\theta_{\mathrm{DS}},\theta_{\mathrm{TC},\mathbb
Z_2},\theta_{\mathrm{TC}}\right)=
\vcenter{\hbox{\includegraphics[width=.10\textwidth]{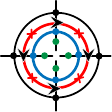}}}\
.
\end{equation}
Here, the four outside legs correspond to the virtual indices, while the
four inside legs are the physical indices.  The rings (and the green dots)
describe MPOs all of which mutually commute: 
\begin{itemize}
\item
The outermost black ring is the MPO of the $D(\mathbb Z_4)$ quantum
double, $\sum_g U_g^{\otimes 4}$, $U_g=X^g$, with $X$ the generator of
the regular representation of $\mathbb Z_4$.
\item
The red ring describes a deformation towards the MPO projector for the
$\mathbb Z_4\boxtimes\mathbb Z_2$ double semion model, where 
\[
\vcenter{\hbox{\includegraphics[width=.06\textwidth]{figs/ds_protensor_2}}}=(X^2)^{i}(Z^2)^{i+j},
\quad i,j=0,1\ ,
\]
with $Z$ the generator of the diagonal representation of $\mathbb Z_4$,
and
$\vcenter{\hbox{\includegraphics[width=.06\textwidth]{figs/z4qd_ds_1}}}=
\mathrm{diag}(\cosh \tfrac{\theta_{\mathrm{DS}}}{2}, \sinh
\tfrac{\theta_{\mathrm{DS}}}{2})$.  For $\theta_{\mathrm{DS}}=\infty$ (and
$\theta_{\mathrm{TC},\mathbb Z_2}=\theta_{\mathrm{TC}}=0$), this gives the
double semion MPO, while for $\theta_{\mathrm{DS}}=0$, it acts trivially. 
\item 
The blue ring describes a deformation towards the $\bm H=\mathbb
Z_2\boxtimes\mathbb Z_1$
Toric Code, where
\[
\raisebox{-1.2em}{\includegraphics[width=3em]{figs/z4qd_ds_3_0}}=
\delta_{ij}\exp((-1)^{i}\theta_{\mathrm{TC},\mathbb Z_2}Z^2),\ i,j=0,1\ .
\]
For $\theta_{\mathrm{TC},\mathbb Z_2}=\infty$, this projects the 
$D(\mathbb Z_4)$ MPO onto a $\mathbb Z_2$ subgroup and thus yields the Toric
Code, while for $\theta_{\mathrm{TC},\mathbb Z_2}=0$, it acts trivially.
\item 
Green circles
$\vcenter{\hbox{\includegraphics[width=3em]{figs/z4qd_ds_2}}}=\exp\left(\theta_{TC}X^2\right)$
describe a deformation towards a $\bm H=\mathbb Z_4\boxtimes \mathbb Z_2$
Toric Code phase: For $\theta_{\mathrm{TC}}=\infty$, this enhances the
symmetry of the $D(\mathbb Z_4)$ MPO to $\bm H=\mathbb Z_4\boxtimes Z_2$,
while for $\theta_{\mathrm{TC}}=0$, it once again acts trivially.
\end{itemize}
The two Toric Code constructions correspond to the two ways of embedding a
``normal'' $\mathbb Z_2\boxtimes \mathbb Z_1\subset \mathbb
Z_2\times\mathbb Z_2$ Toric Code into a $\mathbb Z_4\times \mathbb Z_4$
symmetry described in Sec.~\ref{sec:twdouble-embed-symmetry}.
Note that since all projectors commute with each other, their order does
not matter.

\begin{figure}[t]
\includegraphics[width=0.85\columnwidth]{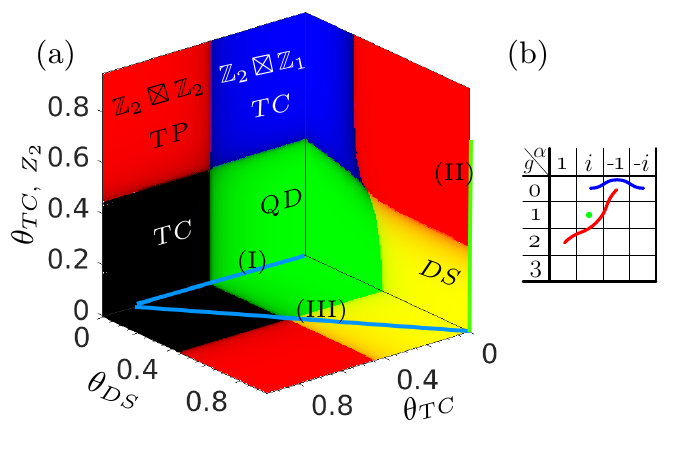}
\caption{
\label{fig:phasediagram}
Three surfaces in the $3$-parameter phase diagram described in
Sec.~\ref{sec:Z4-numerics}. The RGB values of each point give the
expectation value of the string order parameters indicated in panel (b),
cf.~also Fig.~\ref{fig:TC_Ising}: Red
$=\langle\str{2}{1}\otimes\strb{0}{-1}\rangle$, Green
$=\langle\str{1}{i}\otimes\strb{1}{i}\rangle$, Blue
$=\langle\str{0}{i}\otimes\strb{0}{-i}\rangle$, the three of which jointly
allow to discriminate all the phases observed.  QD, TC, DS, and TP denote
the $D(\mathbb Z_4)$ double model, toric code, double semion, and trivial
phase, respectively. Note that the phase diagram exhibits two distinct
toric code phases ($\mathbb Z_2\boxtimes\mathbb Z_1$, blue, and $\mathbb
Z_4\boxtimes \mathbb Z_2$, black), while all three trivial phases have
symmetry $\mathbb Z_2\boxtimes \mathbb Z_2$.  
} 
\end{figure}

\begin{figure}[t]
\includegraphics[width=0.95\columnwidth]{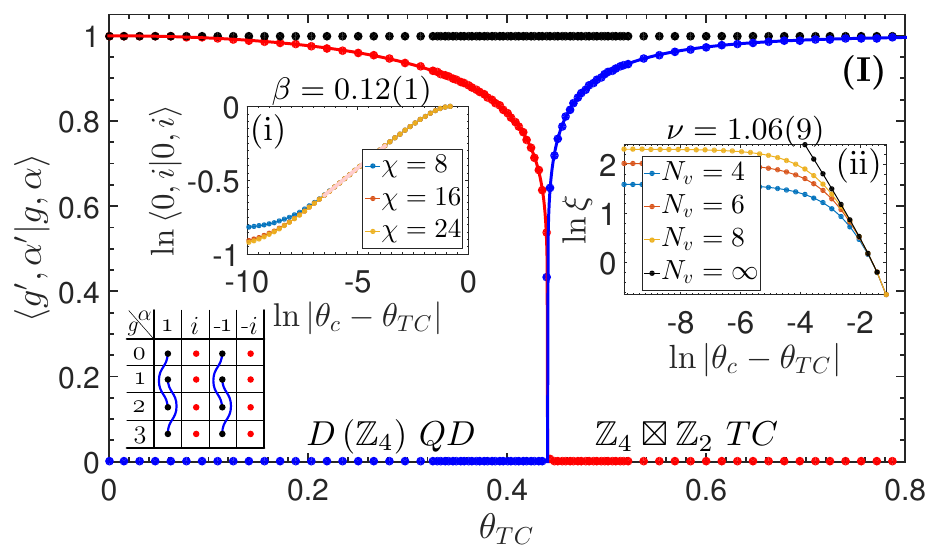}
\caption{
\label{fig:TC_Ising}
String order parameters for condensation and deconfinement for the
interpolation (I) in Fig.~\ref{fig:phasediagram}a, describing a $D(\mathbb
Z_4)$ to Toric Code transition,  obtained by approximating the fixed point
of $\mathbb T$ with an iMPS of bond dimension $\chi=24$.  The type of
string order parameters is encoded by the color, as indicated in the anyon
table in the lower left corner: Dots correspond to the deconfinement
parameter
$\langle\str{g}{\alpha}\otimes\strb{g}{\alpha}\rangle\equiv\langle
g,\alpha|g,\alpha\rangle$, while lines correspond to condensate fractions
$\langle\str{g}{\alpha}\otimes\strb{g'}{\alpha'}\rangle\equiv\langle
g,\alpha|g',\alpha'\rangle$ for the pairs they connect; order parameters
with the same color are (numerically) identical.  Specifically, the
blue line gives the condensate fraction of the $\anyon{2m}=\ex{2}{1}$
magnetic particle, and the red line measures the deconfinement of the
$\anyon{e}=\ex{0}{i}$ particle, which becomes confined if $\anyon{2m}$
condenses.  The solid line is the analytical result from the mapping to
the 2D Ising model, showing excellent agreement.  The upper left inset (i)
shows the scaling of the deconfinement parameter $\langle
\str{0}{i}\otimes\strb{0}{i}\rangle$ in the vicinity of critical point,
with critical exponent $\beta=0.12(1)$.  The right insets (ii) give the
correlation length around the critical point and the corresponding
critical exponent $\nu=1.06(9)$, extracted from exact diagonalization of
the transfer operator on cylinders of diameter $N_v$; the extrapolation
$N_v=\infty$ has been
obtained by fitting with $a\exp\left(-bN_v\right)+C_\infty$.  We find that
the critical exponents are the same on both sides of the phase transition.
}
\end{figure}

We have studied the phase diagram of this family using infinite Matrix
Product States (iMPS) to approximate the fixed point of the transfer
operator, by iteratively applying the transfer operator and truncating the
bond dimension to some given $\chi$, keeping translational symmetry. From
the resulting fixed point iMPS, we can then [using
Eq.~(\ref{eq:sop-in-mps-ring})] immediately compute the order parameters
for condensation, $\langle\str{g}{\alpha}\otimes\strb{0}{1}\rangle$, and
deconfinement, $\langle\str{g}{\alpha}\otimes\strb{g}{\alpha}\rangle$,
respectively, allowing us to distinguish the different topological phases
and map out the phase diagram.  The condensation and deconfinement order
parameters also allow us to study the nature of the phase transitions.
Notably, this gives us non-zero order parameters, and thus critical
exponents $\beta$, for both sides of a condensation-driven phase
transition: in the uncondensed phase, the deconfinement order parameter is
non-zero, while in the condensed phase, the condensate fraction is
non-zero. Note that we use the string operator corresponding to
excitations in the fixed point wavefunction to measure the order parameter
throughout the phase diagram; this is in exact analogy to the use of order
parameters in conventional phase transitions.  In addition to that, we can
further characterize the phase transition by looking at the scaling of the
correlation length $\xi$, which we can extract either from the fixed point
iMPS, or from the finite-size transfer operator and a finite size scaling
(note though that this length does need not be equal to the physical
correlation length, as it includes e.g.\ certain anyon-anyon correlation
functions).

In order to understand the structure of our three-parameter family,
Eq.~(\ref{eq:numerics:def-3par-family}), we have computed the different
condensation and deconfinement order parameters along the three
hyperplanes for which one $\theta_\bullet=0$; the resulting phase diagram
is shown in Fig.~\ref{fig:phasediagram}.  We find that the system exhibits
all phases encoded by the three MPOs in
Eq.~(\ref{eq:numerics:def-3par-family}), as well as a trivial phase with
$\bm H=\mathbb Z_2\boxtimes \mathbb Z_2$, which can be understood
analytically in the limit where two of the $\theta_\bullet$ are taken to
infinity. As expected, the family thus exhibits phase transitions related
to the condensation of anyons from $D(\mathbb Z_4)$ to Toric Code and
Double Semion, and from either to the trivial phase; more notably, though,
the family also exhibits direct phase transitions between the Toric Code
and the Double Semion model, which are not related by anyon condensation.

\begin{figure}[t]
\centering
\includegraphics[width=0.96\columnwidth]{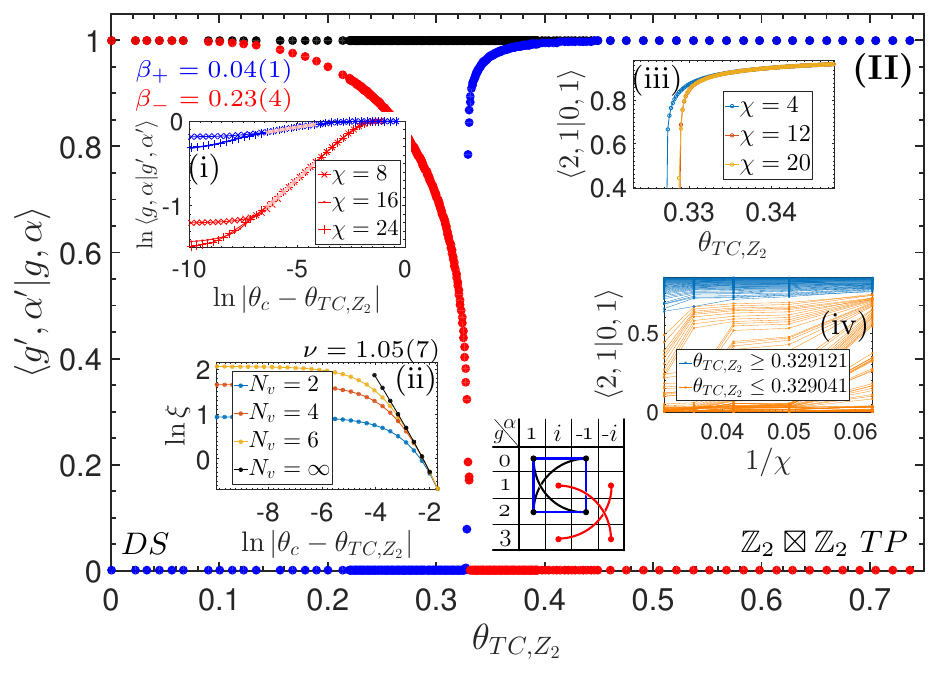}
\caption{
\label{fig:DS-trivial}
Phase transition between double semion model ($\mathbb Z_4\boxtimes Z_2$
symmetry) and trivial phase ($\mathbb Z_2\boxtimes \mathbb Z_2$ symmetry)
along line (II) in Fig.~\ref{fig:phasediagram}. The plot shows the
deconfinement/condensate order parameters (color coded as indicated by the
dots/lines in the anyon table in the inset) vs.\ the interpolation
parameter $\theta_{\mathrm{TC},\mathbb Z_2}$;  see Fig.~\ref{fig:TC_Ising}
for details.  Insets (i) and (ii) are the same as in
Fig.~\ref{fig:TC_Ising}, giving fits for the critical exponents
$\beta_{\pm}$ and $\nu$, respectively.  Inset (iii) shows the behavior
condensate fraction $\langle\str{2}{1}\otimes\strb{0}{1}\rangle \equiv
\langle2,1|0,1\rangle$ (blue in the main plot) close to the phase
transition for different iMPS bond dimensions $\chi$, demonstrating
convergence to a smooth (albeit steep) curve.  In inset (iv), the same
data is plotted against $1/\chi$ for different $\theta_{TC,\mathbb Z_2}$,
reconfirming the second-order nature of the transition [compare with inset
(ii) of Fig.~\ref{fig:1st2ndDSTC}b!], and allowing us to accurately
localize the phase transition.  
} 
\end{figure}

\begin{figure*}[t]
\centering
\includegraphics[width=0.47\textwidth]{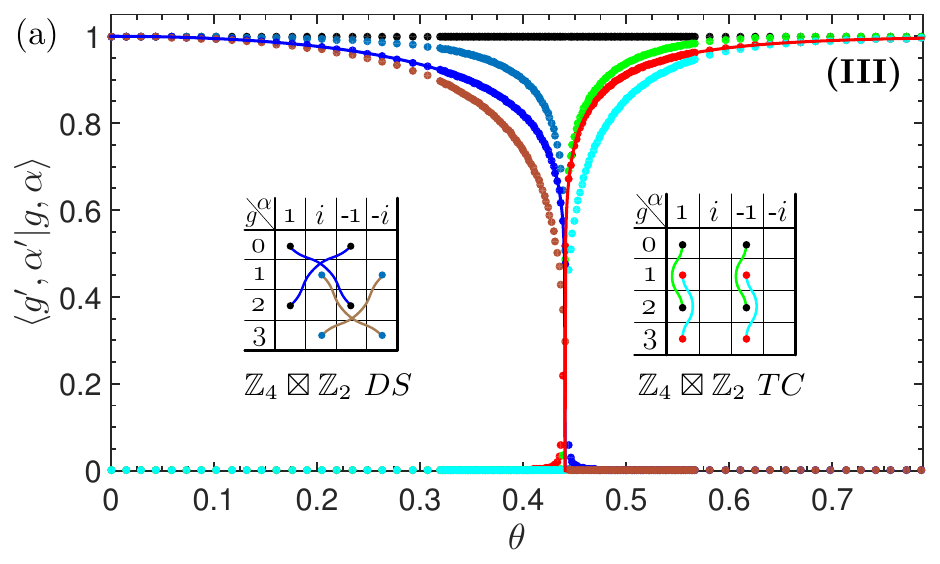}
\hspace*{0.02\textwidth}
\includegraphics[width=0.47\textwidth]{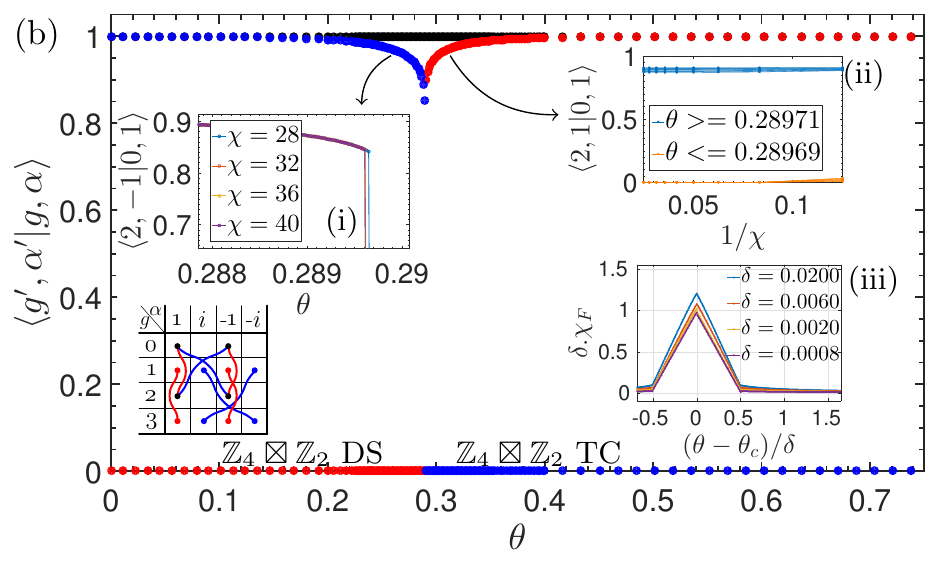}
\caption{
\label{fig:1st2ndDSTC}
Condensation and deconfinement order parameters for two phase transitions
between the Double Semion model (left) and the $\mathbb Z_4\boxtimes
\mathbb Z_2$ Toric Code (right), corresponding to an SPT transition at the
boundary, cf.~Fig.~\ref{fig:TC_Ising}. 
\textbf{(a)} Interpolation along the line (III) in
Fig.~\ref{fig:phasediagram}a, cf.~inset for the color coding.  The order
parameters for condensation of the dyon $\anyon{d}=\ex{2}{-1}$ (blue) and
deconfinement of the magnon $\anyon{m}=\ex{1}{1}$ (red) can be mapped to
the magnetization of the 2D Ising model (solid lines).
\textbf{(b)}
Interpolation obtained by interpolating the per-site transfer operator,
cf.\ text, showing clear signs of a first-order phase transition.  This is
further supported by the analysis in the insets: (i) gives the behavior of
the condensate fraction of the dyon $\anyon{d}=\ex{2}{-1}$ (blue) in the
vicinity of the phase transition for different $\chi$, showing convergence
to a discontinuous curve. (ii) gives the same data as a function of
$1/\chi$ for different $\theta$, and we observe a sharp change of the
behavior at $\theta\approx 0.28970$. [(i) and (ii) should be contrasted
with insets (iii) and (iv) of Fig.~\ref{fig:DS-trivial}.] Finally, inset
(iii) gives the fidelity susceptibility $\chi_F$,
measuring the rate of change of the normalized wave function
$\ket{\hat\psi(\theta)}$ with $\theta$,
$|\langle\hat\psi(\theta)\ket{\hat\psi(\theta+\mathrm{d}\theta)}| =
1-\tfrac12\chi_FN^2(\mathrm{d}\theta)^2$, for
different step sizes $\mathrm{d}\theta=\delta\rightarrow0$, which
approaches a delta peak as expected for a first-order
transition~\cite{zanardi:overlap-curvature,gu:fidelity-phasetransitions}.
}
\end{figure*}

We have studied a number of these phase transitions in more detail; in the
following, we illustrate our findings through a few examples, and refer
the reader for a more detailed analysis to Ref.~\cite{iqbal:preparation}.
First, we have studied the phase transitions in the
$\theta_{\mathrm{TC},\mathbb Z_2}=0$ plane.  Fig.~\ref{fig:TC_Ising} shows
the order parameters along line (I) in Fig.~\ref{fig:phasediagram}, which
describes a $D(\mathbb Z_4)$ to Toric Code transition. Since we have an
analytical mapping to the 2D Ising model for this line, it can serve as a
benchmark, and we find indeed very good agreement with the analytic
predictions.  Further study suggests the existence of an analytical
mapping for the entire $\theta_{\mathrm{TC},\mathbb Z_2}=0$ plane; for the
$\theta_{\mathrm{DS}}=0$ plane, the critical exponents still match those of
the 2D Ising model, though the existence of an exact mapping is unclear.
On the other hand, the transitions in the $\theta_{\mathrm{TC}}=0$ plane
seem to belong to  a different universality class.  As an example,
Fig.~\ref{fig:DS-trivial} shows the transition along the line (II) in
Fig.~\ref{fig:phasediagram}, for which we find critical exponents
$\nu=1.05(7)$ for the correlations in the fixed point of the transfer
operator, and $\beta_+=0.04(1)$ and $\beta_-=0.23(4)$ for the anyon
condensation and deconfinement order parameters, respectively; notably,
the critical exponent $\beta$ is different on the two sides of the
transition. We observe that the critical exponents $\beta_{\pm}$ change
continously as we move along the transition line in the
$\theta_{\mathrm{DS}}=1$ plane towards the $\theta_{\mathrm{TC},\mathbb
Z_2}$ plane, ultimately reaching $\beta_{\pm}=1/8$; a detailed discussion
will be given elsewhere~\cite{iqbal:preparation}.

Finally, let us turn towards the direct Toric Code -- Double Semion
transition, previously only studied with exact diagonalization and on
quasi-1D systems~\cite{morampudi:z2-phase-transition}, whose nature is yet
to be resolved.  As one would assume that interactions generally give
rise to condensation of excitations, one expects that an interpolation
between the two models would typically drive the Toric Code through some
condensation transition, either into a trivial or a more complex phase
[such as the $D(\mathbb Z_4)$ model], and from there through another
condensation-driven transition to the Double Semion model, and a direct
transition would at least require some fine-tuning of interactions.

We can identify one such fine-tuned transition between the ($\bm H=\mathbb
Z_4\boxtimes \mathbb Z_2$) Toric Code and Double Semion phase in our phase
diagram in the $\theta_{\mathrm{TC},\mathbb Z_2}=0$ plane at
$\left(\theta_{\mathrm{DS}}^c,\theta^c_\mathrm{TC}\right)=
\left(\tfrac12\ln(1+\sqrt 2), \ \tfrac12\ln(1+\sqrt 2)\right)$;
this is a multi-critical point adjacent to all four phases which goes away
as one perturbs away from $\theta_{\mathrm{TC},\mathbb Z_2}=0$, separating
the Toric Code from the Double Semion phase.  Fig.~\ref{fig:1st2ndDSTC}a
shows the transition through this point along line (III) in
Fig.~\ref{fig:phasediagram}, and we find that it is a second order phase
transition, driven by two ``counterpropagating'' condensation and
de-condensation transitions, thus preserving the total number of anyons;
like all transitions in that plane, it is again in the 2D Ising universality
class.  Note however that this is a phase transition between two phases
with an identical $\bm H=\mathbb Z_4\boxtimes \mathbb Z_2$ symmetry at the
boundary,  and therefore corresponds to an SPT phase transition at the
boundary in the absence of symmetry breaking, and can therefore only be
detected by string order parameters rather than conventional local order
parameters.  Note however that it has been shown that in certain cases
string order parameters can be mapped to local order parameters through a
duality mapping~\cite{duivenvoorden:stringorder-symbreaking}.

As it turns out, there is another way of obtaining a direct phase
transition between the $\bm H=\mathbb Z_4\boxtimes \mathbb Z_2$ Toric Code
and Double Semion phase, namely by interpolating between the on-site
transfer operators $A^\dagger A$ of the two fixed point models, rather
than the tensors $A$ themselves. Since such an interpolation $\mathbb
E(\theta)=\theta A_0^\dagger A_0 + (1-\theta) A_1^\dagger A_1$ yields a
positive semidefinite $\mathbb E(\theta)\ge0$, we can construct a
continuous path $A(\theta)$ of PEPS tensors by decomposing
$\mathbb E(\theta)=A(\theta)^\dagger A(\theta)$.  This interpolation
yields again a direct transition between the two phases, and a thorough
analysis of the order parameters, shown in Fig.~\ref{fig:1st2ndDSTC}b,
gives compelling evidence that the phase transition is first order.  Thus,
in order to understand the nature  of a generic Toric Code -- Double
Semion phase transition (given it can even be realized in a robust way)
requires further study. In this context, it is an interesting question
whether imposing specific symmetries on the system allows one to
generically obtain a direct transition between these two phases, rather
than requiring fine-tuning of the interactions.

\section{\label{sec:conclusions}Conclusions and outlook}

In this paper, we have studied anyon condensation in Projected Entangled
Pair State models, and have derived conditions governing the condensation and
confinement of anyons.  In order to do so, we have related the behavior of
anyons to string order parameters and thus symmetry protected order
in the fixed point of the transfer operator, this is, the entanglement
spectrum of the system.  We have derived four conditions: Two characterize
the possible symmetry breaking and SPT phases consistent with positivity
of the entanglement spectrum, while the other two related these symmetry
breaking and SPT patterns to the condensation and confinement of anyons.
Specifically, we found that there are topological phases which cannot be
distinguished through their symmetry breaking pattern, but solely through
the SPT structure of their entanglement spectrum, and which describe
phases not related by anyon condensation.  For the case of cyclic groups,
this classification allowed to construct all twisted doubles by condensing
non-twisted double models.

We have exemplified our discussion with the $\mathbb Z_4$ quantum double,
which can give rise to both Toric Code and Double Semion phases which form
an example of phases with identical symmetry breaking pattern but
inequivalent SPT order in the entanglement spectrum. We have also provided
numerical results for the phase diagram and the phase transitions of the
model.  To this end, we have used that the concepts developed in this paper
allow us to measure order parameters for condensation and deconfinement
and thus extract critical exponents for the order parameter. In
particular, we found that this model can realize direct phase transitions
between the Toric Code and Doubled Semion models which are not related by
anyon condensation, and for which we found both first and second order
transitions.

A natural question is the interpretation of symmetry broken and SPT phases
in the fixed point of the transfer operator in terms of physical
properties of the entanglement spectrum and/or edge
physics~\cite{yang:peps-edgetheories}: Symmetries $U_g\rho
U_g^\dagger=\rho$ imply that the entanglement spectrum $\rho$ is
block-diagonal, i.e., it originates from a symmetric Hamiltonian. An
additional single-layer symmetry $U_g\rho=\rho$ implies that the density
operator must live in the trivial irrep sector, while a broken symmetry
and the resulting dependence on distant boundary conditions implies the
existence of a non-local anomalous term in the entanglement Hamiltonian
which depends on distant boundaries and encodes a topological
superselection rule~\cite{schuch:topo-top}.  The implications of SPT order
on the entanglement spectrum, on the other hand, are much less clear, and
it would be very interesting to identify the features of the entanglement
spectrum which would allow to distinguish e.g.\ Toric Code and Doubled Semion order.

It is likely that our results generalize to the case of non-abelian
groups, and beyond that to general Matrix Product Operator
symmetries~\cite{bultinck:mpo-anyons}. An obstacle is that the one-to-one
correspondence between string order parameters and SPTs breaks
down~\cite{pollmann:spt-detection-1d}: While it is known that non-abelian
SPTs are still characterized by group cohomology, we have used SPT phases
to classify the behavior of string order parameters rather than the other
way around, and are thus looking for a classification of the
behavior of non-abelian string order parameters instead. Let us note, however, that a major
simplification might come from the fact that for non-abelian double
models, the irrep at the end of a string must be an irrep of its
normalizer, so it might well be possible that the problem can be
abelianized to an extent which allows to yet again relate it to SPT order.
A related question is the generalization of our results to the
case of non-hermitian transfer operators, or even PEPS which
encode a corresponding global symmetry in a non-trivial way.  In that
case, string order parameters are evaluated between non-identical left and
right fixed points, and the analogy to expectation values in physical
states, and thus the correspondence of string order parameters with SPT
phases, breaks down; for instance, it is not even clear whether the
projective symmetry representation for pairs of left and right fixed
points must be equal.

Finally, the maybe most important question, which goes far beyond the
scope of this work, is a rigorous justification of our main technical
assumption, namely that the structure of the fixed point space of a
transfer operator for a PEPS in a gapped phase is well described by Matrix
Product Operators.  While this is well motivated due to the short-range
nature of the correlations in the system, and is well-tested numerically
through numerous PEPS simulations using contraction schemes which model the
boundary as an MPO, it has withstood rigorous assessment up to now. A
better understanding of this question would lead to a
number of important insights regarding the structure of gapped phases, the
nature of the entanglement spectrum, or the convergence of numerical
methods, just to name a few.

\acknowledgements

We acknowledge helpful conversations with 
M.~Barkeshli,
N.~Bultinck,
M.~Marien, 
B.~Sahinoglu,
C.~Xu,
and B.~Yoshida.
This work has received support by the EU through the FET-Open project QALGO
and the ERC Starting Grant No.~636201 (WASCOSYS), 
the DFG through Graduiertenkolleg 1995,
and the J\"ulich Aachen Research Alliance (JARA) through JARA HPC
grant jara0092 and jara0111.

\appendix

\section{Construction of explicit endpoints 
\label{app:explicit-endpoints}}

In this appendix, we provide an explicit construction for all anyons
$\ex{g}{\alpha}$ which are either condensed or deconfined following
Condition~\ref{condition:3-alpha-nu_g}, i.e.\ $\bm \alpha\vert_{\bm
H}=\nu_{\bm g}$.  To this end, we proceed in two steps: First, we
generalize the construction of Eq.~(\ref{eq:explicit_Ralpha}) to obtain
$\bm{R}_{\bm \alpha}$ which transform as irreps $\bm \alpha$ of $\bm G$
rather than only $\bm H$.  Second, we show that for the case of
condensation, $\bm g=(g,e)$ and $\bm\alpha=(\alpha,1)$, and for the case of
deconfinement, $\bm g =(g,g)$ and $\bm\alpha=(\alpha,\alpha)$, these
$\bm R_{\bm \alpha}$ allow to construct actual anyons, i.e., single-layer
endpoints, for which $\langle\strk{g}{\alpha}\rangle\ne0$; this is exactly
what is also required in
Section~\ref{sec:subsection-anyon-condensation-rules}, where we derive the
anyon condensation rules from
Conditions~\ref{condition:1-structure-of-H}--\ref{condition:4-proj-reps-commute}.

\subsection{Construction of $\bm R_{\bm\alpha}$ for irreps of $\bm G$}

Let $\bm g\in \bm H$, and $\bm\alpha$ an irrep of $\bm G$ such that
$\bm\alpha\vert_{\bm H}=\nu_{\bm g}$. The idea of
Eq.~(\ref{eq:explicit_Ralpha}) was to use injectivity of the MPS tensor
$M$ to define $\bm R_{\bm\alpha}$ such that 
\[
\raisebox{-3em}{\includegraphics[scale=0.7]{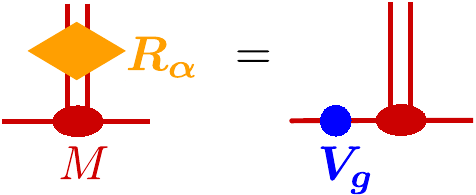}}\quad.
\]
The tensor $M$ describes one symmetry-broken sector (with residual
symmetry group $\bm H$) only. In order to construct some $\bm
R_{\bm\alpha}$ which transforms as an irrep of $\bm G$, we therefore first
need to construct an MPS which does not break the symmetry.  To this end,
choose representants $\bm f_{\mathfrak a}\in \bm G$ of every
symmetry-broken sector $\mathfrak{a}\in \bm G/\bm H$, such
that 
\[
\bm G=\bigoplus_{\mathfrak a\in \bm G/\bm H} \bm f_{\mathfrak a}\bm H\ ;
\]
by starting from the generators of the quotient group $\bm G/\bm H$, it is
possible to pick $\bm f_{\mathfrak a}$ such that $\bm
f_{\mathfrak{ab}}=\bm f_{\mathfrak a}\bm f_{\mathfrak b}$.  Now define
\[
\includegraphics[scale=0.7]{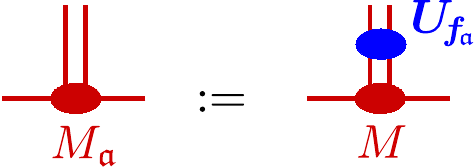}
\]
and $\mathcal M^i = \bigoplus_{\mathfrak a} M^i_\mathfrak{a}$; clearly,
$\mathcal M^i$ is block-injective (i.e., injective on the space of
block-diagonal matrices).  Given $\bm k\in \bm G$, there is a unique
decomposition $\bm k=\bm f_{\mathfrak a}\bm h$, $\bm h\in \bm H$, and thus 
\[
\raisebox{-2.6em}{\includegraphics[scale=0.7]{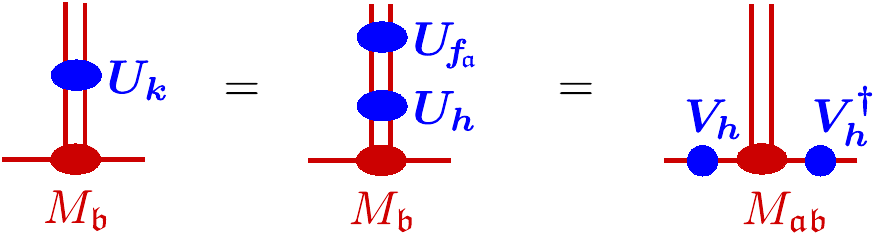}}\quad;
\]
this is, the virtual action of $\bm U_{\bm k}$ is 
\[
\mathcal V_{\bm k}:=\left(\bigoplus \bm V_{\bm h}\right)\Pi_{\mathfrak a}\ ,
\]
where $\Pi_\mathfrak{a}$ permutes the blocks by virtue of
$\mathfrak{b}\mapsto \mathfrak{a}^{-1}\mathfrak{b}$; note that $\mathcal
V_{\bm k}$ forms a projective representation of $\bm G$ (the trivial
induced projective representation induced by $\bm V_{\bm h}$). Now define 
\[
\mathcal W := \bigoplus_{\mathfrak b}
    {\bm\alpha(\bm f_{\mathfrak b})} \bm V_{\bm g}\ ,
\]
and choose $\bm R_{\bm \alpha}$ such that
\[
\includegraphics[scale=0.7]{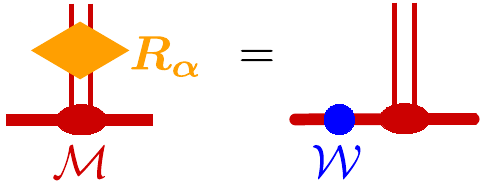}
\]
-- this is always possible since $\mathcal W$ is block-diagonal and $\mathcal M$ is
block-injective. (We use a thick line to indicate the larger ``direct
sum'' virtual space.) We now have that
\begin{align*}
\mathcal V_{\bm k}\mathcal W\mathcal V_{\bm k}^\dagger 
& = 
\left[\Big(\bigoplus \bm V_{\bm h}\Big)\Pi_{\mathfrak a}\right]
    \Big[\bigoplus_{\mathfrak b}
    {\bm\alpha(\bm f_{\mathfrak b})} \bm V_{\bm g}\Big]
    \left[\Pi_{\mathfrak a}^\dagger\Big(\bigoplus \bm V^\dagger_{\bm h}\Big)\right]
\\
& = 
\left[\Big(\bigoplus \bm V_{\bm h}\Big)\right]
    \Big[\bigoplus_{\mathfrak b'}
    {\bm\alpha(\bm f_{\mathfrak a\mathfrak b'})} \bm V_{\bm g}\Big]
    \left[\Big(\bigoplus \bm V_{\bm h}^\dagger\Big)\right]
\\
& = 
\nu_{\bm g}(\bm h)\Big[\bigoplus_{\mathfrak b'}
    {\bm\alpha(\bm f_{\mathfrak a\mathfrak b'})} \bm V_{\bm g}\Big]
= 
\nu_{\bm g}(\bm h)\bm\alpha(\bm f_{\mathfrak a}) \mathcal W 
\\
&= 
\bm \alpha(\bm k) \mathcal W\ ,
\end{align*}
where we have used $\nu_{\bm g}(\bm h)=\bm\alpha(\bm h)$. It immediately
follows that
\[
\includegraphics[scale=0.7]{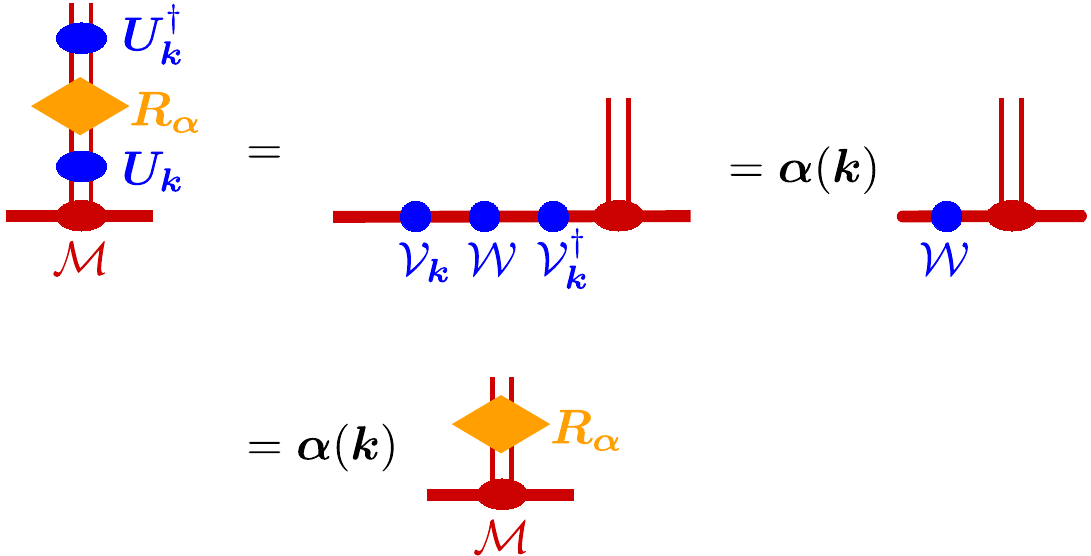}
\]
i.e., $\bm R_{\bm \alpha}$ indeed transforms as the irrep $\bm \alpha$ of
$\bm G$.

\subsection{Explicit construction of condensed anyons}

Let us now show that we can explicitly construct condensed anyons
$\ex{g}{\alpha}$: Given a $\bm R_{(\alpha,1)}$ for which
$\langle\strsemik{(g,e)}{(\alpha,1)}\rangle\ne0$, we show how to construct
a single-layer anyon (i.e., an endpoint to a string of $g$'s transforming
like $\alpha$) with non-zero expectation value 
$\langle \str{g}{\alpha}\otimes\strb{e}{1}\rangle\ne0$, where the endpoint
in the bra layer is trivial. To this end, we start by decomposing
\[
\bm R_{(\alpha,1)} = \sum X^s_\alpha \otimes\bar{Y}^s_1\ ,
\]
where $X_\alpha$ and $Y_1$ transform like $\alpha$ and trivially,
respectively.  Since $R_{(\alpha,1)}$ gives a non-zero expectation value
$\langle\strsemik{(g,e)}{(\alpha,1)}\rangle\ne0$, there must be at least
one $s_0$ for which this also holds; we thus obtain a \emph{separable}
endpoint $X^{s_0}_\alpha\otimes \bar{Y}^{s_0}_1\equiv
X_\alpha\otimes \bar{Y}_1$ with 
$\langle\strsemik{(g,e)}{(\alpha,1)}\rangle\ne0$; however, $Y_1$ can still
be different from the identity.  In order to make the endpoint in the bra
layer entirely trivial, we use that
\[
\includegraphics[scale=0.7]{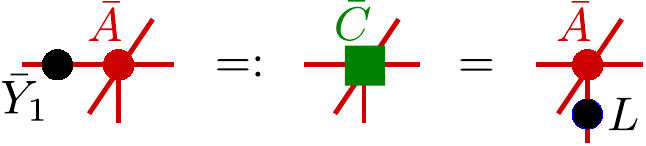}
\]
since $A$ is $G$-injective (and $C$ is $G$-invariant), and thus, 
\[
\includegraphics[scale=0.7]{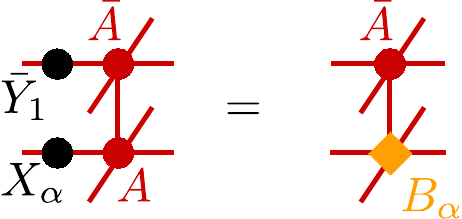}
\]
with the endpoint 
\[
\includegraphics[scale=0.7]{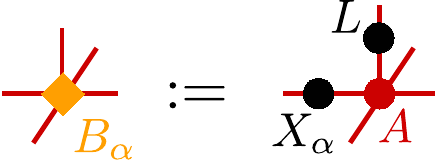}
\]
for the condensed anyon $\ex{g}{\alpha}$.

Note that a simple application of Cauchy-Schwarz yields that any condensed
anyon is also deconfined.

\subsection{Explicit construction of deconfined anyons}

Similar to the preceding section, in this scenario we start from some $\bm
R_{(\alpha,\alpha)}$ s.th.\ 
$\langle\strsemik{(g,g)}{(\alpha,\alpha)}\rangle\ne0$, corresponding to a
deconfined anyon $\ex{g}{\alpha}$, and want to construct \emph{identical}
endpoints $Z_\alpha$ for the ket and bra layer such that 
$\langle \str{g}{\alpha}\otimes\strb{g}{\alpha}\rangle\ne0$.
We again start by decomposing 
\[
\bm R_{(\alpha,\alpha)} = \sum X^s_\alpha \otimes\bar{Y}^s_\alpha\ .
\]
Let us define the shorthand $\langle X_\alpha\otimes Y_\alpha\rangle:=
\langle \str{g}{\alpha}\otimes\strb{g}{\alpha}\rangle$, where $S\otimes
\bar S$ has endpoints $X\otimes Y$.  Now pick $s_0$ such that $\langle
X_\alpha\otimes \bar{Y}_\alpha\rangle\equiv\langle X^{s_0}_\alpha\otimes
\bar{Y}^{s_0}_\alpha\rangle \ne 0$.  If 
$\langle X_\alpha\otimes \bar{X}_\alpha\rangle\ne 0$ or
$\langle Y_\alpha\otimes \bar{Y}_\alpha\rangle\ne 0$, we can choose
$Z_\alpha:=X_\alpha$ (or $Z_\alpha:=Y_\alpha$), and have found the desired
non-vanishing identical ket and bra endpoint $\langle Z_\alpha\otimes
Z_\alpha\rangle\ne0$.  Let us now consider the case where
both are zero. Let $\phi$ such that $\langle X_\alpha\otimes e^{-i\phi}\bar
Y_\alpha\rangle>0$, and define $Z_\alpha := X_\alpha+e^{i\phi}Y_\alpha$. Then, 
\begin{align*}
\langle Z_\alpha \otimes \bar Z_\alpha\rangle
&=
\langle X_\alpha\otimes \bar X_\alpha\rangle + 
\langle Y_\alpha\otimes \bar Y_\alpha\rangle  \\
&\quad\qquad+\langle X_\alpha\otimes e^{-i\phi}\bar Y_\alpha\rangle + 
\langle e^{i\phi} Y_\alpha\otimes \bar X_\alpha\rangle 
\\
&= 2\mathrm{Re}\,\langle X_\alpha\otimes e^{-i\phi}\bar Y_\alpha\rangle >
0\ ,
\end{align*}
thus again yielding identical endpoints $Z_\alpha$ for ket and bra with
non-vanishing expectation value.

\section{Generalization to dressed endpoints
\label{sec:app:dressed}}

Let us now show that the no-go results of
Conditions~\ref{condition:1-structure-of-H}--\ref{condition:4-proj-reps-commute}
derived in Sec.~\ref{sec:classification} equally hold for general
endpoints; the explicit construction for any endpoint compatible with all
the conditions has already been provided in
Appendix~\ref{app:explicit-endpoints}. Let us recall that a general anyon
is of the form 
\[
\raisebox{-0.8cm}{\includegraphics[scale=0.6]{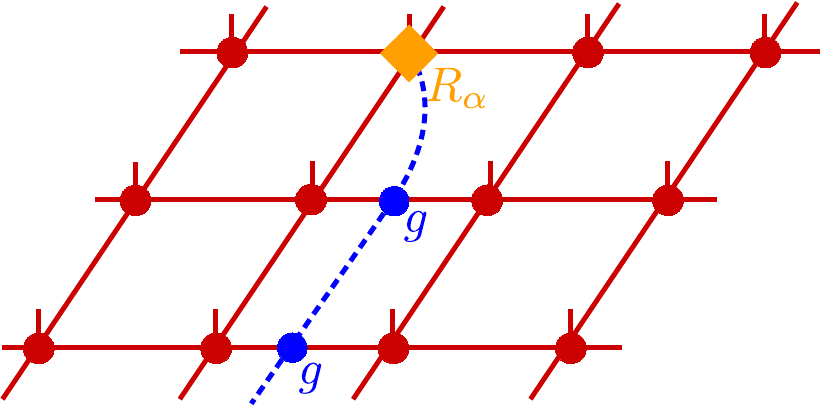}}\ .
\]
For deriving the no-go results, we generally need to consider joint
ket-bra objects; we thus define
\[
\raisebox{-0.55cm}{\includegraphics[scale=0.7]{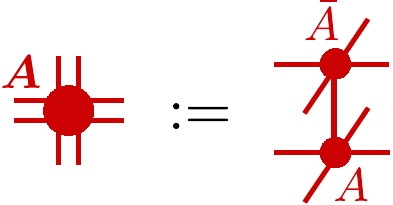}}
\mbox{\quad and\quad} 
\raisebox{-0.55cm}{\includegraphics[scale=0.7]{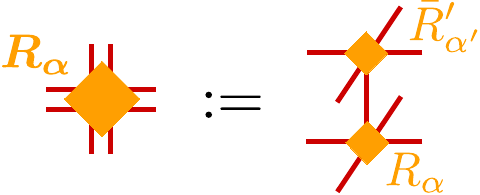}}\ .
\]
The generalization of Eq.~\eqref{eq:sec3:Occ'-def}, describing a general
string order parameter for a ket and bra anyon pair, evaluated in a pair
of fixed points $\vket{\rho_{\bm c}}$ and $\vbra{\rho_{\bm c'}}$, is thus
of the form
\begin{equation}
\label{eq:app2-Occ}
O_{\bm c}^{\bm c'}:=
    \raisebox{-1.3cm}{\includegraphics[scale=0.7]{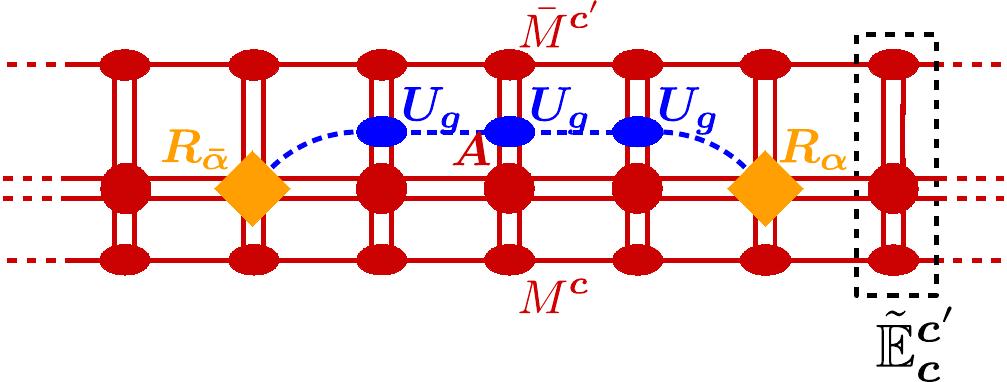}}
\quad.
\end{equation}
Just as in Sec.~\ref{sec:classification}, a central role will be played by
the (mixed) transfer operator $\tilde{\mathbb E}_{\bm c}^{\bm c'}$; we
will therefore analyze its structure in detail in the following.  

\subsection{Structure of $\tilde{\mathbb E}_{\bm c}^{\bm c'}$}

The major complication as compared to the discussion in
Section~\ref{sec:classification} is that for an $M\equiv M^{\bm c}$
describing an injective MPS $\vket{\rho}$ which is a fixed point of the
transfer operator, the tensor 
\[
D:=\quad\raisebox{-1.6em}{\includegraphics[scale=0.7]{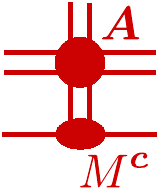}}
\]
describing the MPS obtained after applying $\mathbb T$ only needs to be
\emph{proportional} to $\vket{\rho}$, with a possibly size-dependent
proportionality constant.  This has two
consequences~\cite{perez-garcia:mps-reps,cirac:mpdo-rgfp}: First, $D$ can
consists of several diagonal blocks $D_{s,s}$, $s=1,\dots,S$ each of which
describes a copy of the original MPS, i.e., $D_{s,s}=\tilde M^i_s$,  where
each $\tilde M^i_s$ is equal to $M^i$ up to a block-dependent gauge
transform, $\tilde M^i_s = \gamma_s X_s M^i X_s^{-1}$ with some
left-invertible
$X_s$.  Second, there can in addition be off-diagonal blocks $D_{s,t}$
coupling blocks $s$ and $t$, which however---up to reordering of
blocks---must be upper triangular, i.e., $D_{s,t}\equiv 0$ if $s>t$.  This
implies that any product $D^{i_1}D^{i_2}\cdots D^{i_L}$ can contain each
off-diagonal block $D^i_{s,t}$ at most once, and in particular contains
only a finite number of off-diagonal blocks. W.l.o.g., we will assume that
$D$ is normalized such that the largest $|\gamma_s|=1$, with
the normalization of the $M^i$ as before.

Let us now consider what this implies when taking large powers $\mathbb
(\tilde{\mathbb E}^{\bm c'}_{\bm c})^K$, $K\rightarrow\infty$. In that
case, there will be large contiguous blocks of the form $F_s:=\sum D_s^{i,\bm
c'}\otimes \bar M^{i,\bm c}= \gamma_s\sum X_s M_s^{i,\bm
c'}X_s^{-1}\otimes \bar M^{i,\bm c}$ (specifically, there will be at least
one block with length at least $K/S$), which will therefore converge to
the fixed point of the corresponding original transfer operator $\mathbb
E^{\bm c'}_{\bm c}$, up to normalization and a gauge transform.  In
particular, this implies for $\bm c\ne \bm c'$ that $(\tilde{\mathbb
E}_{\bm c}^{\bm c'})^K$ decays exponentially in $K$.  For $\bm c=\bm c'$,
pick the largest contiguous block $F_s$ within $(\tilde{\mathbb E}_{\bm
c}^{\bm c})^K$, and notice that it converges to a rank-$1$ projector onto
its non-degenerate leading eigenvectors, which therefore transform
trivially under the group action. Since $\tilde{\mathbb E}_{\bm c}^{\bm
c}$ commutes with the symmetry action, further applications of
$\tilde{\mathbb E}_{\bm c}^{\bm c}$ to this rank-$1$ projector do not
change the irrep label of the fixed point.  (The symmetry actions on the
different blocks are related by the corresponding gauge transform $X_s$
and label irreducible representations in the same way; note that we only
care about the symmetry action on the bond degree of freedom of the MPS to
the extent they are related to order parameters, i.e., the symmetry action
on the ``physical'' degrees of freedom.) $(\tilde{\mathbb E}^{\bm c}_{\bm
c})^K$ will generally be a sum over terms in which different
blocks $F_s$ with $|\gamma_s|=1$ converge to their fixed point, and thus,
$(\tilde{\mathbb E}^{\bm c}_{\bm
c})^K\rightarrow\sum_i\vket{\sigma_R^i}\vbra{\sigma_L^i}$, where all
$\sigma_\bullet^i$ transform trivially under the symmetry.  (Though it is
not relevant in what follows, it is worth noting that terms containing more
than one block $F_s$ which converges to the fixed point cannot appear in
any expectation value, since their weight grows linearly with the system
size, whereas in the normalization only single blocks can show up.)

\subsection{Application to dressed endpoints
\label{sec:app:dressed:nogo}}

Let us verify that the modified expectation value Eq.~(\ref{eq:app2-Occ})
satisfies the same Conditions as before.

Condition~\ref{condition:1-structure-of-H} is only about the symmetry
breaking pattern (and does not involve anyon strings), and is thus
entirely unaffected.

The off-diagonal terms in the expectation value Eq.~(\ref{eq:app2-Occ})
again vanish, since the corresponding large power of the off-diagonal
transfer operator will decay as the largest eigenvalue of the mixed
transfer operator $\mathbb E^{\bm c}_{\bm c'}$, and thus faster than the
diagonal terms, as we will see. 

Also, from Eq.~(\ref{eq:app2-Occ}) one can immediately infer that the
expectation value is independent of the fixed point chosen, using again
the same argument as in Eq.~(\ref{eq:sec3:Occ-indep-of-c}).

Next, in analogy to Eq.~\eqref{eq:sop-in-mps}, let us consider what
happens when we separate a pair of anyons. If $\bm g\not\in\bm H$, we again
obtain a mixed transfer operator and thus the corresponding expectation
value vanishes, yielding
Condition~\ref{condition:2-domainwall-strings-vanish}.
If $\bm g\in \bm H$, we can again move the symmetry action to the bond degree
of freedom of the MPS, and are thus left with 
\begin{equation}
\label{eq:app2:string-expval}
    \raisebox{-1.3em}{\includegraphics[scale=0.6]{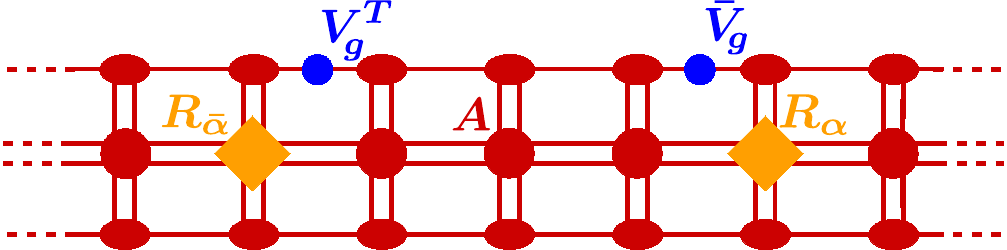}}
\quad.
\end{equation}
As discussed above, we have that the fixed point space of the transfer
operator is of the form
\[
    \raisebox{-1.3em}{\includegraphics[scale=0.7]{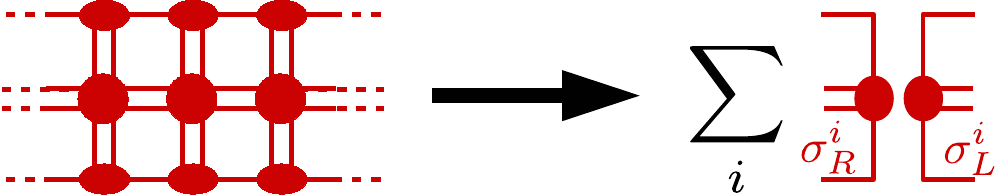}}
\]
where the $\sigma_\bullet^i$ transform trivially under the symmetry
action.  While the endpoints don't decouple any more, we still have that 
the expectation value of Eq.~\eqref{eq:app2:string-expval}
converges to an average over products of expectation values
\[
\raisebox{-1.4em}{
\includegraphics[scale=0.6]{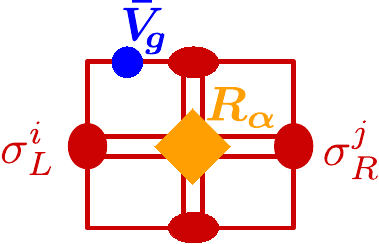}}\quad .
\]
We can now follow the same reasoning as before: Using that 
\[
\includegraphics[scale=0.6]{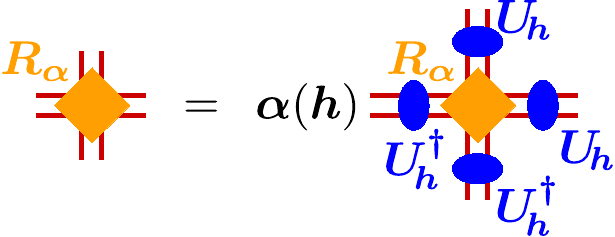}\quad ,
\]
we have that 
\[
\raisebox{-2em}{\includegraphics[scale=0.6]{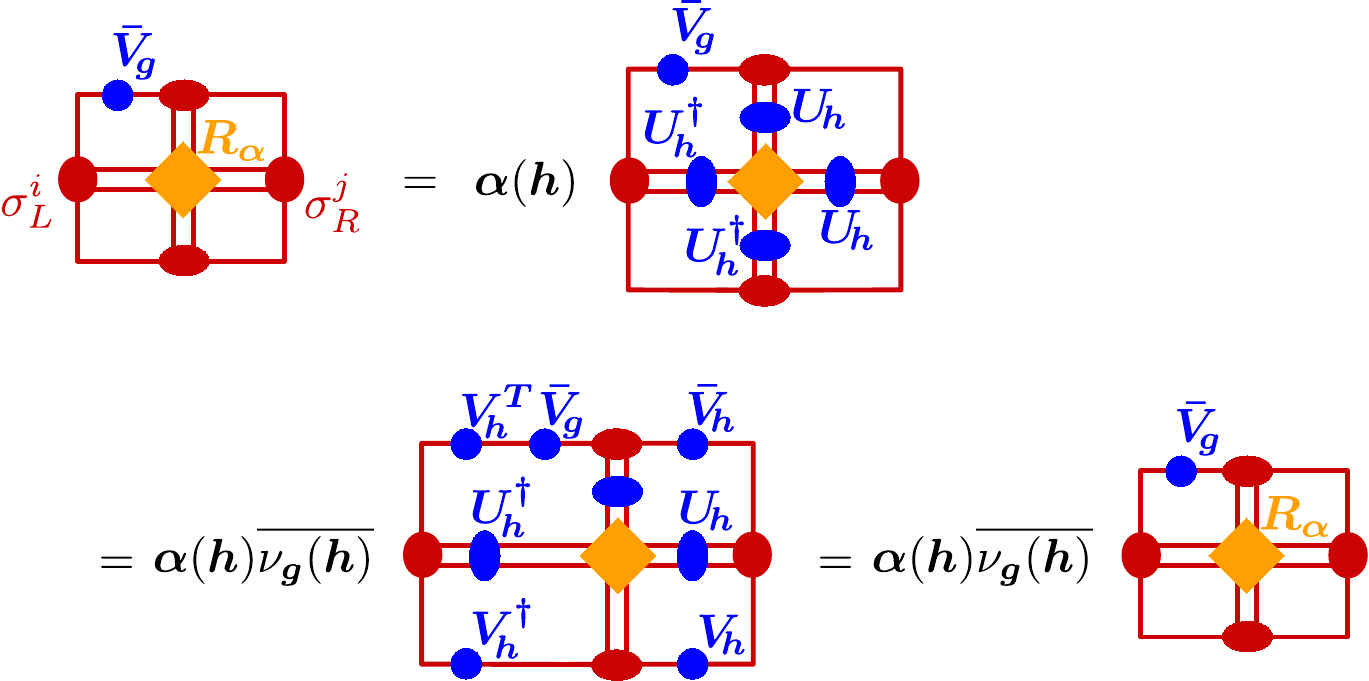}}
\ ,
\]
which shows that Eq.~(\ref{eq:app2:string-expval}) can only be
non-vanishing if $\bm\alpha(\bm h)=\nu_{\bm g}(\bm h)$, yielding
Condition~\ref{condition:3-alpha-nu_g}.  Note that the converse --
that there exists a suitable $\bm R_{\bm \alpha}$ whenever the condition is
satisfied -- has already been shown in
Appendix~\ref{app:explicit-endpoints}.

Finally, the proof of Condition~\ref{condition:4-proj-reps-commute} does
not make explicit reference to the form of the anyons, but just uses the
restrictions on $\langle\strkDblSep{g}{\alpha}{\ell}\rangle$,
Eq.~(\ref{eq:app2:string-expval}), obtained in
Condition~\ref{condition:3-alpha-nu_g}.

\section{\label{appendix:twisted-Zt}
Realization of all twisted $\mathbb Z_t$ double models through
condensation from $D(\mathbb Z_N)$}

In this Appendix we will discuss PEPS tensors which describe
$\mathbb{Z}_t$ twisted quantum
doubles~\cite{buerschaper:twisted-injectivity,bultinck:mpo-anyons,propitius:phd-thesis} 
 with twist $r \in [0,t-1]$.  We will
show that the PEPS tensors have symmetry $\mathbb{Z}_N$, where $N=qt$ and
$q=t/\text{gcd}(t,r)$. We will explicitly construct the fixed points of
the transfer matrix and show that their residual symmetry is given by
$\mathbb{Z}_{qt}\times \mathbb{Z}_q$ and that their entanglement structure corresponds to
the second cohomology class $\frac{rq}{t}$.\\

We start of with defining a right and left handed building blocks for a matrix product operator, referred to as MPO tensors, $M(a)_{ij}$ and $N(a)_{ij}$, i.e. for each $a\in[0,t-1]$ and $i,j\in[0,q-1]$ we define a $qt\times qt$ dimensional matrix,. The non-zero matrix elements of $M(a)_{g_0h_0}$ are given by $[M(a)_{g_0h}]_{\{g,h\},\{g+a,h+a\}}  = \sqrt{\frac{q}{t}} \omega(a,g,h-g_0)\delta_{g_0\equiv g}$, with $g\in[0,t-1]$ and $g_0,h\in[0,q-1]$. Here $\omega$ is a 3-cocycle which we define below and $\delta_{g_0\equiv g}$ is unity if $g_0 = g$ mod $q$, zero otherwise. Note that the subscripts $g+a$ and $h+a$ can be greater than $t$ and $q$, respectively. Here and in the following we will implicitly use modulo $t$ or $q$ when calculating indices which can only take values smaller than $t$ or $q$ respectively. We will use subscript $0$ to distinguish between a variable modulo $t$ and $q$ if both are used in the same equation, i.e. as in the definition of $[M(a)_{g_0h}]_{\{g,h\},\{g+a,h+a\}}$ for $g$. The left and right handed MPO tensors are related by $N(a)_{h_0g_0} = \overline{M(a)_{g_0h_0}}$ (bar denotes complex conjugation). The non-zero values of $M(a)_{g_0h}$ and $N(a)_{hg_0}$ can also be depicted graphically by:
\begin{align} \label{eq:app_MPO1}
M(a)_{g_0h} =\sqrt{\frac{q}{t}}& 
 \begin{split}
 \includegraphics{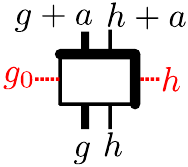}
 \end{split}\ \omega(a,g,h-g_0)\ \ , \\ \label{eq:app_MPO2}
 N(a)_{hg_0} = \sqrt{\frac{q}{t}}& \begin{split}
 \includegraphics{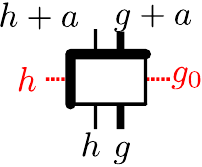}
 \end{split}\ \overline{\omega(a,g,h-g_0)}\ \ ,
\end{align}
where at the r.h.s. $g$ can take any value satisfying $g_0 = g$ mod $q$. The horizontal (red, dotted) legs correspond to the indices of $M(a)$ and $N(a)$ , the vertical legs correspond to the indices of the matrices $M(a)_{g_0h_0}$  and $M(a)_{g_0h_0}$. The thick leg ($t$ dim) and the thin leg ($q$ dim) together form a $qt$ dimensional space. The thick edges of the box indicate its orientation and is also used to distinguish $M$ from $N$. The 3-cocycle $\omega$ is defined by:
\begin{align}\label{eq:app_cocyle}
 \omega(a,g,d)  = \exp[\frac{2\pi i r d}{t^2}(a+g-\lfloor a+g \rfloor)]  \ \ ,
\end{align}
where $\lfloor \cdot \rfloor$ denotes modulo $t$ and $r\in [0,t-1]$
specifies the specifies the class of the 3-cocycle. Note that this gauge
differs (by a co-boundary) from the one defined in
Ref.~\onlinecite{propitius:phd-thesis}. This cocycle has the following
invariant $\omega(a,g,d) = \omega(a,g,d+q)$, and satisfies the following
cocycle condition: \begin{align}\label{eq:app_cocylecondition}\nonumber
 \omega&(g_1,g_2,g_3)\omega(g_1,g_2+g_3,g_4)\omega(g_2,g_3,g_4) \\
 &= \omega(g_1+g_2,g_3,g_4)\omega(g_1,g_2,g_3+g_4) \ \ .
\end{align}
for any set of $g_i$'s. We use two copies of $N(a)$ and $M(a)$ to construct the map $A(a) = \sum_{ijkl} M(a)_{ij}\otimes M(a)_{jk}\otimes N(a)_{kl}\otimes N(a)_{li}$. The PEPS tensor is linear combination of these maps: $A =\frac{t}{q^2} \sum_a A(a)$ which can be graphically represented by 
\begin{align}\label{eq:app_PEPS}
\sum_a\begin{split}
 \includegraphics{figs/plaatjes4-crop} \ \ .
 \end{split}
\end{align}
The inner legs correspond to a physical site and the outer four groups of two legs correspond to the four auxiliary sites.

\subsection{\label{app:twdbl:subsec-equiv-twdbl}%
Twisted Quantum double}
The above defined PEPS tensor can also be obtained by starting from MPO
tensors for the twisted double defined in Ref.~\onlinecite{bultinck:mpo-anyons}:
\begin{align} M(a)_{gh} =& 
 \begin{split}
 \includegraphics{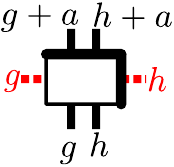}
 \end{split}\ \omega(a,g,h-g)\ \ ,\\
 N(a)_{hg} =& \begin{split}
 \includegraphics{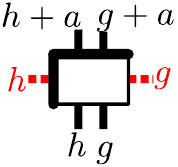}
 \end{split}\ \overline{\omega(a,g,h-g)}\ \ .
\end{align}
All legs (thick) correspond to a $t$ dimensional space. As indicated by expression \eqref{eq:app_PEPS} these MPO tensors can also be used to create a PEPS tensor. We will now discuss two unitaries which can be used to relate the state described by the above MPO tensors to the state described by the MPO tensors given by Eqs.~\eqref{eq:app_MPO1} and \eqref{eq:app_MPO2}. First consider $U$ acting on a $t^5$ dimensional space with non-zero matrix entries $U_{ijpqm,ijklm} = \delta_{p,k-(x-x_0)}\delta_{q,l-(m-m_0)}$, where $x = i+l-j$. Acting with this unitary on the following physical sites
\begin{align}
 \begin{split}
    {U} \begin{split} \includegraphics{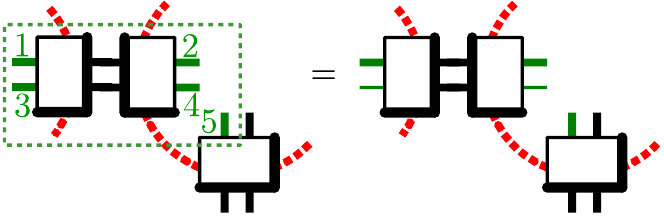}\end{split} \ \ ,
 \end{split}
\end{align}
gives a state whose reduced density matrix for the physical legs indicated by a thin leg has support on only a $q$ dimensional space, i.e. spanned by the first $q$ vectors of the computational basis. Acting with multiple copies of this unitary (one for each PEPS link) one can effectively reduce the dimension of the on-site Hilbert space from $t^8$ to $(qt)^4$. Note that the order does not matter since $U$ acts diagonally on overlapping sites. In the following three steps we successively reduce the entanglement space  indicated by the green arrow: 
\begin{align}
 \begin{split}
  \includegraphics{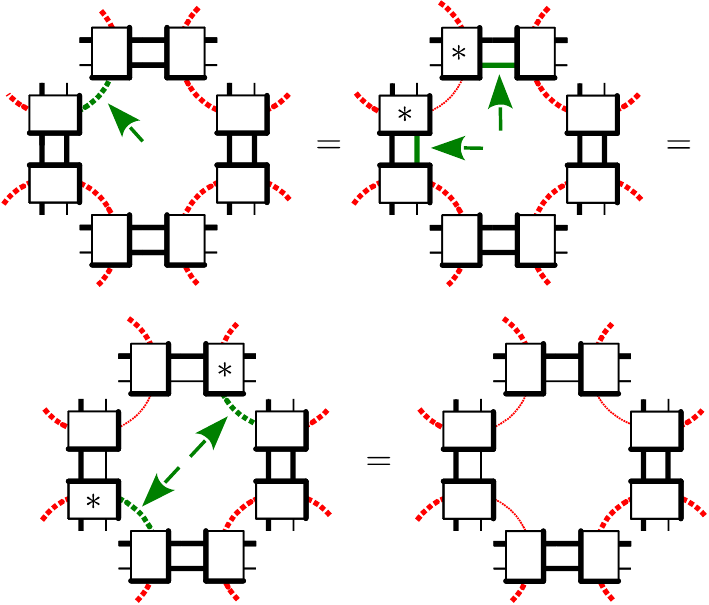}\ \ .
 \end{split}
\end{align}
The first reduction is valid since these indices are also coupled through the remaining 6 MPO tensors (we could actually have removed this link completely). In the second and third step we make use of the invariance of the 3-cocylce $\omega(a,g,d) = \omega(a,g,d+q)$. The MPO tensor labeled by $*$, only depends on the indicated index modulo $q$. Applying this reduction to all plaquettes almost gives the model arising from the PEPS defined in Eq.~\eqref{eq:app_PEPS}, except for the entanglement space between the upper and left MPO tensors of the PEPS tensor still being $t$-dimensional. A second unitary $\tilde{U}$ will reduce this entanglement. It acts on a $t^3q$ dimensional space as:
\begin{align}
 \begin{split}
\includegraphics{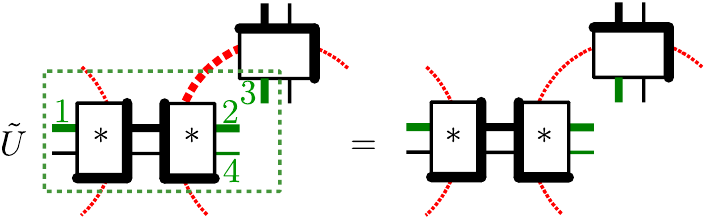}\ \ ,
 \end{split}
\end{align}
 and has non-zero matrix entries $\tilde{U}_{\{g_1+kq,g_2+kq,g_3,g_4\},\{g_1,g_2,g_3,g_4\}} = F(\frac{\lfloor g_2+kq-g_3\rfloor}{q},\frac{\lfloor g_2-g_3\rfloor}{q}) \omega(g_3,kq,g_4-g_3) \overline{\omega(g_1,kq,g_4-g_3)}$, where again $\lfloor \cdot \rfloor$ denotes modulo $t$ and $F(a,b) = \sqrt{\frac{q}{t}} \exp[\frac{2abq\pi i }{t}]$. $\tilde{U}$ is unitary since $F$ is a Fourier transform in the difference between second and third index (mod $q$). Before applying this unitary the reduced density matrix $\rho_{23}$ of the state, corresponding to the sites labeled by 2 and 3 in the above equation, is a maximally mixed state, whereas after applying this unitary, $\rho_{23}$ has Schmidt rank $q$. The corresponding 3-cocycles in the  definition for $\tilde{U}$ ensure that after disentangling, the MPO tensors labeled by $*$ in the above equation, still have the right phase factor. Indeed we have that $\tilde{U} {\omega(a+h-h',h',g-h)}\overline{\omega(a,h,g-h)} \ket{h',h,h,g} = \sum_k\alpha_k\ket{h'+kq,h+kq,h,g}$ where $\alpha_k$ is given by:
\begin{align}\nonumber
 \alpha_k =  &\overline{\omega(h',kq,g-h)}{\omega(a+h-h',h',kq+g-h)}\cdot \\ \nonumber
 &\omega(h,kq,g-h)\overline{\omega(a,h,kq+g-h)} \\\nonumber
 = &\overline{\omega(a+h,kq,g-h)}{\omega(a+h-h',h'+kq,g-h)} \cdot \\ \nonumber
&{\omega(a+h-h',h',kq)} \cdot \\ \nonumber
& \omega(a+h,kq,g-h)\overline{\omega(a,h+kq,g-h)}\overline{\omega(a,h,kq)} \cdot\\\nonumber
=& \overline{\omega(a,h+kq,g-h)}{\omega(a+h-h',h'+kq,g-h)} \ \ ,
\end{align}
for any integer $a$. The second equation follows after applying the cocycle condition, Eq.~\eqref{eq:app_cocylecondition}, twice: once with $g_1 = a+h-h'$, $g_2= h'$, $g_3=kq$ and $g_4=g-h$ and once with $g_1 = a$, $g_2= h$, $g_3=kq$ and $g_4=g-h$. The first and third equality follow from the invariance $\omega(a,g,d) = \omega(a,g,d+q)$.

\subsection{Properties}
We will shows that the tensor constructed is a projector: $A^\dagger = A^2 = A$. Both properties can be studied on the level of the MPO tensors $M(a)$ and $N(a)$. First of we have that $M(-a)_{ij}$ and $M(a)_{ij}^\dagger$ are related by a gauge transformation:
\begin{align}\label{eq:app_Mtranspose}
M(a)_{ij}^\dagger = \sum_{kl} Q(a)_{ik}M(-a)_{kl}\overline{Q(a)_{jl}} \ \ . 
\end{align}
The matrix entries of $Q(a)$ are given by the cocycle defined in Eq.~\eqref{eq:app_cocyle}: $Q(a)_{ij} = \omega(-a,a,i)\delta_{i+a_0,j}$. The above relation follows from the cocycle condition: non-zero entries on the l.h.s. are $\overline{\omega(a,g,j-i)}$ (for matrix indices $[\{g+a,j+a\},\{g,j\}]$ with $g_0=i$), the corresponding matrix entries on the r.h.s. are $\omega(-a,a,i)\omega(-a,a+g,j-i)\overline{\omega(-a,a,j)}$. These are equal by Eq.~\eqref{eq:app_cocylecondition} using $g_1=-a$, $g_2=a$, $g_3=g$ and $g_4=j-i$, and from the fact that in the chosen gauge for $\omega$ we have that $\omega(0,g_3,g_4) = 1$. The same equation can also be derived for $N(a)$. Since $Q$ obeys $\sum_j\overline{Q(a)_{ij}}Q(a)_{kj} = \delta_{ik}$ it follows that the tensor $A$ is Hermitian.\\

The product of two MPO tensors $M(a)$ and $M(b)$ is related to the MPO
tensor $M(a+b)$ by a gauge transformation:  $\sum_{mn}
Z(a,b)_{i,mn} (M(b)_{mk}\cdot M(a)_{nl}) = \sqrt{q/t}\sum_{j}
M(a+b)_{ij} Z(a,b)_{j,kl}$ where $Z(a,b)$ is a $q\times
q^2$ matrix whose non-zero entries are given by $Z(a,b)_{i,kl} =
\omega(a,b,i)\delta_{i,k}\delta_{i+b_0,l}$. This equation can also be
represented graphically as:
\begin{align}\label{eq:app_zipper0}
 \begin{split}
  \includegraphics{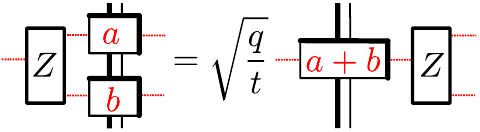}
 \end{split} \ \ .
\end{align}
 This relation follows again from the cocycle condition, Eq.~\eqref{eq:app_cocylecondition}: i.e. one can verify that $\omega(a,b,i)\omega(b,g,k-i)\omega(a,b+g,k-i) = \omega(a+b,g,k-i)\omega(a,b,k)$, being the entry-wise equation for the above relation for matrix entries $[\{g,k\}, \{g+a+b,k+a+b\}]$ with $g_0=i$. The zipper $Z$ obeys  $\sum_{kl} Z(a,b)_{i,kl} \overline{Z(a,b)_{j,kl}} =  \delta_{ij}$ and $\sum_{i} \overline{Z(a,b)_{i,kl}} Z(a,b)_{i,mn} = \delta_{mk}\delta_{nl}\delta_{m+b_0,n} $ which can be represented graphically as:
\begin{align} \label{eq:app_zipper1}
 \begin{split}
 \includegraphics{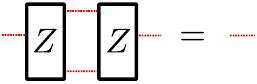}
 \end{split} \ \ ,\\ \label{eq:app_zipper2}
 \begin{split}
 \includegraphics{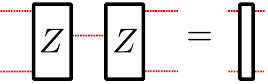}
 \end{split}\ \ .
\end{align}
Note that the product of two zippers, given by Eq.~\eqref{eq:app_zipper2}, is not equal to identity but rather equal to a projector. These equations are used in showing that $A^2=A$. To see this one first uses zippers $Z(a,b)$ to simplify the product $A(b)A(a)$ to $\frac{q^2}{t^2} A(a+b)$ which can best be graphically explained:
\begin{align}
  \includegraphics{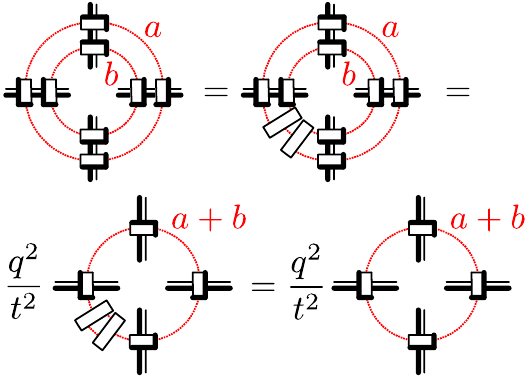}
\end{align}
In the first equation one used Eq.~\ref{eq:app_zipper2} to insert two zippers. Although this product of zippers is a projector rather than identity, this equation is still valid since the support of this projector contains the image of the product of two MPO tensors. Eq.~\ref{eq:app_zipper0} is used to move one of the two zippers along the string of MPO tensors after which Eq.~\ref{eq:app_zipper1} is used to remove the two zippers. Using this equation one can show that $A^2  = \frac{t^2}{q^4}\sum_{ab}A(b)A(a) = \frac{1}{q^2}\sum_{ab} A(a+b) = \frac{t}{q^2}\sum_{c} A(c) = A$.  This motivates the pre-factor of $\frac{t}{q^2}$ in the definition of $A$.\\

The last property of the PEPS-tensor we will discuss in this section is that the corresponding transfer-matrix can be constructed from the MPO tensors: $T = \sum{|r_a)(l_a|}$, where $|r_a)$ and $|l_a)$ are given by
\begin{align}
 |r_a) &= \sum_{\{i_n\}} M(a)_{i_1i_2}\otimes M(a)_{i_2i_3} \otimes \dots \otimes M(a)_{i_Li_1} \ \ ,\\
|l_a) &= \sum_{\{i_n\}} N(a)_{i_1i_2}\otimes N(a)_{i_2i_3} \otimes \dots \otimes N(a)_{i_Li_1} \ \ .
\end{align}
The crucial step in deriving this statement is that the product of two MPO building blocks reduces to a delta function: $\text{Tr} M(a)_{ij}N(b)_{nm}^T  = \delta_{ab}\delta_{in}\delta_{jm}$. Or graphically:
\begin{align}
 \begin{split}
   \includegraphics{figs/plaatjes19-crop}
 \end{split} \ \delta_{ab} \ \ .
\end{align}
This motivates the factor of $\sqrt{\frac{q}{t}}$ in the definition of the MPO tensors. Moreover, it can be used to show that the left and right eigenvectors are orthogonal: $(l_a|r_b) \propto \delta_{ab}$. Using the above equation one can graphically derive the fixed points of T as follows:
\begin{align}
 \begin{split}
   \includegraphics{figs/plaatjes16-crop}
 \end{split} \ \ .
\end{align}

\subsection{Symmetries}

In this section we show that the PEPS tensor $A$ has a $\mathbb{Z}_N$ symmetry, where $N=qt$ (hence the transfer matrix $T$ has a $\mathbb{Z}_N\times \mathbb{Z}_N$ symmetry), and that the fixed points of the the transfer matrix break this symmetry to $\mathbb{Z}_{qt}\times \mathbb{Z}_{q}$. Moreover, the remaining symmetry acts projectively on the auxiliary space of the fixed points (being the MPO-string space). To derive these statements we introduce a unitary $S$ which relates $M(a)$ and $M(a+1)$ up to a gauge transformation: $M(a)_{ij} S = \sum_{kl}U(a)_{ik}M(a+1)_{kl} \overline{U(a)_{jl}}$. Similarly, by combining this with Eq.~\eqref{eq:app_Mtranspose} it follows that  $S^\dagger M(a)$ is related to $M(a-1)$ as $ S^\dagger M(a)_{ij} = \sum_{kl}V(a)_{ik}M(a-1)_{kl} \overline{V(a)_{jl}} $ where $V(a) = \overline{Q(a)}\overline{U(-a)}Q(-a+1)$. Both equations can be represented graphically: 
 \begin{align}
 \begin{split}
 \includegraphics{figs/plaatjes18-crop}
 \end{split} \ \ \text{and} 
 \begin{split}
 \includegraphics{figs/plaatjes17-crop}
 \end{split}\ \ .
\end{align}
The symmetry $S$ is defined by $S_{(i_1,i_2),(j_1,j_2)} = \delta_{i_1+1,j_1}\delta_{i_2+1,j_2}\omega(1,i_1,i_2-i_1$) and the gauge transformation $U$ is defined by $U(a)_{ij} = \delta_{ij}\overline{\omega(1,a,i)}$. Note that $S$ is independent of $a$. The equations follows from the cocycle condition $\omega(1,a+g,d)\omega(a,g,d) = \overline{\omega(1,a,g)}\omega(1+a,g,d)\overline{\omega(1,a,g+d)}$.\\

To see that $S$ is a generator of $\mathbb{Z}_N$ we evaluate $S^t$. It is
a diagonal matrix,  $(S^t)_{(i,i+d),(i,i+d)} = \prod_{j=1}^t \omega(1,j,d)
= \exp[\frac{2\pi idr}{t}] = \exp[\frac{2\pi
id}{q}\frac{r}{\text{gcd}(t,r)}]$, whose matrix entries are $q$-th roots
of unity, which are moreover primitive if gcd($d,q)=1$ (for example,
$d=1$). Thus $S^N = 1$ and $N$ is the smallest exponent for which this is
the case. Both $S$ and $S^\dagger$ are symmetries of the tensor $A$ and of
the transfer matrix. They are not symmetries of the fixed points  $|r_a)$
and $|l_a)$. Only the global action of $S\otimes S^\dagger$ and the global
action of $S^t\otimes \mathbb{I}$ are symmetries of the fixed points, and
they generate the group  $\mathbb{Z}_{qt}\times \mathbb{Z}_q$. Their
action on an MPO tensor is given by $S^\dagger M(a)_{ij}S =
\sum_{kl}{P_1}_{ik}M(a)_{kl}\overline{{P_1}_{jl}}$ and  $M(a)_{ij}S^t =
\sum_{kl}{P_2}_{ik}M(a)_{kl}\overline{{P_2}_{jl}}$ The corresponding gauge
transformations are $P_1(a) = U(a)V(a+1)$ and $P_2(a) =
\prod_{i=a}^{a+t-1} U(i)$. The later of these two gauge transformations is
most easily analyzed since $U(a)$ is diagonal: $P_2(a)_{nn} =
\prod_{m=a}^{a+t-1}\overline{\omega(1,m,n)} = \exp[-\frac{2\pi i
n}{q}\frac{r}{\text{gcd}(t,r)}]$. Hence $P_2(a)$ is independent of $a$ and
is (up to a permutation) the generalized Pauli $Z$ matrix in
$\mathbb{Z}_q$. The non-zero matrix entries of the other gauge
transformation, being $P_1(a)_{n,n+1}$ are all equal, independent of $n$,
due to:
\begin{align}\nonumber
 & \overline{\omega(1,a,n)}\overline{\omega(-a-1,a+1,n)}\\ \nonumber
&\ \ \ \cdot {\omega(1,-a-1,1+a+n)}{\omega(a,-a,a+1+n)}\\ \nonumber
=& \overline{\omega(1,a,n)}\overline{\omega(-a,a+1,n)}\\ \nonumber
&\ \ \ \cdot{\omega(1,-a-1,a+1)}{\omega(a,-a,a+1+n)}\\ \nonumber
=& \overline{\omega(1,a,n)}\omega(a,-a,a+1){\omega(1,-a-1,a+1)}{\omega(a,1,n)}\\ \nonumber
=& \omega(a,-a,a+1){\omega(1,-a-1,a+1)} \ \ .
\end{align}
Here we have used the cocycle condition twice and in the last step we use that in our choice of gauge for $\omega$ we have that $\omega(a,1,n) = \omega(1,a,n)$. Hence, up to a phase $P_1$ is a shift operator which upon conjugation by $P_2$ gives rise to a phase
\begin{align}
 P_1P_2P_1^\dagger P_2^\dagger  = \exp[-\frac{2 \pi i}{q}\frac{r}{\text{gcd}(t,r)}] \ \ .
\end{align}
Thus together $P_1$ and $P_2$ generate a projective representation of $\mathbb{Z}_{qt}\times \mathbb{Z}_q$, and $\frac{r}{\text{gcd}(t,r)}=\frac{rq}{t}$ specifies the corresponding second cohomology class. \\

This family of examples saturates all possible boundary theories of $\mathbb{Z}_N$ invariant PEPS models satisfying the conditions stated in the main text, in which the diagonal symmetry is maximal. In the general case (Condition 1) the residual symmetry is $\mathbb{Z}_{qt}\times \mathbb{Z}_q$ where $qt$ is a merely a divisor of $N$, instead of $qt=N$. However, by increasing the dimension of the auxiliary space with a factor of $x=N/(qt)$ one could simply add extra trivial symmetry to the PEPS tensor which would imply extra symmetry of the transfer matrix. The fixed points will break this extra symmetry because they do not have support on the added auxiliary space, and hence the residual symmetry is still $\mathbb{Z}_{qt}\times \mathbb{Z}_q$.


\begin{thebibliography}{10}

\bibitem{bais:anyon-condensation}
F.~Bais and J.~Slingerland,
\newblock Phys. Rev. B {\bf 79}, 045316 (2009), arXiv:0808.0627.

\bibitem{bais:quantum-sym-breaking}
F.~A. Bais, B.~J. Schroers, and J.~K. Slingerland,
\newblock Phys.Rev.Lett. {\bf 89}, 181601 (2002), hep-th/0205117.

\bibitem{bais:hopf-symmetry-breaking-jhep}
F.~Bais, B.~Schroers, and J.~Slingerland,
\newblock JHEP {\bf 305}, 068 (2003), hep-th/0205114.

\bibitem{kitaev:gapped-boundaries}
A.~{Kitaev} and L.~{Kong},
\newblock Commun. Math. Phys. {\bf 313}, 351 (2012), 1104.5047.

\bibitem{kong:anyon-condensation-tensor-categories}
L.~Kong,
\newblock Nucl. Phys. B {\bf 886}, 436 (2014), arXiv:1307.8244.

\bibitem{verstraete:mbc-peps}
F.~Verstraete and J.~I. Cirac,
\newblock Phys.~Rev.~A {\bf 70}, 060302 (2004), quant-ph/0311130.

\bibitem{buerschaper:stringnet-peps}
O.~Buerschaper, M.~Aguado, and G.~Vidal,
\newblock Phys.\ Rev. B {\bf 79}, 085119 (2009), arXiv:0809.2393.

\bibitem{gu:stringnet-peps}
Z.-C. Gu, M.~Levin, B.~Swingle, and X.-G. Wen,
\newblock Phys. Rev. B {\bf 79}, 085118 (2009), arXiv:0809.2821.

\bibitem{schuch:peps-sym}
N.~{Schuch}, I.~{Cirac}, and D.~{P{\'e}rez-Garc{\'{\i}}a},
\newblock Ann. Phys. {\bf 325}, 2153 (2010), arXiv:1001.3807.

\bibitem{buerschaper:twisted-injectivity}
O.~Buerschaper,
\newblock Ann. Phys. {\bf 351}, 447 (2014), arXiv:1307.7763.

\bibitem{sahinoglu:mpo-injectivity}
M.~B. Sahinoglu {\em et~al.},
\newblock (2014), arXiv:1409.2150.

\bibitem{bultinck:mpo-anyons}
N.~{Bultinck} {\em et~al.},
\newblock (2015), arXiv:1511.08090.

\bibitem{schuch:topo-top}
N.~Schuch, D.~Poilblanc, J.~I. Cirac, and D.~Perez-Garcia,
\newblock Phys. Rev. Lett. {\bf 111}, 090501 (2013), arXiv:1210.5601.

\bibitem{haegeman:shadows}
J.~Haegeman, V.~Zauner, N.~Schuch, and F.~Verstraete,
\newblock Nature Comm. {\bf 6}, 8284 (2015), arXiv:1410.5443.

\bibitem{marien:fibonacci-condensation}
M.~Marien, J.~Haegeman, P.~Fendley, and F.~Verstraete,
\newblock 1607.05296v1.

\bibitem{fernandez:symmetrized-tcode}
C.~Fernandez-Gonzalez, R.~S.~K. Mong, O.~Landon-Cardinal, D.~Perez-Garcia, and
  N.~Schuch,
\newblock Phys. Rev. B {\bf 94}, 155106 (2016), arXiv:1608.00594.

\bibitem{perez-garcia:parent-ham-2d}
D.~Perez-Garcia, F.~Verstraete, J.~I. Cirac, and M.~M. Wolf,
\newblock Quantum Inf. Comput. {\bf 8}, 0650 (2008), arXiv:0707.2260.

\bibitem{kitaev:toriccode}
A.~Kitaev,
\newblock Ann. Phys. {\bf 303}, 2 (2003), quant-ph/9707021.

\bibitem{castelnovo:tc-tension-topoentropy}
C.~Castelnovo and C.~Chamon,
\newblock Phys. Rev. B {\bf 77}, 054433 (2008), arXiv:0707.2084.

\bibitem{cirac:peps-boundaries}
J.~I. Cirac, D.~Poilblanc, N.~Schuch, and F.~Verstraete,
\newblock Phys. Rev. B {\bf 83}, 245134 (2011), arXiv:1103.3427.

\bibitem{chen:1d-phases-rg}
X.~{Chen}, Z.~{Gu}, and X.~{Wen},
\newblock Phys. Rev. B {\bf 83}, 035107 (2011), arXiv:1008.3745.

\bibitem{schuch:mps-phases}
N.~{Schuch}, D.~{Perez-Garcia}, and I.~{Cirac},
\newblock Phys. Rev. B {\bf 84}, 165139 (2011), arXiv:1010.3732.

\bibitem{hastings:arealaw}
M.~Hastings,
\newblock J. Stat. Mech. , P08024 (2007), arXiv:0705.2024.

\bibitem{verstraete:faithfully}
F.~Verstraete and J.~I. Cirac,
\newblock Phys. Rev. B {\bf 73}, 094423 (2006), cond-mat/0505140.

\bibitem{schuch:mps-entropies}
N.~Schuch, M.~M. Wolf, F.~Verstraete, and J.~I. Cirac,
\newblock Phys.\ Rev.\ Lett. {\bf 100}, 30504 (2008), arXiv:0705.0292.

\bibitem{perez-garcia:mps-reps}
D.~Perez-Garcia, F.~Verstraete, M.~M. Wolf, and J.~I. Cirac,
\newblock Quant.\ Inf.\ Comput. {\bf 7}, 401 (2007), quant-ph/0608197.

\bibitem{cirac:mpdo-rgfp}
J.~Cirac, D.~Perez-Garcia, N.~Schuch, and F.~Verstraete,
\newblock (2016), arXiv:1606.00608.

\bibitem{Note1}
Note that this does not imply that the fixed point space is actually spanned by
  the ${\delimiter "026A30C \rho _{\protect \bm {c}})}$. In fact, it is easy to
  see that this would require extra conditions such as rotational invariance,
  since e.g.\ a transfer operator projecting onto a GHZ-type state would have a
  unique fixed point (the GHZ state) which is not an injective MPS.

\bibitem{rispler:peps-symmetrybreaking}
M.~Rispler, K.~Duivenvoorden, and N.~Schuch,
\newblock Phys. Rev. B {\bf 92}, 155133 (2015), arXiv:1505.04217.

\bibitem{perez-garcia:inj-peps-syms}
D.~Perez-Garcia, M.~Sanz, C.~E. Gonzalez-Guillen, M.~M. Wolf, and J.~I. Cirac,
\newblock New J. Phys. {\bf 12}, 025010 (2010), arXiv.org:0908.1674.

\bibitem{fernandez-gonzalez:uncle-long}
C.~Fernandez-Gonzalez, N.~Schuch, M.~M. Wolf, J.~I. Cirac, and D.~Perez-Garcia,
\newblock Commun.\ Math.\ Phys. {\bf 333}, 299 (2015), arXiv:1210.6613.

\bibitem{sanz:mps-syms}
M.~Sanz, M.~M. Wolf, D.~Perez-Garcia, and J.~I. Cirac,
\newblock Phys. Rev. A {\bf 79}, 042308 (2009), arXiv:0901.2223.

\bibitem{pollmann:symprot-1d}
F.~Pollmann, E.~Berg, A.~M. Turner, and M.~Oshikawa,
\newblock (2009), arXiv.org:0909.4059.

\bibitem{chen:spt-order-and-cohomology}
X.~Chen, Z.-C. Gu, Z.-X. Liu, and X.-G. Wen,
\newblock Phys. Rev. B {\bf 87}, 155114 (2013), arXiv:1106.4772.

\bibitem{propitius:phd-thesis}
M.~{de Wild Propitius},
\newblock {\em Topological interactions in broken gauge theories},
\newblock PhD thesis, 1995, arXiv:hep-th/9511195.

\bibitem{Note2}
This can be seen using the cocycle conditions and the fact that the group is
  abelian as follows: \begin {align*} \protect \hspace *{1.5em}\protect \frac
  {\nu _{\protect \bm {h}}(\protect \bm {g}_1) \nu _{\protect \bm {h}}(\protect
  \bm {g}_2)}{\nu _{\protect \bm {h}}(\protect \bm {g}_1\protect \bm {g}_2)}
  &=\protect \frac {\omega (\protect \bm {g}_1,\protect \bm {h})\omega
  (\protect \bm {g}_2,\protect \bm {h}) \omega (\protect \bm {h},\protect \bm
  {g}_1\protect \bm {g}_2)}{ \omega (\protect \bm {h},\protect \bm {g}_1)\omega
  (\protect \bm {h},\protect \bm {g}_2) \omega (\protect \bm {g}_1\protect \bm
  {g}_2,\protect \bm {h})} \protect \tmspace +\thinmuskip {.1667em} \protect
  \frac {\omega (\protect \bm {h} \protect \bm {g}_1,\protect \bm
  {g}_2)}{\omega (\protect \bm {h} \protect \bm {g}_1,\protect \bm {g}_2)} \\
  &=\protect \frac {\omega (\protect \bm {g}_1,\protect \bm {h})\omega
  (\protect \bm {g}_2,\protect \bm {h}) \omega (\protect \bm {h},\protect \bm
  {g}_1\protect \bm {g}_2) \omega (\protect \bm {h} \protect \bm {g}_1,\protect
  \bm {g}_2) }{ \omega (\protect \bm {h}, \protect \bm {g}_1\protect \bm {g}_2)
  \omega (\protect \bm {g}_1,\protect \bm {g}_2) \omega (\protect \bm
  {h},\protect \bm {g}_2) \omega (\protect \bm {g}_1\protect \bm {g}_2,\protect
  \bm {h})} \\ &=\protect \frac {\omega (\protect \bm {g}_1,\protect \bm
  {h})\omega (\protect \bm {g}_2,\protect \bm {h}) \omega (\protect \bm
  {h},\protect \bm {g}_1\protect \bm {g}_2) \omega (\protect \bm {h} \protect
  \bm {g}_1,\protect \bm {g}_2) }{ \omega (\protect \bm {h}, \protect \bm
  {g}_1\protect \bm {g}_2) \omega (\protect \bm {h},\protect \bm {g}_2) \omega
  (\protect \bm {g}_1,\protect \bm {g}_2\protect \bm {h}) \omega (\protect \bm
  {g}_2,\protect \bm {h}) } \\ &=\protect \frac {\omega (\protect \bm
  {g}_1,\protect \bm {h}) \omega (\protect \bm {h} \protect \bm {g}_1,\protect
  \bm {g}_2) }{ \omega (\protect \bm {h},\protect \bm {g}_2) \omega (\protect
  \bm {g}_1,\protect \bm {g}_2\protect \bm {h}) } \\ &=\protect \frac {\omega
  (\protect \bm {g}_1,\protect \bm {h}) \omega (\protect \bm {g}_1 \protect \bm
  {h},\protect \bm {g}_2) }{ \omega (\protect \bm {h},\protect \bm {g}_2)
  \omega (\protect \bm {g}_1,\protect \bm {h}\protect \bm {g}_2) } =1 \ . \end
  {align*}.

\bibitem{pollmann:spt-detection-1d}
F.~Pollmann and A.~M. Turner,
\newblock Phys.\ Rev.\ B {\bf 86}, 125441 (2012), arXiv:1204.0704.

\bibitem{Note3}
Note that the same cannot hold for all abelian groups: Condensing from an
  abelian group gives another abelian model, while twisting an abelian model
  can give rise to non-abelian models~\cite {propitius:phd-thesis}.

\bibitem{iqbal:preparation}
M.~{Iqbal \emph{et al.}},
\newblock in preparation .

\bibitem{zanardi:overlap-curvature}
P.~Zanardi, P.~Giorda, and M.~Cozzini,
\newblock Phys. Rev. Lett. {\bf 99}, 100603 (2007), quant-ph/0701061.

\bibitem{gu:fidelity-phasetransitions}
S.-J. Gu,
\newblock Int. J. Mod. Phys. B {\bf 24}, 4371 (2010), arXiv:0811.3127.

\bibitem{morampudi:z2-phase-transition}
S.~C. Morampudi, C.~von Keyserlingk, and F.~Pollmann,
\newblock Phys. Rev. B {\bf 90}, 035117 (2014), arXiv:1403.0768.

\bibitem{duivenvoorden:stringorder-symbreaking}
K.~Duivenvoorden and T.~Quella,
\newblock Phys. Rev. B {\bf 88}, 125115 (2013), arXiv:1304.7234.

\bibitem{yang:peps-edgetheories}
S.~Yang {\em et~al.},
\newblock Phys. Rev. Lett. {\bf 112}, 036402 (2013), arXiv:1309.4596.

\end{thebibliography}
\end{document}